\documentclass[12pt]{article}
\usepackage{a4wide}
\usepackage[centertags]{amsmath}
\usepackage{amssymb}
\usepackage[sort&compress,numbers]{natbib}
\renewcommand{\baselinestretch}{1.3}
\newlength{\lingap}
\newcommand{\sgap}{\\[-\lingap]}
\newcommand{\preprintno}[3]{\hfill\raisebox{#1}[0cm][0cm]{
\begin{minipage}[t]{#2}\begin{flushright} #3 \end{flushright}\end{minipage}}
\vspace*{-\baselinestretch\baselineskip}}
\newcommand{\half}{\frac{1}{2}}
\newcommand{\CO}{\mathcal{O}}
\newcommand{\CL}{\mathcal{L}}
\newcommand{\CN}{\mathcal{N}}
\newcommand{\CM}{\mathcal{M}}

\newcommand{\p}{\partial}
\newcommand{\apm}{{\alpha^{\prime}}}
\newcommand{\ra}{\rightarrow}
\newcommand{\lr}{\leftrightarrow}
\newcommand{\dr}{\mathrm{d}}
\newcommand{\ir}{\mathrm{i}}
\newcommand{\e}{\mathrm{e}}
\newcommand{\zb}{{\bar z}}
\newcommand{\wb}{{\overline w}}
\newcommand{\cb}{{\bar c}}
\newcommand{\ab}{{\bar a}}
\newcommand{\bb}{{\bar b}}
\newcommand{\bp}{{\bar\partial}}
\newcommand{\CP}{\mathbb{CP}}
\newcommand{\C}{\mathbb{C}}
\newcommand{\Z}{\mathbb{Z}}
\newcommand{\R}{\mathbb{R}}

\title {Supersymmetric states of  $\mathcal{N}=4$
Yang-Mills\sgap from giant gravitons}

\author{Indranil Biswas$^{a}$,
Davide Gaiotto$^{b}$,
Subhaneil Lahiri$^{c,b}$ and
Shiraz Minwalla$^{c,b}$\\
\small{\emph{$^{a}$School of Mathematics,
                   Tata Institute of Fundamental Research,}}\sgap
\small{\emph{Homi Bhabha Rd, Mumbai 400005, India}}\sgap
\small{\emph{$^{b}$Jefferson Physical Laboratory,
                   Harvard University, Cambridge MA 02138, USA}}\sgap
\small{\emph{$^{c}$Department of Theoretical Physics,
                   Tata Institute of Fundamental Research,}}\sgap
\small{\emph{Homi Bhabha Rd, Mumbai 400005, India}} }

\begin{document}

\maketitle

\preprintno{12cm}{6cm}{
    HUTP-06/A0021 \sgap
    \texttt{hep-th/0606087}
}

\begin{abstract}

Mikhailov has constructed an infinite family of $\frac{1}{8}$ BPS
D3-branes in $AdS_5 \times S^5$. We regulate Mikhailov's solution
space by focussing on finite dimensional submanifolds. Our
submanifolds are topologically complex projective spaces with
symplectic form cohomologically equal to $2\pi N$ times the
Fubini-Study K\"ahler class. Upon quantization and removing the
regulator we find the Hilbert Space of $N$ noninteracting Bose
particles in a 3d Harmonic oscillator, a result previously
conjectured by Beasley. This Hilbert Space is isomorphic to the
classical chiral ring of $\frac{1}{8}$ BPS states in $\mathcal{N}=4$
Yang-Mills theory. We view our result as evidence that the spectrum
of $\frac{1}{8}$ BPS states in $\mathcal{N}=4$ Yang Mills theory,
which is known to jump discontinuously from zero to infinitesimal
coupling, receives no further renormalization at finite values of
the \mbox{`t Hooft} coupling.

\end{abstract}

\pagebreak

\tableofcontents


\section{Introduction and summary}\label{intro}

The $U(N)$, $\mathcal{N}=4$ Yang-Mills theories are special in many
ways. These theories occur in fixed lines and are maximally
supersymmetric (enjoying invariance under the 32 supercharge
supergroup $PSU(2,\!2|4)$) for every value of the coupling constant
\cite{Sohnius:1981sn, Howe:1983sr, Brink:1982pd, Seiberg:1988ur}.
They exhibit invariance under $S$ duality \cite{Montonen:1977sn,
Goddard:1976qe, Osborn:1979tq, Sen:1994yi}. In the large $N$ 't
Hooft limit they also admit a dual reformulation as a weakly coupled
string theory, a description which in turn is well approximated by
supergravity on $AdS_5 \times S^5$ at large values of the 't Hooft
coupling $\lambda=g^2N$ \cite{Maldacena:1997re, Gubser:1998bc,
Witten:1998qj, Aharony:1999ti}. Finally the appearance of integrable
structures in recent studies of perturbation theory in the 't Hooft
limit suggests that these theories may be unusually tractable, and
may even be `solvable' in appropriate limits (see, for instance,
\cite{Minahan:2002ve, Beisert:2003yb, Beisert:2004yq,
Staudacher:2004tk, Beisert:2005tm, Janik:2006dc}). In summary
$\mathcal{N}=4$ Yang-Mills are both intensely interesting as well as
unusually tractable and deserve to be thoroughly studied.

The $\mathcal{N}=4$ Yang-Mills theories on $S^3$ possess an infinite
number of special states, distinguished by the fact that they are
annihilated by some fraction of the supersymmetry generators
(together with the Hermitian conjugates of these supercharges). As
the  $\mathcal{N}=4$ theory is special largely because of its high
degree of supersymmetry, the study of its supersymmetric states may
well prove particularly tractable and rewarding.

The simplest supersymmetric states are annihilated by half of the 16
supercharges together with their Hermitian conjugates
\cite{Dolan:2002zh}. In the absence of a phase transition, it
follows immediately from $PSU(2,\!2|4)$ representation theory that
the spectrum of half BPS states cannot change under continuous
variations of a parameter. As a consequence the spectrum of half BPS
spectrum of $\mathcal{N}=4$ Yang-Mills, at all values of the
coupling, may be enumerated by via a simple counting in the free
theory.

Less special supersymmetric states may be annihilated by a quarter
\cite{Ryzhov:2001bp, D'Hoker:2003vf, Beisert:2003tq}, an eighth or a
sixteenth of the supercharges and their Hermitian conjugates. The
spectrum of these states can, in general, vary as a function of the
coupling. Indeed the spectrum of one fourth, one eight and one
sixteenth BPS states in $U(N)$, $\mathcal{N}=4$ Yang-Mills at
infinitesimal coupling is discontinuously smaller than the same
spectrum in the free theory. It was suggested by the authors of
\cite{index}, that this spectrum receives no further renormalization
at finite coupling. According to this conjecture $U(N)$,
$\mathcal{N} =4$ Yang-Mills theories have the same spectrum of
supersymmetric states at every finite, nonzero value of the
Yang-Mills coupling.

The authors of \cite{index} also computed an exact (finite $N$)
formula for the finite coupling partition function over the
classical one eighth BPS chiral ring of $U(N)$ Yang-Mills theory.
The formula in \cite{index} turns out to reproduce the spectrum of
one eighth BPS multi-gravitons in $AdS_5 \times S^5$ (a description
of the theory that is accurate at large $\lambda$) at energies much
smaller than $N$, but disagrees with this spectrum at energies of
order $N$ or higher. Nonetheless the authors of \cite{index}
proposed that their classical chiral ring partition function was
equal, at all finite nonzero values of the 't Hooft coupling,  to
the partition function over one eighth BPS states in $U(N)$,
$\mathcal{N}=4$ Yang Mills theory. This proposal is not in conflict
with the AdS/CFT correspondence. Gravitons propagating with energies
of order $N$ in $AdS_5\times S^5$ blow up into puffed up $D3$
branes; so called giant gravitons \cite{Myers:1999ps,
McGreevy:2000cw, Grisaru:2000zn, Hashimoto:2000zp}. As a consequence
supersymmetric spectrum of IIB theory on $AdS_5 \times S^5$ at
energies of order $N$ or higher, cannot be determined by the
Kaluza-Klein reduction of IIB theory on $AdS_5 \times S^5$, but
instead requires a detailed study of the moduli space of giant
gravitons; the content of this paper.

In a beautiful paper written over five years ago, Mikhailov
\cite{mikhailov} constructed a large class of classical solutions
for $\frac{1}{8}$ BPS giant gravitons in $AdS_5 \times S^5$, which
we will now briefly review. Consider (generically 3 dimensional)
surfaces defined by the intersection of the zero set (i.e.\
pre--image of zero) of any holomorphic function in $\C^3$ with the
five dimensional unit sphere $\sum_{i=1}^3 |z^i|^2 =1$. Mikhailov
demonstrated that D3-branes that wrap any such surface on the $S^5$
of $AdS_5 \times S^5$, with the world volume field strength set to
zero, are supersymmetric. Four years ago, Beasley \cite{Beasley}
already made an insightful conjecture about the quantization of
these solutions. In this paper we perform a detailed study of the
quantization of moduli space of holomorphic surface in $\C^3$ that
intersect the unit sphere, with respect to the symplectic form
determined by the world volume action of the D3-brane; our results
are in complete agreement with Beasley's conjecture. In particular
we demonstrate that the Hilbert space so obtained exactly reproduces
the (appropriate restriction of) the partition function over the
classical chiral ring in \cite{index}. In our opinion, this result
provides significant evidence in support of the finite $N$, finite
$\lambda$ partition function over one eighth BPS states proposed in
\cite{index}.

We highlight physically interesting aspects of our results in the
rest of this subsection, postponing a summary of our technical
results to the next subsection. As we have described above, we have
quantized a manifold of supersymmetric solutions. Now the
quantization of an arbitrary submanifold of solution space certainly
does not, in general,  produce any subspace of the full Hilbert
space. However there is circumstantial evidence (see, for instance,
\cite{Mandal:2005wv, Grant:2005qc, Maoz:2005nk}) that supersymmetric
solutions are special; that the the Hilbert Space obtained from the
quantization of the set of all supersymmetric solutions of a system
is equal to the restriction to supersymmetric states of the full
Hilbert Space. The results of this paper may be taken as further
evidence for this statement. It would be useful and interesting to
have better understanding of this issue.

Mikhailov's solutions do not exhaust the full set of one eighth BPS
configurations of D3-branes for two separate reasons. First there
exists a disconnected branch of one eighth BPS giant D3-brane
configurations, the so called dual giant gravitons. As we will
outline in \S\ref{disc}, as yet unpublished work of G. Mandal and N.
Suryanarayana \cite{Nemani} strongly suggests that the partition
function of \cite{index} may also be obtained by quantizing the
manifold of one eighth BPS dual giant gravitons! Thus the familiar
(but still slightly mysterious) duality between half BPS giants and
dual giants appears to extend to one eighth BPS states. It would be
certainly be interesting to understand this better.

The second incompleteness in Mikhailov's space of supersymmetric
solutions has to do with one eighth BPS D3-brane configurations
which involve world volume fermions and gauge fields. Mikhailov's
construction describes only those $\frac{1}{8}$ BPS D-brane
configurations with worldvolume fermions and gauge fields set to
zero and so the quantization of these solutions produces only the
part of the $\frac{1}{8}$ BPS cohomology constructed from the three
chiral Yang-Mills scalar fields. In order to recover all other
states in the $\frac{1}{8}$ BPS cohomology, one could complete the
construction of all supersymmetric classical solutions with
worldvolume fermions and gauge fields turned on (see
\cite{Kim:2005mw} for work in this direction) and quantize this
manifold of solutions. Although we have not carried out this
procedure, for our purposes it seems almost redundant; as we will
now describe, supergroup representation theory almost guarantees
that this procedure will reproduce all the remaining states in the
cohomology of \cite{index}.

The full set of one eighth BPS states appear in representations of
$SU(2|3)$ subgroup that commutes with the supercharges that
annihilate these states. In Appendix \ref{susy} we demonstrate that
the full $\frac{1}{8}$ BPS cohomology may be obtained by acting on
states built out of the chiral scalars with the generators of the
$SU(2|3)$ subgroup of $PSU(2,\!2|4)$ of supercharges that commute
with those that are annihilated on $\frac{1}{8}$ cohomology. As a
consequence, any quantization procedure that respect $SU(2|3)$
invariance is almost guaranteed to fill out the remaining states in
the cohomology listed in \cite{index}.\footnote{This argument makes
the reasonable assumption that the states obtained from the
quantization of supersymmetric D3-branes with world volume fermions
and gauge fields turned off are all $SU(2|3)$ primaries, and,
moreover, are the only such primaries. A direct demonstration of
this point would allow us to remove the word `almost' in this
paragraph.}

We have quantized the supersymmetric motion of the $D3$-brane on the
$S^5$ of $AdS_5 \times S^5$. We would like to emphasize that the
restriction to the supersymmetric sector retains several of the
complications that have plagued previous attempts at quantizing $p
\geq 2$ branes. In particular smooth variations of the parameters of
the holomorphic polynomial can cause the surface of the $D3$ brane
to undergo topology changing transitions. Our quantization procedure
manages to deal with these transitions in a smooth way (\cite{fuzzy}
made similar remarks in another context). It is possible that a
detailed study of the quantization in the supersymmetric sector will
throw up lessons of relevance to quantization of higher dimensional
surfaces in general.

In this paper we have demonstrated that the quantization of giant
gravitons yields the same BPS spectrum as weakly coupled Yang-Mills.
It would be fascinating if we could see Mikhailov's holomorphic
surfaces emerge more directly from an analysis of the gauge theory
(see \cite{Berenstein:2005aa, Berenstein:2005ek, Berenstein:2005jq,
Berenstein:2006yy} for related work).

Finally, it would be natural, and extremely interesting, to attempt
to extend our work to the quantization of $\frac{1}{16}$
supersymmetric giant gravitons. The potential payoffs of such an
extension are large, as $\frac{1}{16}$ BPS states in $\mathcal{N}=4$
Yang-Mills are much richer and much less well understood than their
$\frac{1}{8}$ BPS counterparts. In particular there exist smooth
$\frac{1}{16}$ BPS black holes in $AdS_5 \times S^5$ (see
\cite{Gutowski:2004ez, Gutowski:2004yv, Chong:2005da, Chong:2005hr,
Kunduri:2006ek} and references therein), whose entropy has not yet
successfully been accounted for (see \cite{index, Berkooz:2006wc}
for a recent discussion).

\subsection{Technical aspects}

In this subsection we will briefly summarize our technical
constructions and results.

We first describe how the D3-brane surfaces described by Mikhailov
may be algebraically parameterized. By the Weierstrass approximation
theorem any surface $f(z^i)=0$ that is holomorphic in an open shell
surrounding the unit sphere in $\C^3$ may be approximated to
arbitrary accuracy (in that neighbourhood) by a sequence of surfaces
$P_m(z^i)=0$ where $P_m(z^i)$ are polynomials in variables $z^i$
($i=1,2,3$). As a consequence the set of all supersymmetric
configurations of D3-branes is generated by the intersections of
arbitrary polynomial surfaces $P(z^i)=0$ with the unit sphere. We
will find it useful, in this paper,  to regulate this set of
surfaces by studying the linear set of polynomials, $P_C$, generated
by arbitrary linear combinations of $n_C$ arbitrarily specified
monomials (we denote the set of 3 tuples $(n_1,n_2,n_3)$ by $C$, so
the set of monomials is $\left\{(z^1)^{n_1}(z^2)^{n_2}(z^3)^{n_3}
\mid \vec{n}\in C\right\}$ and $P_C$ is its linear span)
\footnote{For example $P_{C_k}$ could be the set of all polynomials
in three variables of degree at most $k$; in this case
$n_C=\binom{k+3}{3}= (k+3)(k+2)(k+1)/3!$).}. In this paper we
determine the Hilbert Space $\mathcal{H}_C$ obtained by quantizing
$P(z)=0$ with the unit sphere for $P \in P_C$.

We will now describe the set of intersections of $P(z^i)=0$ (for
$P(z)$ in $P_C$) with the unit sphere, and the associated Hilbert
Space $\mathcal{H}_C$ in more detail. Note that the zero sets of the
polynomials in $P_C$ are left unchanged by an overall rescaling of
the coefficients in the polynomial; as a consequence these zero sets
are in one to one correspondence with $\CP^{n_C-1}$, as already
noted by Beasley \cite{Beasley}. However, not all of these complex
surfaces $\{(z_1,z_2,z_3)\mid P(z_1,z_2,z_3)=0\}$, $P\in P_C$,
intersect the unit five sphere $S^5$. Polynomials $P(z)$ that give
surfaces that do not intersect with $S^5$ do not belong to the space
to be quantized; as a consequence the space of interest is the
projective space of polynomial coefficients  with holes eaten out.
Furthermore even those surfaces, $P(z^i)=0$, that do intersect the
unit 5-sphere are not parameterized in a one to one fashion by
polynomials $P(z^i)$; there exist degenerate families of
polynomials, all of whose surfaces have the same (nonzero)
intersection with the unit $S^5$ \footnote{The `holes' referred to
above are a special case; polynomials that do not intersect the unit
$S^5$ all have equal (empty) intersections with it. }. In addition,
the symplectic form also has singularities along certain surfaces
(e.g.\ where the topology of the intersection changes). Despite all
these apparent complications, we show in sections \ref{symfch} and
\ref{phasespace} below that the space of intersections is extremely
simple. In particular we demonstrate that the space in question is
topologically $\CP^{n_C-1}$ and that, in an appropriate sense, the
symplectic form is well defined, $U(3)$ invariant, everywhere
invertible, and in the cohomology class of $(2\pi
N)\omega_\mathrm{FS}$, where $\omega_\mathrm{FS}$ is the
usual\footnote{Note that the Fubini-Study metric defined in
\cite{gsw2} is a factor of $2\pi$ larger than that in, e.g.\
\cite{Demailly}. We use the latter normalization.} Fubini-Study form
on $\CP^{n_C-1}$. In sections \ref{linearsection}, \ref{singvar} and
\ref{hompoly} we illustrate and illuminate aspects of the
mathematically abstract arguments of sections \ref{symfch} and
\ref{phasespace}  by an independent direct and detailed study of
three specially chosen subfamilies of polynomials $P_C$ (and their
quantization, see below).

It follows almost immediately (see \S\ref{quant}) that
$\mathcal{H}_C$ is isomorphic to the Hilbert space obtained from
holomorphic quantization of $\CP^{n_C-1}$ equipped with the
symplectic form $(2\pi N)\omega_\mathrm{FS}$. It is well known that
the latter may be identified with the set of homogeneous
polynomials, of degree $N$ of the $n_C$ projective coordinates
$\{w_{\vec{n}}\,|\, \vec{n} \in C \}$ of $\CP^{n_C-1}$
($w_{\vec{n}}$ is related to the coefficient of
$(z^1)^{n_1}(z^2)^{n_2}(z^3)^{n_3}$ in the set of polynomials and,
in particular, has the same $U(3)$ transformation properties). After
quantization, the charge operators are $\sum_{{\vec{n}}\in C} n_m
w^{\vec{n}}\partial_{w_{\vec{n}}}$, where $n_m $ is the charge of
$w_{\vec{n}}$ under the $U(1)$ rotation in the $m^{th}$ independent
two-plane $(m=1 \ldots 3)$  in $\R^6=\C^3$. Consequently the charge
of a degree $N$ monomial in $w_{\vec{n}}$'s is simply the sum over
the charges $n_m$, over the $N$ factor $w_{\vec{n}}$'s of the
monomial. Thus $\mathcal{H}_C$ is isomorphic to the Hilbert space of
$N$ identical noninteracting bosons with an $n_C$ dimensional single
particle Hilbert space whose states have charges ${\vec{n}}$.

Note that the regulated Hilbert Spaces ${\cal H}_C$ have the
following inclusivity property. If two 3 tuple sets obey $C_{k_1}
\subset C_{k_2}$ then $\mathcal{H}_{k_1} \subset \mathcal{H}_{k_2}$.
Now consider a sequence of 3 tuple sets $C_{k_i}$ chosen so that
$C_{k_1} \subset C_{k_2}$  for $k_1< k_2$ and with the property that
any particular polynomial $P(z)$ is contained in $P_{C_k}$ for some
large enough $k$. The completion of the direct limit, namely
$\mathcal{H}=\lim_{k \to \infty} \mathcal{H}_k$, is a unique, well
defined (indeed familiar) infinite dimensional Hilbert space, which
may thus be regarded as the Hilbert space of Mikhailov's giant
gravitons.

The maximal Hilbert Space $\mathcal{H}$ is especially familiar. In
this case the single particle Hilbert Space is simply the Hilbert
space of a three dimensional harmonic oscillator, with the 3 $U(1)$
charges identified with the excitation number operators of the three
oscillators. As a consequence, the quantization of Mikhailov's
solutions yields the Hilbert space of $N$ identical bosons in a 3d
harmonic oscillator, and is identical to that part of $\frac{1}{8}$
BPS Hilbert space of $\mathcal{N}=4$ Yang Mills in the conjectured
formula of \cite{index}, that is made up entirely out of the three
holomorphic scalar fields (see \cite{Beasley} for related remarks).

\section{The symplectic form and charges}\label{symfch}

\subsection{Classical $\frac{1}{8}$ BPS solutions}\label{clssols}

As we have described in the introduction, Mikhailov\cite{mikhailov}
has demonstrated that the intersection of the zero set of the
polynomial
\begin{equation}\label{intpol}
 P(z^i)=
\sum_{n_1 n_2 n_3 } c_{n_1 n_2 n_3} \e^{-\ir (n_1+n_2+n_3)t }
(z^1)^{n_1} (z^2)^{n_2} (z^3)^{n_3}
\end{equation}
with the unit 5 sphere $\sum_i |z^i|^2=1$ describes the (time
dependent) world volume of a $\frac{1}{8}$ BPS giant graviton (see
also \cite{Kim:2005mw}). The restriction of the full symplectic form
to these solutions yields a symplectic form on the manifold of
Mikhailov's solutions; the variables $c_{{\vec n}}$ constitute a set
of (projective) coordinates on this manifold.

\subsection{The symplectic form for spatial motion of D3 branes }

The phase space of a classical system may be identified with the
space of solutions to the equation of motion of that system.
Canonical quantization yields a symplectic form on phase space, and
(by restriction) on appropriate sub-classes of solution space
\cite{Crnkovic:1986ex, zuckerman}.

In this section we study the symplectic form on the world volume of
$D$3-branes. In particular we study the motion of $D$3-branes in a
geometrical space with metric $\widetilde{G}_{\mu\nu}$ and 4 form
potential $A_{\mu_1 \mu_2 \mu_3 \mu_4}$, and derive a formal
expression for the symplectic form restricted to motions that are
purely spatial, i.e.\ solutions in which the worldvolume field
strength $F_{\mu \nu}$ is identically zero.

The action on the world volume of a $D$3-brane is given by
\begin{equation}\label{CSaction}
\begin{split}
S=& S_\mathrm{BI}+S_\mathrm{WZ} \\
& =\frac{1}{(2 \pi)^3 (\apm)^2 g_s}
   \int\!\!\dr^4\sigma\, \sqrt{-\tilde{g}}  +
\int\!\! \dr^4 \sigma\: A_{\mu_0 \mu_1 \mu_2 \mu_3}
  \frac{\p x^{\mu_1}}{\p\sigma^{\alpha_0}}
  \frac{\p x^{\mu_1}}{\p\sigma^{\alpha_1}}
  \frac{\p x^{\mu_2}}{\p\sigma^{\alpha_2}}
  \frac{\p x^{\mu_3}}{\p\sigma^{\alpha_3}}
  \frac{\epsilon^{\alpha_0 \alpha_1 \alpha_2 \alpha_3}}{4!}
\\ & =\frac{1}{(2 \pi)^3 (\apm)^2 g_s}
    \int\!\!\dr^4\sigma\, \sqrt{-\tilde{g}}  +
\int\!\!\dr t\,\dr^3\sigma\:
 A_{\mu_0\mu_1\mu_2\mu_3}\, \dot{x}^{\mu_0}\,
  \frac{\p x^{\mu_1}}{\p\sigma^1}
  \frac{\p x^{\mu_2}}{\p\sigma^2}
  \frac{\p x^{\mu_3}}{\p\sigma^3}\, ,
\end{split}
\end{equation}
where
\begin{equation}\label{indmetbig}
\tilde{g}_{\alpha \beta}= \widetilde{G}_{\mu \nu}
    \frac{\partial x^\mu}{\partial \sigma^\alpha}
    \frac{\partial x^\nu}{\sigma^\beta}\, ,
\end{equation}
The symplectic form of interest is given by
\begin{equation}\label{sympforminto}
\begin{split}
\omega_\mathrm{full} &=\omega_\mathrm{BI}+ \omega_\mathrm{WZ}
= \int_\Sigma\!\! \dr^3 \sigma\:
 \delta\! \left( (p_\mu)_\mathrm{BI} + (p_{\mu})_\mathrm{WZ} \right)
\wedge \delta  x^\mu
\\ & =\int_\Sigma\!\! \dr^3 \sigma\:
\delta\! \left( \frac{1}{(2 \pi)^3 (\apm)^2 g_s } \left(
\sqrt{-\tilde{g}} \tilde{g}^{0 \alpha}
 \frac{\partial x^\nu }{\partial \sigma^\alpha}
 \widetilde{G}_{\mu \nu}\right) +
A_{\mu\mu_1\mu_2\mu_3}
 \frac{\p x^{\mu_1}}{\p\sigma^1}
 \frac{\p x^{\mu_2}}{\p\sigma^2}
 \frac{\p x^{\mu_3}}{\p\sigma^3} \right)
 \wedge \delta x^\mu.
 \end{split}
\end{equation}
The forms $\omega_\mathrm{BI}$ and $\omega_\mathrm{WZ}$ are both
manifestly closed. In fact $\omega_\mathrm{BI}$ is also exact; the
same is not true of $\omega_\mathrm{WZ}$ as the 4 form gauge field
$A_{\mu_1 \mu_2 \mu_3 \mu_4}$ is not globally well defined. Equation
\eqref{sympforminto} may be massaged into (see Appendix
\ref{sympdetWZ})
\begin{multline}\label{sympformintt}
\omega_\mathrm{full} =\frac{1}{(2 \pi)^3 (\apm)^2 g_s }
\int_\Sigma\!\! \dr^3 \sigma\:
\delta\!
\left( \sqrt{-\tilde{g}} \tilde{g}^{0 \alpha}
 \frac{\partial x^\nu}{\partial \sigma^\alpha}
   \widetilde{G}_{\mu \nu}\right) \wedge \delta x^\mu  +\\
\int_\Sigma\!\!\dr^3\sigma\:
  \frac{\delta x^{\lambda}
  \wedge \delta x^{\mu}}{2} \left(
  \frac{\p x^{\nu}}{\p\sigma^1}
  \frac{\p x^{\rho}}{\p\sigma^2}
  \frac{\p x^{\sigma}}{\p\sigma^3}\right)
  F_{\lambda\mu\nu\rho\sigma}\, ,
\end{multline}
where $F_{\lambda\mu\nu\rho\sigma}$ is the 5 form field strength.

We are interested in the motion of a D3-brane on the $S^5$ of
$AdS_5\times S^5$. In this background the 5-form is given by
$F=\frac{2\pi N}{\pi^3} \epsilon$ where $\epsilon$ is the volume
form on $S^5$ and $\pi^3$ is the total volume of the unit 5-sphere.
Moreover ${\widetilde G}_{\mu\nu} = \sqrt{(4\pi(\apm)^2 g_s N)}
G_{\mu\nu}$ where $G_{\mu\nu}$ is the metric on unit radius $AdS_5
\times S^5$. Plugging in we find
\begin{multline}\label{sympformintth}
\omega_\mathrm{full} = \omega_\mathrm{BI} +\omega_\mathrm{WZ}
= \frac{N}{2\pi^2} \int_\Sigma\!\! \dr^3 \sigma\:
 \delta\! \left( \sqrt{-g} g^{0\alpha}
 \frac{\partial x^\nu}{\partial \sigma^\alpha}
 {G}_{\mu \nu}\right) \wedge \delta x^\mu\\
+\frac{2N}{\pi^2}\int_\Sigma\!\!\dr^3\sigma\:
  \frac{\delta x^{\lambda}\wedge \delta x^{\mu}}{2} \left(
   \frac{\p x^{\nu}}{\p\sigma^1}
   \frac{\p x^{\rho}}{\p\sigma^2}
   \frac{\p x^{\sigma}}{\p\sigma^3}\right)
  \epsilon_{\lambda\mu\nu\rho\sigma}\, .
\end{multline}
where $g_{\alpha \beta}$ is now defined by
\begin{equation}\label{indmet}
g_{\alpha \beta}= G_{\mu \nu}
 \frac{\partial x^\mu}{\partial \sigma^\alpha}
 \frac{\partial x^\nu}{\partial \sigma^\beta}\, .
\end{equation}
The integrals in \eqref{sympforminto}--\eqref{sympformintth} are all
taken over surfaces of constant $\sigma^0$; the invariance of the
symplectic form under a coordinate redefinition follows from the
equations of motion \cite{Lee:1990nz}.

\subsection{Symplectic form on Mikhailov's solutions}

In the previous subsection, we have derived a general expression for
the symplectic form $\omega_\mathrm{full} =
\omega_\mathrm{WZ}+\omega_\mathrm{BI}$ on the space of spatial
motions of D3 branes on $S^5$. While the expression for
$\omega_\mathrm{WZ}$ is simple and geometrical, the expression for
$\omega_\mathrm{BI} =\dr \theta_\mathrm{BI}$ is rather complicated.
It turns out to be possible to find a relatively simple geometrical
expression for that the restriction of $\theta_\mathrm{BI}$ to the
submanifold of Mikhailov's solutions (see Appendix \ref{sympdetBI})
\begin{equation}\label{exptheta}
\begin{split}
 \theta_\mathrm{BI}&= \frac{N}{2\pi^2} \int_\Sigma\!\!\dr^3\sigma\,
      \sqrt{-g}\, g^{0\alpha}
      \frac{\partial x^\mu}{\partial \sigma^\alpha}\,
      G_{\mu \nu}\,\delta x^\mu
\\
    &= \frac{N}{\pi^2}
     \int_S\!\!\dr^4\sigma\:
        \epsilon_{\mu_1\cdots\mu_6}
        \left[
        \frac{\p x^{\mu_1}}{\p\sigma^1}
        \cdots
        \frac{\p x^{\mu_4}}{\p\sigma^4}
        \right]
        e_\perp^{\mu_5}\,\delta x^{\mu_6}
        \,\delta\!\left(|z^i|^2-1\right)\,,
\end{split}
\end{equation}
where the integral is taken over the four dimensional spatial volume
of the holomorphic surface $S$ at constant time and $e_\perp$ is the
unit position vector in $\C^3$ (the vector from the origin to the
point in question, normalized to have unit norm, so $e_\perp
(\underline{w})= \underline{w}/\Vert \underline{w}\Vert$ for all
$\underline{w}\in \C^3$).

\subsection{Geometrical description of the symplectic form} \label{geom}

The contraction of $\omega_\mathrm{WZ}$ with arbitrary infinitesimal
vectors $v_1$ and $v_2$ at a threefold $\Sigma$ in $S^5$ may be
given a simple geometrical interpretation: it is proportional to the
5 volume formed from out of $\Sigma$ and the two vector fields $v_1$
and $v_2$ supported on $\Sigma$; the details are given in the next
subsection.

Let us now turn to $\theta_{\rm BI}$. The contraction of $\theta_{\rm BI}$
with an infinitesimal vector $v_1$ is given by the following
slightly elaborate geometrical construction. Consider the volume,
$\delta V$, of formed out of that part of the (real) 4 surface
$P(z)=0$ that intersects a shell of thickness $\delta r$ around the
unit sphere, the unit normal vector field $\e_\perp$, and $v_1$. The
contraction of $v_1$ with $\theta_\mathrm{BI}$ is proportional to
$\frac{\delta V}{\delta r}$.

We will now reword the constructions above more formally; this will
allow us to establish certain smoothness properties of these forms
in the next subsection.

We have seen
\begin{equation} \label{sumomeg}
\omega_\mathrm{full}= \dr\theta_\mathrm{BI} + \omega_\mathrm{WZ}\,.
\end{equation}
Let $\CN$ denote the moduli space of intersections of holomorphic 2
surfaces with $S^5$ (we will have a lot more to say about $\CN$
below). We will now define $\CM$, a real codimension 2 real
submanifold in $S^5 \times \CN$. The submanifold $\CM$ consists of
all points $(x, z)\in S^5 \times \CN$ with the property that the
point $x\in S^5$ is in the holomorphic two surface corresponding to
$z$. Therefore, $\CM$ is the fibration over $\CN$ whose fibre over a
point in $\CN$ is the intersection of $S^5$ with the holomorphic
2-surface in question. The dimension of the real manifold $\CM$ is
$\dim_{\R} \CN +3$.

The volume form $\epsilon_5$ on $S^5$ has a natural lift to a 5-form
on the space $S^5 \times \CN$ (i.e.\ to a 5-form that contracts in
the usual way with vectors on $S^5$, but to zero with vectors on
$\CN$). Upon restriction this form yields a 5-form on $\CM$ (recall
that $\CM$ is a submanifold of $S^5 \times \CN$). Integrating this
5-form over the (generically 3 dimensional) fibres in $\CM$ yields a
2-form on $\CN$. The two form obtained via this process is
proportional to $\omega_\mathrm{WZ}$.

The form $\theta_\mathrm{BI}$ has a similar description. We define
$\CM'$, a real codimension two (complex codimension one) submanifold
in $\C^3 \times \CN$, as the fibration of those points in $\C^3$
that lie in the holomorphic 2 surface \footnote{More precisely it is
the set of points that lie on any holomorphic 2-surface that has the
appropriate intersection with $S^5$; which holomorphic surface is
chosen does not matter. } labeled by the base point $\CN$.

Consider a (distributional or current) 5 form  in $\C^3$ defined by
\begin{equation} \label{deffform}
\begin{split}
\epsilon_5'=& (\iota_{e_\perp} \epsilon_6) \: \psi_\tau(1-|z|^2)\,,\\
\psi_\tau(x)&= \left\{  \begin{array}{ll}
                           1, & 0 \leq x \leq \tau, \\
                           0, & \text{otherwise,} \\
                         \end{array}\right.
\end{split}
\end{equation}
where $e_\perp$ is the unit normal vector in $\C^3$ and the symbol
$\iota$ denotes contraction of differential forms with a vector
field. As above $\epsilon'_5$ defines a 5-form on $\CM'$ by pulling
back. Integrating this 5-form over the 4 real dimensional fibres in
$\CM'$ produces a one form in $\CN$. The form $\theta_\mathrm{BI}$
is proportional to its derivative with respect to $\tau$ at
$\tau=0$.

\subsection{Smoothness of $\omega_\mathrm{WZ}$ and $\theta_\mathrm{BI}$}
\label{current}

We will now use the construction of the previous subsection to
demonstrate that the restriction of $\omega_\mathrm{WZ}$ and
$\theta_\mathrm{BI}$ to any compact, finite ($n_C$) dimensional
subspace $\CN_C$ is a `current'. A current on $\CN_C$ is, by
definition, a form whose singularities (if any) are mild enough to
permit well defined integration against genuine forms of degree
$n_C-d$ on $\CN_C$. We will show that $\omega_\mathrm{WZ}$ is a
current of degree two and $\theta_\mathrm{BI}$ is a current of
degree one. See Ch.1, \S2 of \cite{Demailly} for a discussion of the
properties of currents.

As in the previous subsection, let $\CM_C$ denote the (real)
codimension 2 real submanifold in $S^5 \times \CN_C$ consists of all
points $(x, z)\in S^5 \times \CN_C$ such that the point $x\in S^5$
is in the holomorphic 2 surface corresponding to $z$. Consider a
smooth $n_C-2$ form $\beta$ on $\CN_C$. From the definition of
$\omega_\mathrm{WZ}$ we have
\begin{equation*}
\int_{\CN_C} \omega_\mathrm{WZ}\wedge \beta  \propto
\int_{\CM_C} f^*(\epsilon_5) \wedge g^{*}(\beta)\, ,
\end{equation*}
where $g: \CM_C\rightarrow \CN_C$ is the projection defined by $(x,
z)\mapsto z$, and $f: \CM_C\rightarrow S^5$ is the projection
defined by $(x, z)\mapsto x$. The above identity shows that
$\omega_\mathrm{WZ}$ is a current on $\CN_C$ of degree two.

We will now give a similar description of $\theta_{\rm BI}$.

For any $\tau\in (0,1)$, let
\begin{equation*}
S_\tau := \{(z_1,z_2,z_3)\in \C^3 \mid 1-\tau
\leq |z_1|^2 +|z_2|^2 + |z_3|^2\leq 1\} \subset \C^3
\end{equation*}
be the solid shell in $\C^3$ of width $\tau$. Let $\CM^\tau_C$ be the
fibration over $\CN_C$ of real dimension four defined by all $(x,
z)\in S_\tau  \times \CN$ such that the point $x$ lies in the complex 2
surface (in $\C^3$) corresponding to $z$.

Take any smooth form $\beta$ on $\CN_C$ of degree $n_C-1$. It
follows from the description of $\theta_{\rm BI}$ in the previous
subsection that
\begin{equation}\label{howta}
\int_{\CN_C} \theta_\mathrm{BI} \wedge \beta \propto
\left.\frac{\dr}{\dr\tau}\int_{\CM^\tau_C}
  \phi^*(\iota_{e_\perp} \epsilon_6)\wedge \gamma^*(\beta)
  \right\vert_{\tau=0}\, ,
\end{equation}
where $\phi: \CM^\tau_C \rightarrow \C^3$ is the projection defined
by $(x, z)\mapsto x$ (recall that $\CM^\tau_C \subset S_\tau \times
\CN \subset \C^3\times \CN$), $\gamma : \CM^\tau_C \rightarrow
\CN_C$ is the projection defined by $(x, z)\mapsto z$, and
$\iota_{e_\perp} \epsilon_6$ is the contraction of the standard
volume form $\epsilon_6$ on $\C^3$ by the radial vector field
$e_\perp$ on $\C^3\setminus \{0\}$ that assigns any
$x\in \C^3\setminus \{0\}$ the tangent
vector $x/\Vert x\Vert\in T^{1,0}_x \C^3 = \C^3$.

From \eqref{howta} it follows immediately that $\theta_\mathrm{BI}$
is a current on $\CN_C$ of degree one.

This conclusion is important for the following reason. As we will
see below, the manifold $\CN_C$ contains several points of
degeneration; for instance points of topology change at which the
number of connected components of the holomorphic 2 surface changes.
Intuitively one might expect the symplectic form to develop
singularities at such points, and in certain coordinate systems this
is indeed the case (see Appendix \ref{degenerate}). The results of
this subsection guarantee that these singularities are tame enough
to be dealt with, i.e.\ to permit geometric quantization, as all of
the necessary structures associated with forms can also be defined
for currents.

\subsection{$U(3)$ charges}\label{rcharges}

Let $L^m$ denote the generators of $U(3)$ on $\C^3$. We choose our
basis in the space of $U(3)$ generators such that $L^1, L^2, L^3$
are the generators corresponding to three $U(1) \in U(3)$ charges
that generate the rotations $z^i \rightarrow \e^{-\ir \alpha} z^i$
where $i = 1 \ldots 3$.  Let $\xi^{m\alpha}$ denote the
infinitesimal variations of the phase space coordinates $x^{\alpha}$
under the generator $L^m$. Every symplectic form $\omega$ we study
in this paper is  $U(3)$ invariant; this means that
\begin{equation}\label{invomeg}
\mathcal{L}_{\xi} \omega =0\, , \; \: \: \mathrm{i.e.} \: \: \:
\xi^\alpha\partial_\alpha \omega_{\beta \gamma}
  + \omega_{\alpha \gamma} \partial_\beta \xi^\alpha
  + \omega_{\beta \alpha} \partial_\gamma \xi^\alpha =0\, .
\end{equation}
(Here $\mathcal{L}_{\xi}$ denotes the Lie derivative with respect to the
vector field $\xi$.) Locally $\omega =\dr\theta$. It is always
possible to choose $\theta$ so that it is also $U(3)$
invariant,\footnote{This may be proved as follows. Given any $\theta$
such that $\dr\theta=\omega$, the quantity $\theta'$
\begin{equation} \label{invpot}
\theta'= \int\!\! \dr U\, \phi_{U}^*(\theta)
\end{equation}
(where $\phi_{U}$ represents the diffeomorphism generated by the
$U(3)$ element $U$, and $\phi_{U}^* (\eta)$ denotes the pull back
of an arbitrary form $\eta$ under this diffeomorphism) also obeys
this equation and moreover is $U(3)$ invariant.} and we make this
choice in what follows.

The Noether procedure yields the formula
\begin{equation}\label{noetherlocal}
    L^m = \iota_{\xi^m}\theta = \xi^{m\alpha}\theta_\alpha
\end{equation}
for the conserved charges corresponding to the generators $L^m$.
Notice that\footnote{The LHS of \eqref{noethercharges} is a
manifestly closed one form; consistency demands the same is true of
the RHS. That this is indeed the case follows upon using
\eqref{invomeg}. }
\begin{equation}\label{noethercharges}
\dr L^m = -\iota_{\xi^m}\omega\, , \; \; \; \mathrm{i.e.} \:\:\:
\partial_{\alpha} L^m= \omega_{\alpha \beta} \xi^{m\beta}\, .
\end{equation}

Upon quantization, the functions $L^m$ are promoted to operators.
According to the rules of geometric quantization (see Appendix
\ref{geoquant})
\begin{equation*}
    \hat{L}^m =
\omega^{\alpha\beta}\partial_\alpha L^m \left(\ir \partial_\beta +
\theta_\beta \right) + L^m \,  =-\ir\xi^{m\alpha}\p_\alpha\,.
\end{equation*}
As expected, $\hat{L}^m$ is simply the generator of $U(3)$ acting
on functions of the phase space coordinates. Note that the final
expression for $\hat{L}^m$ is independent of the symplectic form.

\section{The topology and symplectic geometry of phase space}
\label{phasespace}

We now turn to a study of the distinct intersections with the unit
5-sphere of the equations $P(z)=0$ for Polynomials $P(z)$ in $P_C$
labeled by a given set, $C$,  of $n_C$ 3-tuples $(n_1,n_2,n_3)$ and
defined as
\begin{equation}\label{polyset}
    P_C=\left.\left\{ P(z^i)=\sum_{\vec{n}\in C}
           c_{\vec{n}}\,(z^1)^{n_1}(z^2)^{n_2}(z^3)^{n_3}
        \right\vert c_{\vec{n}}\in \C\right\}.
\end{equation}
Points in the projective space $\CP^{n_C-1}$, whose projective
coordinates are the coefficients $c_{\vec{n}}$, label these
intersections. This labeling, however, suffers from a flaw; it is
many to one. The space of distinct intersections of holomorphic 2
surfaces with $S^5$ is obtained by performing the appropriate
identifications $\CP^{n_C-1}$. In this section we study the phase
space obtained from this process.

\subsection{A hole in $\CP^{n_C-1}$}\label{hole}

Let us first study the subset of $\CP^{n_C-1}$ that labels the empty
intersection. On any surface $P(z^i)=
\sum_{\vec{n}\in C} c_{\vec{n}}
\,(z^1)^{n_1}(z^2)^{n_2}(z^3)^{n_3}=0$, there will be some points
that are nearest to the origin $z^i=0$. Let us define a distance
function $\rho(c,\bar{c})$ to be the distance to these nearest
points:
\begin{equation}\label{distfun}
    \rho(c,\bar{c})=\min\left\{
           \sum_{i=1}^3 |z^i|^2 \; \;
           \left\vert P(z) = \sum_{\vec{n}\in C} c_{\vec{n}}
           \,(z^1)^{n_1}(z^2)^{n_2}(z^3)^{n_3}=0
        \right.\right\}
\end{equation}
(below we will sometimes use the alternate symbolic notation
$\rho(P(z))$ for \eqref{distfun}).
The surface will only intersect the sphere if $\rho\leq 1$.
Consequently, the set of points $c_{\vec{n}}$ such that $\rho(c,
\bar{c})>1$ all yield $P(z)$ with the same (namely empty)
intersection with the unit 5-sphere. All these points have to be
contracted away in our phase space; we will sometimes refer to the
set of these points as a hole (in $\CP^{n_C-1}$).

\subsection{$\omega$ at the boundary of the hole}\label{bound}

Consider polynomials $P(z)$ such that $\rho(P(z))=1$. Such
polynomials lie at the boundary of the hole described above; for
these polynomials the surface $P(z)=0$ skims the unit sphere without
cutting. In Appendix \ref{touching}, we demonstrate that, when
$\rho(P(z))=1$,  the intersection of any holomorphic surface
$P(z)=0$ with the unit sphere is of real dimension $\leq 2$.
\footnote{More generally, for any polynomial $P$ in $n$ variables
with $\rho(P(z))=1$, the hypersurface $P(z)=0$ of $\C^n$
intersects the unit sphere on a real surface whose real
dimension is at most $n-1$. For example, the surfaces $(z^1)^2=1$,
$(z^1)^2+(z^2)^2=1$, \ldots, $\sum_{i=1}^n (z^i)^2=1$ touch the unit
sphere at a point, a line, \ldots, an $n-1$ sphere respectively.}
This fact has immediate implications for the restriction of the
symplectic form to the boundary of the hole.

Recall from \S\S\ref{geom} that the contraction of
$\omega_\mathrm{WZ}$ with vectors $v_1$ and $v_2$ is proportional to
the 5 volume formed from out of the intersection surface, $v_1$ and
$v_2$. It follows immediately that $\omega_{\mathrm{WZ}}=0$ when the
intersection of $P(z)=0$ is less than 3 dimensional. In particular,
$\omega_{\mathrm{WZ}}$ vanishes when restricted to the boundary of
the hole.

Recall, also from \S\S\ref{current}, that the current $\theta_{\rm BI}$
satisfies the identity in \eqref{howta}. From \eqref{howta} it
follows that the restriction of $\theta_\mathrm{BI}$ to the boundary
vanishes whenever it is defined. We emphasize that
$\theta_\mathrm{BI}$ is a current which need not be a smooth
differential form, hence it may not be well defined everywhere (see
Appendix \ref{touching} for more on this).

\subsection{Contracting away the hole}\label{conthole}

We have argued in the previous subsection that $\omega_\mathrm{full}
= \omega_\mathrm{WZ}+ \dr\theta_\mathrm{BI}$ vanishes when
restricted to the boundary of the hole. This suggests that in the
correct coordinates all points within and at the boundary of the
hole are identified.\footnote{ As an analogy, flat $d$ dimensional
space, when written in polar coordinates, appears to have a boundary
$S^{d-1}$ at $r=0$. However the vanishing of the metric restricted
to the $S^{d-1}$ is a clue that all of the `boundary' is in fact a
single bulk point. We will see a better example of this phenomenon
-- involving symplectic forms rather than metrics -- in section
\ref{linearsection}. } Of course this interpretation is consistent
if and only if all points in the hole (including the boundary) may
continuously be contracted to a single point. We will now show that
this is indeed the case.

Given $n_C$ monomials in three variables, there is a hole
the corresponding projective space $\CP^{n_C-1}$ formed
out of linear combinations of them contains a hole if and only
if the constant function $1$ is one of the $n_C$ monomials. Assume
that $1$ is among the $n_C$ monomials.

Notice that the distance function $\rho$ has the following
homogeneity property
\begin{equation}\label{distscale}
    \rho\left(\lambda^{n_1+n_2+n_3} c_{n_1, n_2, n_3} ,
           \lambda^{m_1+m_2+m_3} \bar{c}_{m_1, m_2, m_3}
    \right)=
    \lambda^{-1} \rho \left(c_{n_1, n_2, n_3} ,
                      \bar{c}_{m_1, m_2, m_3} \right)
\end{equation}
In order to demonstrate that the hole is contractible
(in fact it is diffeomorphic to a ball), let
$\lambda(t)$ be a decreasing function on $[0, 1]$ such that
$\lambda(0)=1$ and $\lambda(1)=0$. According to \eqref{distscale}
the map $c_{n_1, n_2, n_3} \rightarrow c_{n_1, n_2, n_3}
\lambda(t)^{m_1+m_2+m_3}$ continuously maps every point in the hole
to its `center' $c_{\vec{n}}=0$ (the constant function $1$).

It follows immediately that the space obtained by contracting away
the hole is $\CP^{n_C-1}$ itself. We will now demonstrate this
(intuitively obvious) fact. In order to identify the space obtained
after this contraction, let $h(\rho)$ be any nondecreasing function
defined on $[0, \infty)$ satisfying $h(0)=1$ and $h(\rho)=0$ for
$\rho\geq 1$. The coordinate change
\begin{equation}\label{distcoord}
    w_{\vec{n}}=[h(\rho)]^{n_1+n_2+n_3}\,c_{\vec{n}}
\end{equation}
yields a continuous map from $\CP^{n_C-1}$ minus the hole to
$\CP^{n_C-1}$, the last space being parameterized by the $w$
coordinates. Notice that this map takes all points in the hole,
including the boundary $\rho=1$ to a single point, the `origin' in
the new space. It follows that our original space, with the hole
shrunk to a point is topologically $\CP^{n_C-1}$.

The distinguished new point -- the `origin' in $w$ coordinates
represents all configurations that graze the unit ball together with
those configurations that do not intersect the unit ball at all. All
of these a priori distinct physical configurations map to the same
point in physical phase space.

\subsection{Characterization of distinct intersections}\label{distinct}

Now let us turn to a consideration of manifolds $P(z)=0$ that cut
(and don't just graze) the unit $S^5$, i.e.\ polynomials for which
$\rho(P(z)) <1$.

Suppose $S = \{z \in \C^3 | f(z)=0\}$ and $ T = \{z \in \C^3 |
g(z)=0\}$ are two irreducible hypersurfaces (this means that neither
of $f$ and $g$ factorize) such that both $S$ and $T$ cut the
unit sphere $S^5$ (in $\C^3$), and also, $S\bigcap S^5 = T\bigcap S^5$.
Note that  $S\cap T$ is a complex variety. Since the real dimension
of $S\bigcap S^5$ is $3$ (recall that $S$ cuts $S^5$ and not just
grazes it), any complex variety containing $S\bigcap S^5$
must be of complex dimension at least two. Since the complex
dimensions of $S$ and $T$ are two, these imply that $S$ and $T$
share a common open subset. Given that $S$ and $T$ are both
irreducible, from this it follows that $S = T$. Hence $f
=c.g$, where $c$ is some nonzero complex number.

Therefore, two polynomials
$P_1(z)=0$ and $P_2(z)=0$ have identical, three dimensional,
intersections with the unit 5-sphere if and only if $P_1(z)=q(z)
r_1(z)$ and $P_2(z)=q(z) r_2(z)$ where $r_1(z)$ and $r_2(z)$ are
polynomials with distance functions $\geq 1$.

Thus physical phase space -- the phase space of distinct
intersections of $P(z)=0$ with the unit 5-sphere -- is the
$\CP^{n_C-1}$ space parameterized by polynomial coefficients,
subject to the following identification: polynomials of the form
$f(z) g(z)$ are to be identified with $f(z)$ when $\rho(g(z)) \geq
1$.

\subsection{The topology of distinct intersections}\label{topspace}

We will now argue that the space described in the previous
subsection is, in fact, diffeomorphic to $\CP^{n_C-1}$. This result
follows almost immediately from general considerations that we now
briefly review.

Consider a smooth real manifold $M$ of dimension $\ell$. Let $S$ be
a closed submanifold of $M$, and let $f : S \rightarrow Q$ be a
smooth projection, such that each fibre of the projection $f$ is
diffeomorphic to the unit ball in $\R^{\ell}$ (we note that $S$ is
allowed to have dimensions smaller than that of $M$). Then it is
always possible to find a map $D$ from $M \times [0,1]$ to $M$ such
that $D(-,t)$ is a diffeomorphism on $M$ for $t$ in $(0,1)$ with
$D(-,0)$ being the identity map of $M$, and furthermore, $D(-,1)$
reduces to a contraction when restricted to any of the fibres of of
the projection $f$; see \cite[Theorem 5.8]{Milnor}. This result may
be worded more pithily; a manifold $M$ retains its diffeomorphism
type upon contracting away any embedded family of balls of arbitrary
dimension; the contraction is done along the direct of the balls, or
in other words, each individual ball in the family is contracted to
a single point, but distinct balls are contracted to distinct
points. If there are finitely many disjoint embedded fibre bundles
of the above type, then we contract one by one (of this finite
collection). After each step of the contraction, the resulting
manifold remains diffeomorphic to the one in previous step.
Therefore, the final manifold remains diffeomorphic to the original
one.

We will see how these results apply to our situation in an example.
Let $P_C$ denote the set of polynomials of degree at most two; this
space is some contraction of $\CP^{9}$. The only new identification
(apart from the hole in this space) is $p(z) q(z) \sim p(z)$ where
$p$ and $q$ are each degree one polynomials, and $\rho(q(z)) \geq
1$. Contracting away the hole in $\CP^9$ already identifies all
polynomials  $p(z) q(z)$ where $\rho(p) \geq 1$ and $\rho(q) \geq 1$
with a single point (the `origin' in $\CP^9$ which corresponds to
the constant function $1$). The arguments of subsection
\ref{conthole} establish that the set of points in $\CP^9$
corresponding to polynomials $p(z) q(z)$ with a fixed $p(z)$
but varying over all $q(z)$ such that $\rho(q(z)) \geq 1$, is
contractible, in fact it is diffeomorphic to a ball in a Euclidean
space. Associate this set, which is diffeomorphic to a ball, to
$p(z)$.  Now varying over all $p(z)$ with $\rho(p(z)) < 1$ gives
a family of disjoint balls. According to the theorem quoted at the
beginning of this subsection, the space obtained after contracting
away these sets continues to have the topology of $\CP^9$.

In the general case one may proceed similarly, first shrinking away
the hole, then dealing with polynomials with 2, 3, \ldots factors.
At every stage in this process we always contract away disjoint sets
of balls (of arbitrary dimensions), and so the theorem quoted above
guarantees that the space we obtain at the end of this whole process
is $\CP^{n_C-1}$. This $\CP^{n_C-1}$ space may parameterized by a
set of projective coordinates $w_{\vec{n}}$ where $\vec{n}$ belongs
to the set $C$. The coordinates $w_{\vec{n}}$ may be thought of as
functions of the polynomial coefficients $c_{\vec{n}}$. Since each
component that is contracted to a point is left invariant by the
standard action of $U(3)$ on $\CP^{n_C-1}$, the quotient space (the
space obtained after all the contractions are done) is equipped with
an action of $U(3)$ which is induced by the action of $U(3)$ on
$\CP^{n_C-1}$.

As the distance function $\rho$ is $U(3)$ invariant, it is possible
to perform all contractions in a $U(3)$ invariant manner (this can
be checked for the variable change \eqref{distcoord} that
contracts away the hole) \footnote{In this paragraph we assume that
the set $C$ consists of full $U(3)$ multiplets. If this is not the
case, the statements in this paragraph continue to hold with $U(3)$
replaced by $U(1)^3$ } so that
\begin{equation}\label{coordchange}
w_{\vec n}= c_{\vec n} f_{|n|} (c, {\bar c} )
\end{equation}
where $f_{|n|}$ are $U(3)$ invariant functions and $|n|\equiv \sum_i
n_i$.

\subsection{Cohomology class of the symplectic form} \label{cohom}

Recall from \S\S\ref{geom} that $\omega_\mathrm{WZ}$ is the fibre
integral of a closed 5 form, and so is closed. The form
$\omega_\mathrm{BI}$ is also closed (it is exact). It is known that
$H^2(\CP^m, \Z) =\Z$ for all $m>0$ (the generator $1$ in $H^2(\CP^m,
\Z)$ is given by the Poincar\'e dual of a linear hyperplane of
(complex) codimension one in $\CP^m$). As before, let
$\omega_\mathrm{FS}$ be the usual Fubini-Study form on
$\CP^{n_C-1}$. The cohomology class $[\omega_\mathrm{FS}]\in
H^2(\CP^m, {\mathbb C})$ of the closed form $\omega_\mathrm{FS}$
coincides with $1\in H^2(\CP^m, {\mathbb Z})\subset H^2(\CP^m,
{\mathbb C})$. Therefore, $[\omega]=M_C [\omega_\mathrm{FS}]$ for
some complex number $M_C$, and the equality holds within the second
cohomology.

We will now demonstrate that the number $M_C$ is independent of the
set $C$. Consider any two sets of 3-tuples, $C_1$ and $C_2$, such
that $C_1 \subset C_2$. In the auxiliary $\CP^{n_C-1}$ space (from
which physical phase space may be obtained by performing appropriate
contractions) the restriction from $C_2$ to $C_1$ is very simple; it
is achieved by setting to zero the projective coordinates
$c_{\vec{n}}$ for those $\vec{n}$ that belong to $C_2$ but not to
$C_1$. However, under this restriction, the Fubini-Study form in
$\CP^{n_{C_2}-1}$ gives the Fubini-Study form in $\CP^{n_{C_1}-1}$.
The same is true of cohomology classes. In other words, the
homomorphism
\begin{equation*}
\Z = H^2(\CP^{n_{C_2}-1}, \Z)\longrightarrow
H^2(\CP^{n_{C_1}-1}, \Z) =\Z\,,
\end{equation*}
induced by the inclusion map $\CP^{n_{C_1}-1}\hookrightarrow
\CP^{n_{C_2}-1}$, is the identity map of $\Z$. Therefore, the
constant $M_C$ may be determined from the study of a single
convenient example for $C$. In section \ref{linearsection} we will
study the example $C_0= (0,0,0), (1,0,0), (0, 1, 0), (0, 0, 1)$ in
extensive detail, and will find $M_{C_0}=2\pi N$ (where $N$ is the
rank of the gauge group of the dual gauge theory, or the number of
units of flux through the $S^5$). Now, given any $C$, we define
$C'=C\cup C_0$. The restrictions of $C'$ to $C$ and $C_0$
respectively yield the equations $M_{C'}=M_C$ and $M_{C'}=2\pi N$;
it follows that $M_C= M_{C_0} = 2\pi N$ for every $C$. This result
was anticipated in \cite{Beasley} using physical reasoning.

As the map from the auxiliary phase space $\CP^{n_C-1}$ to physical
phase space is continuous, it follows that the symplectic form is in
the cohomology class of $(2\pi N)\omega_\mathrm{FS}$ on physical
phase space as well.

\section{Quantization of phase space}

\subsection{Quantization}\label{quant}

In the previous section we have argued that the phase space $\CN_C$
is $\CP^{n_C-1}$ equipped
with a symplectic form in the cohomology class of
$(2\pi N)\omega_\mathrm{FS}$. The projective coordinates of this
space, $w_{\vec n}$, transform under the $U(1)^3$ transformation
$z^i \rightarrow \e^{-\ir \alpha_i } z^i$ as $w_{\vec n} \rightarrow
\e^{\ir n_i \alpha_i} w_{\vec n}$.

The Hilbert Space that follows from geometric quantization is the
space of polarized sections of the symplectic line bundle (the line
bundle whose curvature is the symplectic form). To get our
discussion started, let us first suppose that the symplectic form
$\omega_\mathrm{full}$ is $(1,1)$ with respect to the $w_{\vec n},
\wb_{\vec{n}}$ complex structure. We have derived the resultant
Hilbert Space in Appendix \ref{cpholquant}; we briefly review the
logic here. When $\omega_\mathrm{full}$ is $(1,1)$ it is possible to
choose complex polarization $D_{\wb_{\vec n}} \phi =0$ and to choose
the symplectic potential such that $\theta_{{\wb}_{\vec n}}=0$ (such
a potential is said to be adapted to our polarization); the
polarization condition is solved when $\phi$ is an analytic function
of $w_{\vec n}$; further $\phi$ is forced to be a function of degree
$N$ in the coordinates $w_{\vec n}$ in order that it be a globally
well defined section. In summary the Hilbert Space $\mathcal{H}_C$
may be identified with the space of degree $N$ Holomorphic
Polynomials of $w_{\vec n}$. \footnote{This result is very close to
Beasley's conjecture for a Hilbert Space consisting of degree $N$
homogeneous polynomials of the variables $c_{{\vec n}}$ .} The 3
$U(1)$ charges $L^m$ are implemented by the operator $L^m= n^m
w_{\vec n}
\partial_{w_{\vec n}}$.

The partition function is
\begin{equation}\label{part}
\mathrm{Tr}_{\mathcal{H}_C} \e^{-\beta_m L^m} = Z^C_N(\beta_m)
\end{equation}
where
\begin{equation} \label{genfunc}
\sum_{r=0}^\infty Z^C_r(\beta_m)  p^r = \prod_{{\vec n} \in C}
\frac{1}{1-p\, \e^{-n_m \beta_m} }
\end{equation}
In particular the total number of states in this
Hilbert Space is $\binom{n_C-1}{N}$.

The Hilbert space $\mathcal{H}_C$ is isomorphic to another Hilbert
space that is more familiar to most physicists. Consider a
collection of $N$ identical, noninteracting bosons whose single
particle Hilbert Space consists of the vectors $|{\vec n} \rangle$
whose $U(1)^3$ charges are $L^m|{\vec n} \rangle = n^m |{\vec n}
\rangle$. \footnote{Note in particular that when $C$ is chosen to be
maximal, this single particle Hilbert space is that of the 3d
harmonic oscillator, with $L^m$ equal to the number operator of the
$m^{th}$ oscillator.} The Hilbert Space of this system is isomorphic
to $\mathcal{H}_C$; the function $\prod_{{\vec n} \in C}
(w_{{\vec{n} }})^{m_{{\vec n}}}$ maps to the state in which  the
occupation number of the single-particle state $| {\vec n} \rangle$
is $m_{{\vec{n}}}$. Equation \eqref{part} applies equally well to
the partition function over the Hilbert space of these
noninteracting bosons.

Equation \eqref{part} was derived under the assumption that the
$\omega_\mathrm{full}$ is $(1,1)$. The pre-quantum line bundle is
not changed by the addition of an exact form $\dr\psi$ to
$\omega_\mathrm{full}$. However $\psi$ does constrain the choice of
polarization. In particular, if the full symplectic form were not
(1,1) we would not be allowed to choose a holomorphic polarization.
As a consequence the derivation of \eqref{part} outlined above does
not go through unchanged for arbitrary $\omega_\mathrm{full}$;
nonetheless the final result \eqref{part} continues to apply as we
now demonstrate.

Recall that the infinitesimal change of a closed 2 form $\omega$
under an infinitesimal coordinate transformation parameterized by
the vector field $\zeta$ is equal to the Lie derivative
\begin{equation*}
\CL_\zeta\omega = \dr\! \left( \iota_{\zeta} \omega \right)
+ \iota_{\zeta}\dr\! \left(\omega \right) =
 \dr\! \left( \iota_{\zeta} \omega \right)\, .
\end{equation*}
As $\omega$ is invertible, it follows that any infinitesimal
deformation, $\dr\psi$, of a closed two form that is also exact may
be undone by a coordinate change. Denoting the coordinates by
$x^\alpha$ the coordinate change in question is given in components
by
\begin{equation*}
\delta x^\alpha = \omega^{\alpha \beta} \psi_\beta\,.
\end{equation*}

Now suppose $\omega$ is $(1,1)$ and $\psi$ is infinitesimal but
otherwise arbitrary. It follows from the paragraph above that the
Hilbert Space obtained by quantizing $\omega + \dr\psi$ with
holomorphic polarization in the new coordinates is identical to the
Hilbert Space obtained from quantizing $\omega$ with holomorphic
polarization in the old coordinates. This argument may be iterated
step by step by integration and so it holds also for finite $\psi$.
As $\omega_\mathrm{full}$ and the one form $\psi$ in our case are
each $U(3)$ invariant, the new coordinates have the same $U(3)$
transformation properties as the old coordinates, and \eqref{part}
applies to all $\omega_\mathrm{full}$ in the cohomology class of
$(2\pi N)\omega_{\mathrm{FS}}$.

Notice that we have derived \eqref{part} knowing only the cohomology
class and $U(3)$ invariance of the symplectic form; in particular,
our result was insensitive to the detailed form of
$\omega_\mathrm{BI}$ (which is an exact form, see \cite{Beasley} for
related remarks). In that sense, our quantization is analogous to
the quantization of the lowest Landau level, as in
\cite{Simons:2004nm, Gaiotto:2004pc, Gaiotto:2004ij}.

\subsection{Semiclassical quantization}\label{semiclass}

Consider any set of polynomials, $P_C$, as described in
\eqref{polyset}: those built from monomials
$(z^1)^{n_1}(z^2)^{n_2}(z^3)^{n_3}$ with $(n_1,n_2,n_3)$ in a given
set, $C$, of 3-tuples. As a check of the slightly formal arguments
that have led to \eqref{part}, in this section we directly quantize
the intersections of $P=0$ (for $P \in P_C$) with the unit
$5$-sphere, in the semiclassical approximation. We demonstrate that
the density of states graded by the three $U(3)$ Cartan charges in
this system, to leading order in the effective Plank constant
$\frac{1}{N}$, is identical to the density of states obtained, in
the same approximation, from the quantization of $\CP^{n_C-1}$ with
symplectic form $(2\pi N)\omega_\mathrm{FS}$ and with $U(1)^3$ charge
operators as described in the previous subsection.

In \S\S\ref{cohom} we saw that, in the auxiliary $\CP^{n_C-1}$ with
the $c_{\vec{n}}$s as projective coordinates, the symplectic form
$\omega_\mathrm{full}$ is in the same cohomology class as $(2\pi
N)\omega_\mathrm{FS}$, i.e.\ we can write
\begin{equation} \label{difference}
(2\pi N)\omega_\mathrm{FS} -\omega_\mathrm{full} = \dr \psi\, .
\end{equation}
where $\psi$ is a well defined 1-form current on $\CP^{n_C-1}$.
Using the transformation in \eqref{invpot}, we can choose to have
$a$ invariant under $U(1)^3$ (i.e.\ have vanishing Lie derivative
with respect to $\xi^m$, the vector fields that generate $U(1)^3$)

Now consider the one parameter set of closed, $U(1)$ invariant
differential forms
\begin{equation}\label{onepar}
\omega(x)=\omega_\mathrm{full} + \dr (x \,\psi)\, .
\end{equation}
Following \eqref{noetherlocal}, we should also define
\begin{equation}\label{noetpar}
    L^m(x)= L^m_0 + x\,\iota_{\xi^m}\psi\, ,
\end{equation}
with $L^m_0$ the original Noether charges. With this definition,
\eqref{noethercharges} holds for all $x$. Note that
$\omega(0)=\omega_\mathrm{full}$ and that $\omega(1)=(2\pi
N)\omega_\mathrm{FS}$; similar relations hold for $L^m(x)$.

We will now demonstrate that the classical partition function,
\begin{equation} \label{clpfx}
Z(\beta_i)= \int\!\! \frac{\omega(x)^{n_C-1}}
                      {(n_C-1)!(2\pi)^{n_C-1}}\;
                 \e^{-\beta_m L^m(x)}\, ,
\end{equation}
is independent of $x$. Differentiating with respect to $x$, we find:
\begin{multline}\label{clpfdif}
  \frac{\dr Z}{\dr x} \propto \int\!\left(
                            (n_C-1)\omega(x)^{n_C-2}\wedge \dr \psi
                            -\beta_m\omega(x)^{n_C-1}\iota_{\xi^m}\psi
                            \right)\e^{-\beta_m L^m(x)}\\
   =\int\!\left\{
      \dr\!\left[(n_C-1)\omega(x)^{n_C-2}\wedge \psi\,\e^{-\beta_m L^m(x)}\right]
      -\beta_m \iota_{\xi^m}\!\!\left[ \omega(x)^{n_C-1}\wedge \psi \right]
                \e^{-\beta_m L^m(x)}
      \right\}\, .
\end{multline}
The first term vanishes because we are in a compact space with no
boundary. The second term vanishes because $\omega^{n_C-1}$ is a top
form.

We have thus demonstrated that the two symplectic forms
$\omega_\mathrm{full}$ and $(2\pi N)\omega_\mathrm{FS}$ generate
identical densities of states, graded with respect to the $U(3)$
Cartan charges, in the semiclassical approximation. In particular,
Taking the $\beta_i\rightarrow 0$ limit, we find that the total
number of states, $\Omega$, in Bohr-Sommerfeld quantization is equal
to
\begin{equation}\label{states}
 \Omega=\frac{N^{n_C-1}}{(n_C-1)!}\, .
\end{equation}
Note that
\begin{equation}\label{approxstates}
\lim_{N \to \infty} \frac{\Omega}
                         {\frac{(n_C-1+N)!}{(n_C-1)! N!}} =1\, .
\end{equation}
Consequently, in the large $N$ (semiclassical) limit $\Omega$
reproduces the exact number of states, $\frac{(n_C-1+N)!}{(n_C-1)!
N!}$, that results from the K\"ahler quantization of $\CP^{n_C-1}$
with symplectic form $(2\pi N)\omega_\mathrm{FS}$.

\section{Linear polynomials} \label{linearsection}

\subsection{Preview of the rest of the paper}

The partition function \eqref{part}, the main result of this paper,
applies to the quantization of the intersections of $P(z)=0$ with
the unit sphere, for the set of polynomials  $P(z) \in P_C$, where
$C$ is an arbitrary collection of 3 tuples, and $P_C$ denotes the
corresponding linear set of polynomials. As very general arguments
that led to \eqref{part} have been slightly formal in character, we
devote the rest of the paper to a more explicit and detailed study
of the quantization of special linear sets of polynomials $P_C$.

In this section we work through the quantization, in gory and
explicit detail, for $C$ chosen as the set $(0,0,0), (1,0,0),
(0,1,0), (0,0,1)$; i.e. for $P_C$ chosen as the set of polynomials
of degree not larger than one.  Our results, are all fully
consistent with results of the previous sections, and illuminate
aspects of the arguments presented above. Our explicit results also
allow us determine the coefficient of the symplectic form (this was
a loose end in \S\S\ref{cohom}).

In section \ref{singvar} we choose $C$ to be the collection $(m, 0,
0)$ for $m=0, 1, 2 \ldots k$. Physically, the  polynomials generated
by this set of $C$ describe the motion of up to $k$ identical half
BPS giant gravitons. Using this interpretation we are able to
independently compute the partition function over $\mathcal{H}_C$ in
this case. We find agreement with \eqref{part}; we view this as a
nontrivial consistency check of \eqref{part}.

In section \ref{hompoly} we study the quantization of homogeneous
polynomials of degree $k$; (i.e.\ for $C$ chosen as the set $(n_1,
n_2, n_3)$ such that $\sum_i{n_i}=k$). The analysis of \ref{hompoly}
uses the formal arguments of the same flavour as those used in
sections \ref{phasespace} and \ref{quant}; however the relative
simplicity of the space of homogeneous polynomials permits us to be
more explicit, and to verify the result \eqref{part} with a greater
degree of mathematical rigour.

\subsection{Linear polynomials: Setting up the problem}

In this section we study the quantization of the linear polynomials
\begin{equation}\label{compcurve}
  c_i z^i-1=0\, ,
\end{equation}
where $i = 1 \ldots 3$. In order to perform this quantization we
need the symplectic \eqref{sympformintth} restricted to Polynomials
of the form \eqref{compcurve}. This form must respect $U(3)$
invariance and must also be closed; these conditions constrain the
symplectic form to be of the form
\begin{equation}\label{formofform}
 \omega =f(|c|^2)\,            \frac{\dr\cb^i\wedge\dr c_i}{2\ir}
      + f'(|c|^2)\,{\cb^ic_j}\,\frac{\dr\cb^j\wedge\dr c_i}{2\ir}\, ,
\end{equation}
for some function $f(|c|^2)$. In Appendix \ref{linear} we have
explicitly evaluated contribution of the Wess-Zumino and the
Born-Infeld terms to the symplectic form \eqref{sympformintth}
(restricted to linear polynomials). Our results are of the form
\eqref{formofform} with
\begin{equation}\label{linfs}
 \begin{split}
    f_\mathrm{BI}(|c|^2)
         &= {2N}\left(\frac{1}{|c|^4}-\frac{1}{|c|^6}\right) , \\
    f_\mathrm{WZ}(|c|^2)
         &= {2N}\left(\frac{1}{|c|^2}-
\frac{2}{|c|^4}+\frac{1}{|c|^6}\right) , \\
    f_\mathrm{full}(|c|^2)
         &= {2N}\left(\frac{1}{|c|^2}-\frac{1}{|c|^4}\right) .
 \end{split}
\end{equation}

Further, (with no summation over the index $m$)
\begin{equation}\label{charges}
\omega_{{\bar \imath} j}(\ir  c^m \delta^j_m)=
 \half f c^m \delta^i_m + \half f' c^i |c^m|^2=
\half\partial_{{\bar c}_{i}} \left( |c^m|^2 f(|c|^2) \right)\, .
\end{equation}
Comparing with \eqref{noethercharges} we identify the Noether
charges:
\begin{equation}\label{noetherf}
L^m= \half |c^m|^2 f\, .
\end{equation}
Notice that $L^1+L^2+L^3$ evaluated using $f_\mathrm{full}$ and
\eqref{noetherf} yields \eqref{energy} as expected on general
grounds.

\subsection{Semiclassical quantization}\label{linsemi}

We will now discuss the semiclassical quantization of the space of
linear polynomials, with respect to the three symplectic form
$\omega_\mathrm{BI}$, $\omega_\mathrm{WZ}$ and
$\omega_\mathrm{full}$ and associated charges. The Bohr-Sommerfeld
density of states $\frac{\omega \wedge \omega \wedge \omega}{3! (2
\pi)^3}$ is easily evaluated; we find \footnote{The algebra leading
to this result may be processed as follows. Define
\begin{equation*}
\alpha=\frac{\dr\cb^i\wedge\dr c_i}{ 2\ir}\, ,\;\;\;\;
\beta=\frac{\cb^ic_j}{|c|^2}\frac{\dr\cb^j\wedge\dr c_i}{2\ir}\, ,
\end{equation*}
and use the identities:
\begin{equation*}
\beta\wedge\beta=0\, , \;\;\;
3\alpha\wedge\alpha\wedge\beta=\alpha\wedge\alpha\wedge\alpha=6\epsilon_6\, ,
\;\;\;(a\alpha+b\beta)^3=6a^2(a+b)\epsilon_6\, ,
\end{equation*}
(where $\epsilon_6$ is the usual volume form on $R^6$). Substitute
$a= f, \;\;\;b= |c|^2 f'$, and $\epsilon_6=\frac{\Omega_5}{2}
|c|^4 \dr(|c|^2) $ and use the variable $x$ for $|c|^2$. }

\begin{equation}\label{intoverden}
  \int \frac{\omega \wedge \omega \wedge \omega}{3! (2\pi)^3}
     = \frac{N^3}{2} \int E(x)^2 E'(x) \dr x
     = \frac{N^3}{2} \int{E^2}\dr E
     = N^3 \left(\frac{E^3(\infty)}{6} -\frac{E^3(1)}{6} \right) ,
\end{equation}
where
\begin{equation}
E(x)=\frac{1}{2N} x f(x)\, , \; \; \; \: \:
x=|c|^2\, ,
\end{equation}
is the Noether charge $L^1+L^2+L^3$. Substituting in formulae for
$f(x)$ we conclude that semiclassical quantization with
$\omega_\mathrm{WZ}$ and $\omega_\mathrm{full}$ each yield a Hilbert
space with $N^3 /6$  expectations from and density of states (graded
with respect to $E$) proportional to $E^2$; in perfect agreement
with the results of \ref{semiclass}. Note that the semiclassical
quantization using $\omega_\mathrm{BI}$ yields no states; this is a
direct consequence of the fact that $\omega_\mathrm{BI}$ is exact.

\subsection{Exact quantization}\label{linexact}

In this subsection we will proceed to perform an exact quantization
of Mikhailov's linear polynomials. Notice that polynomials with
coefficients s.t. $|c|^2 <1$ fail to intersect the unit sphere; this
is the hole of \S\S\ref{hole}. Notice also that
$\omega_\mathrm{full}$, $\omega_\mathrm{BI}$ and
$\omega_\mathrm{WZ}$ all vanish on the boundary of the hole (see
\eqref{linfs} ), as expected from \S\S\ref{bound}. According to
\S\S\ref{conthole} we should deal with the hole by finding a $U(3)$
invariant coordinate change that maps $|c^2|=1$ to the origin of the
new coordinate system. Under a $U(3)$ invariant variable change
(i.e.\ a variable change of the form $c^i=w^i g(|w|^2)$ for any
function $g$), \eqref{formofform} retains its form, with the
function $f(|c|^2)$ being replaced by a new function ${\tilde
f}(|w|^2)$ defined by the equation\footnote{More explicitly
\begin{equation}\label{newfex}
{\tilde f}(|w|^2) = \left|\frac{c(w)}{w}\right|^2 f(|c(w)|^2)
=\left|g\!\left(|w|^2 \right)\right|^2
  f\!\left( |w|^2 \left|g\! \left(|w|^2 \right)\right|^2 \right)\, .
\end{equation} }
\begin{equation}\label{newf}
|w|^2 {\tilde f}(|w|^2) = |c|^2 f(|c|^2)\, .
\end{equation}
We will choose our coordinate $w$ to satisfy the equation
\begin{equation}\label{eqsolvecp}
2N\frac{|w|^2}{1+|w|^2}= |c|^2 f(|c|^2)\, .
\end{equation}
We choose to work in the variable $w$ defined by \eqref{eqsolvecp}
so that the in the new $w$ variables the symplectic form is simply
$2\pi N$ times the Fubini-Study form on $\CP^3$ (in the gauge in
which one of the coordinates is set to unity)! \footnote{Note that
\eqref{eqsolvecp} ensures that ${\tilde f}(w^2)=2 N
\frac{1}{1+w^2}$; plugging into \eqref{formofform} $\omega$ turns
into $2\pi N$ times the Fubini-Study form.}

Provided that $|c|^2 f(|c|^2)$ increases monotonically from zero at
$|c|=1$ to $2N$ at $|c|=\infty$, the coordinate change defined by
\eqref{eqsolvecp} is legal; i.e. it is one one and maps the region
$|c|>1$ to $\C^3$; the 5 sphere $|c|=1$ to the origin. In other
words, provided $|c|^2 f(|c|^2)$ decreases monotonically between the
two limits of interest, (this is true of $\omega_\mathrm{full}$ as
well as  $\omega_\mathrm{WZ}$ but not of $\omega_\mathrm{BI}$)
\footnote{Explicitly, the variable change for $\omega_\mathrm{full}$
is
\begin{equation}\label{cpcoord}
c_i=\sqrt\frac{|w|^2+1}{|w|^2}\:w_i\, ,
\end{equation}
i.e.\ choosing $h(\rho)=\sqrt{1-\rho^2}$ in \eqref{distcoord}.},
the space \eqref{formofform} is simply $\CP^3$ with symplectic form
equal to $2\pi N$ times the Fubini-Study form, written in perverse
coordinates, exactly as anticipated in \S\S\ref{conthole},
\S\S\ref{topspace} and \S\S\ref{cohom}.

The quantization of the space of intersection of linear polynomials
with the $S^5$, with symplectic form given either by
$\omega_\mathrm{WZ}$ or $\omega_\mathrm{full}$  is now
straightforward (see \S\S\ref{quant} and Appendix \ref{geoquant}).
The Hilbert space is given holomorphic polynomials , of degree $N$,
of the four variables $1, w^1, w^2, w^3$. On this Hilbert space the
charge operators are simply $L^1=w^1\partial_{w^1},
\;\;\;L^2=w^2\partial_{w^2}, \;\;\;L^3= w^3\partial_{w^3}$. As
explained in the introduction, this is precisely the Hilbert space
of a system of $N$ identical noninteracting bosons, whose single
particle Hilbert space consists of four states with charges $(L^1,
L^2, L^3)$ equal to $(0,0,0), \;(1,0,0), \;(0,1,0) \;\text{and}\;(0,
0, 1)$.

The partition function over this Hilbert Space is, of course, given
by \eqref{part} with the choice of $C$ listed above. In particular
upon setting $\beta_1=\beta_2=\beta_3=\beta$ we find
\begin{equation}
Z_{\mathrm{Lin}}=\sum_{k=0}^N \binom{k+2}{2} \e^{-\beta k}
\end{equation}

In the large $N$ limit the summation over $k$ may be replaced by an
integral over a continuous variable $E$ and we find
\begin{equation}
Z_{\mathrm{Lin}}\approx \int_0^N \dr E\;  \frac{E^2}{2}\;
\e^{-\beta E}
\end{equation}
in perfect agreement with the semiclassical results of \S\S\ref{linsemi}.

\section{Polynomials of a single variable}\label{singvar}

\subsection{Expectations from \S\S\ref{quant}}\label{pred}

In this subsection we study the quantization of degree $k$
polynomials of a single variable $z$. According to the general
arguments of \S\S\ref{quant}, this submanifold of solutions
should be topologically $\CP^{k}$;  the Hilbert space obtained by
quantizing these solutions should be given by degree $N$ polynomials
of the variables $v_0$, $v_1$ \ldots $v_k$ respectively where the
subscript denotes the $L^3$ charges of these variables. The
partition function Tr $\e^{-\beta L^3}$ over this Hilbert space is
predicted to be \eqref{part} ${\widetilde Z}_N$ where
\begin{equation}\label{pfn}
\sum_{m=1}^\infty q^m {\widetilde Z}_m =
   \prod_{r=0}^k \frac{1}{1-q \e^{-\beta r}}\, .
\end{equation}

\subsection{Direct Evaluation} \label{eval}

We will now test all these predictions against an independent direct
study of this space and its quantization. Any degree $k$ polynomial
of a single variable is proportional to
\begin{equation}\label{arbitpol}
P(z) =\prod_{a=1}^k
\left(\frac{\sqrt{1+|w^a|^2}}{|w^a|}\: w^a z-1\right)\, .
\end{equation}
for some choice of $w^a$.  The intersection of $P(z)=0$ with the
unit sphere describes a gas of noninteracting half BPS giant
gravitons; i.e.\ the corresponding D3-brane surfaces consist of a
set of disconnected, parallel $S^3$s of squared radius
$\frac{|w^a|^2}{1+|w^a|^2}$. \footnote{Half BPS giant gravitons have
been studied extensively, see for instance \cite{Das:2000fu,
Jevicki:2000ty, Das:2000st, Ho:2000qh, Balasubramanian:2001nh,
Corley:2001zk, Balasubramanian:2002sa, Berenstein:2004kk,
Lin:2004nb, Mandal:2005wv, Dhar:2005qh, Grant:2005qc, Maoz:2005nk,
Dhar:2005fg, Dhar:2005su}. The discussion below overlaps with these
studies.}

The $U(1)$ charges of these branes are additive. Geometrically the
space parameterized by the coordinates $w^a$ is  $(\CP^{1})^k/ S_k$
(where $S_k$ is the permutation group) and the symplectic form on
this space is
\begin{equation}\label{sympform}
\omega= \sum_a \omega_a\, ,
\end{equation}
where
\begin{equation}\label{omeg}
\omega_a= \frac{2N}{ 1+|w^a|^2}
\left( \frac{\dr\wb^a \wedge \dr w^a}{2\ir}
      -\frac{w^a \wb^a}{1+|w^a|^2}
            \frac{\dr\wb^a \wedge \dr w^a}{2\ir}
\right)
\end{equation}
is $2\pi N$ times the Fubini-Study form on $\CP^1$.

The Hilbert space obtained from the quantization of \eqref{omeg} is
the set of symmetric holomorphic polynomials of degree $N$ or less
in each variable $w^a$ (the symmetry of polynomials is forced from
the fact that the polynomials $P(z)$ in \eqref{arbitpol} are
invariant under permutations of $w^a$). The partition function
weighted by $e^{-\beta H}$ (where $H$ is the $U(1)$ rotation of $z$)
over this Hilbert space is given by $Z_k$ where
\begin{equation}\label{pfsc}
\sum_{k} p^k Z_k =
    \prod_{m=0}^N \frac{1}{1-p \e^{-\beta m}}\, .
\end{equation}

\subsection{Comparison} \label{comp}

We will now demonstrate that the geometrical spaces and Hilbert
Spaces described in subsections \ref{pred} and \ref{eval} are the
same.

The quotient space $(\CP^{1})^k/ S_k$ may be holomorphically
identified with $\CP^{k}$ in the following way: Let $(x^a,y^a)$ be
projective coordinates for the $a^{th}$ $\CP^1$. Let $z^j$
($j=0,\ldots,k$) be projective coordinates for $\CP^k$. Let $S_j$ be
the set of subsets of $\{1,\ldots,k\}$ of cardinality $j$. The map
\begin{equation*}
    z^j = \sum_{\tau\in S_j}\left[ \prod_{a\in\tau}x^a
                             \prod_{b\in\tau^c}y^b \right]
\end{equation*}
provides the identification between $(\CP^{1})^k/ S_k$ and
$\CP^{k}$. Furthermore, if the $x^a$'s have charge 1 and the $y^a$'s
have charge 0, then $z^j$ will have charge $j$.


The cohomology class of $(\CP^1)^k/S_k$ defined by $\omega$
coincides with the cohomology class of $\CP^{k}$ given by the
Fubini-Study form on $\CP^{k}$ (after identifying $(\CP^{1})^k/ S_k$
with $\CP^{k}$). Indeed, this follows from the facts that the degree
of the projection $(\CP^{1})^k\, \rightarrow\, (\CP^{1})^k/ S_k$ is
$k!$, and
\begin{equation*}
\int_{(\CP^{1})^k} \omega^k \, =\, k!\, .
\end{equation*}
All of this implies that the Hilbert Space obtained from the
quantization of $\frac{(\CP^{1})^k }{S_k}$ is isomorphic to that
obtained from the holomorphic quantization of $\CP^{k}$ whose
projective coordinates have charges $(0,1,\ldots,k)$; i.e. implies
that
\begin{equation}\label{identtoprove}
  \sum_{N=0}^\infty q^N \prod_{m=0}^N
    \frac{1}{1-p \e^{-\beta m}}
= \sum_{k=0}^\infty p^k \prod_{r=0}^k
    \frac{1}{1-q \e^{-\beta r}}\, ,
\end{equation}

This is indeed a true identity. It can be made manifest by writing
the partition function in a way that is explicitly symmetric in $p$
and $q$, as a sum over Young Tableaux.

\begin{equation}\label{identtoprove2}
  \sum_{N=0}^\infty q^N \prod_{m=0}^N
    \frac{1}{1-p \e^{-\beta m}}
=  \sum_{N=0}^\infty q^N \sum_{n_0=0}^\infty \sum_{n_1=0}^\infty
\cdots \sum_{n_N=0}^\infty p^{\sum_m n_m} \e^{-\beta \sum_m m n_m}
\end{equation}

If $n_m$ is the number of rows of length $m+1$ in a tableau $R$
(ignoring the first and longest, of length $l(R)= N+1$), the number
of rows of the tableau is $h(R)= 1+ \sum_m n_m$ and the number of
boxes in the tableau is  $n(R) = N + 1 + \sum_m (m+1) n_m$. Then the
sum can be written in the symmetric form:

\begin{equation}\label{identtoprove3}
\sum_{N=0}^\infty q^N \sum_{n_0=0}^\infty \sum_{n_1=0}^\infty
\cdots \sum_{n_N=0}^\infty p^{\sum_m n_m} \e^{-\beta \sum_m m n_m}
= \sum_R q^{l(R)-1} p^{h(R)-1} \e^{-\beta (n(R)-l(R)-h(R)+1)}.
\end{equation}


\subsection{Giant Gravitons as Subdeterminants}

We end this subsection we describe, in more detail, the map between
giant gravitons and operators. We have argued that single giant
gravitons (states obtained from the quantization of linear
polynomials) of energy $k$ (i.e.\ the wavefunction
$\psi=(w_{001})^k(w_{000})^{N-k}$) correspond to the state in the
auxiliary counting problem of \cite{index} that has $k$ of the
identical bosons in the first excited state of the harmonic
oscillator, and all others in the ground state; this, in turn,
corresponds to the gauge invariant function of the operator $Z$ that
is equal to $\sum_\mathrm{subsets} z^{i_1} z^{i_2} \ldots z^{i_k}$
(where $z^1, z^2 \ldots z^N$ represent the various eigenvalues of
$Z$, and the sum is over all subsets $\{i_1,i_2,\ldots,i_k\}$ of $\{
1, \ldots, N\}$ with $k$ elements). This, however, is precisely the
character polynomial of $U(N)$ in its $k^{th}$ antisymmetric
representation (representation with $k$ boxes in the first column of
a Young tableau). We have thus reproduced the standard map from
giant gravitons to operators!

The considerations of the last paragraph are easily generalized. A
basis for operators that correspond to $m$ giant gravitons of
arbitrary charge (polynomials of degree exactly $m$) is given by the
character polynomials of the operator $Z$ in representations $R$
whose corresponding Young tableaux have boxes only in the first $m$
columns \cite{Balasubramanian:2001nh, Corley:2001zk}. See
\cite{Beasley} for a beautiful further generalization of this map to
arbitrary $\frac{1}{8}$ BPS operators.


\section{Homogeneous polynomials}\label{hompoly}

In this section we determine the Hilbert Space that follows from
the quantization of homogeneous polynomials of degree $k$. We denote
this submanifold of phase space by $M_k$. The restriction to
homogeneous polynomials of a given degree results in several
simplifications as compared to the general case, the
two most important
of these are the facts that the surfaces $P(z)=0$ always intersect the
sphere for homogeneous $P$, so that the space of intersections of
such surfaces with the unit sphere is a projective space with no
holes, and also there are no identifications among the polynomials.
While the symplectic form in the natural projective variables
is not everywhere nonsingular, in this section we demonstrate
that the singularities that exist are tame enough to control and
that the holomorphic quantization of this space yields that is in
precise accord with the conjecture at the beginning of this section.
The rest of this section employs mathematical language and
technology. We have attempted to make this section as self contained
as possible.

The zero set of a homogeneous polynomial of any degree is invariant
under the rescaling $z^i \rightarrow \lambda z^i$ for any complex
number $\lambda$. The group $S^1$ acts freely on $S^5$ with $\CP^2$
as the quotient; the $S^1$ action corresponds to the Killing vector
$(\ir z^j\partial_{z^j} -\ir\zb^j \partial_{\zb^j})$ on $S^5$.
Therefore, the intersection of holomorphic homogeneous polynomials
of degree $k$ with the unit $S^5$ is a principal $S^1$-bundle over a
degree $k$ curve in $\CP^2$ (this means that $S^1$ acts freely
transitively on the intersection). \footnote{The genus of a
smooth curve
in $\CP^{2}$ of degree $k$ is $\frac{(k-1)(k-2)}{2}$.} The
topological isomorphism classes of a principal $S^1$-bundle over a
complex
curve $X$ are parameterized by $H^2(X, \Z) =\Z$. We note that the
topological isomorphism class of the above $S^1$-bundle over a
degree $k$ curve is $k$.

As a consequence \eqref{sympformintth} induces a symplectic form on
the space of degree $k$ curves in $\CP^{2}$. The nature of this
symplectic form is not difficult to determine. We first note that
the D3-brane surfaces defined by homogeneous polynomials do not
evolve in time (the time dependence in \eqref{intpol} may be
absorbed into a time dependent overall rescaling). It follows as a
consequence that $\omega_\mathrm{BI}=0$ for such surfaces (this
follows from \eqref{exptheta}, on noting that $e_\perp$ is tangent
to these surfaces). Therefore, we have
$\omega_\mathrm{full}=\omega_\mathrm{WZ}$.

The symplectic form $\omega_\mathrm{WZ}$ is constructed using the
volume form on $S^5$. We will relate volume form $\epsilon_5$ on
$S^5$ with the volume form on $\CP^{2}$ given by the Fubini-Study
metric.

We will denote by $\omega_2$ the Fubini-Study K\"ahler form on
$\CP^{2}$. Therefore, $\omega_2 \wedge \omega_2$ is a volume form on
$\CP^{2}$; its total volume is $1$. Let $f: S^5 \rightarrow \CP^{2}$
be the quotient map for the action of $S^1$. We will denote by
$\dr\theta$ the relative one form on $S^5$ for the projection $f$
given by the form $\dr\theta$ on $S^1$, where $\theta$ is the angle.
(This is the Maurer--Cartan relative $1$-form for the principal
$S^1$-bundle.) Note that
$f^*(\omega_2\wedge\omega_2)\wedge\dr\theta$ is a well-defined
5-form on $S^5$ (the wedge product is independent of the choice of
the extension of the relative form $\dr\theta$ to a differential
$1$-form on $S^5$). It is easy to see that
\begin{equation}\label{volidt}
\epsilon_5 = \frac{\pi^2}{2}
f^*(\omega_2\wedge\omega_2)\wedge\dr\theta
\end{equation}
where $\epsilon_5$ is the volume form of $S^5$.

Let $V_k$ denote the complex vector space of homogeneous polynomials
of degree $k$ in three variables. The corresponding complex
projective space will be denoted by $\mathbb{P}V_k$; so
$\mathbb{P}V_k$ parameterizes all complex lines in the complex vector
space $V_k$. Therefore, $\mathbb{P}V_k$ is biholomorphic to the
complex projective space of complex dimension $(k+1)(k+2)/2-1$. We
have a real submanifold
\begin{equation}\label{incl}
 \mathcal{Y} \subset S^5\times \mathbb{P}V_k
\end{equation}
of real dimension $(k+1)(k+2)+1$ whose fibre over any point $p\in
\mathbb{P}V_k$ is the intersection $S^5\bigcap
\{P(z^1,z^2,z^3)=0\}$, where $P$ is any homogeneous polynomial of
degree $k$ giving $p$ (recall that points of $\mathbb{P}V_k$
parameterize lines in $V_k$). Let
\begin{equation}\label{projphi}
\phi : \mathcal{Y}\rightarrow S^5
\end{equation}
be the projection to the first factor.

The 2-form $\omega_\mathrm{WZ}$ on $\mathbb{P}V_k$ is the
integral\eqref{sympformintth}
\begin{equation}\label{fiveformonp}
\omega_\mathrm{WZ} = \frac{2N}{\pi^2}
\int_{\mathcal{Y}/\mathbb{P}V_k} \phi^*\epsilon_5\,
,
\end{equation}
where $\phi$ is the map in \eqref{projphi} and
$\int_{\mathcal{Y}/\mathbb{P}V_k}$ is the integral of differential
forms on $\mathcal{Y}$ along the fibres of the projection
$\mathcal{Y}\rightarrow \mathbb{P}V_k$.

On the other hand, there is a complex submanifold
\begin{equation}\label{incll}
\mathcal{Z} \subset \CP^{2}\times \mathbb{P}V_k
\end{equation}
of complex dimension $(k+1)(k+2)/2$ whose fibre over any point $p\in
\mathbb{P}V_k$ is the intersection $\CP^{2}\bigcap
\{P(z^1,z^2,z^3)=0\}$, where $P$ is any homogeneous polynomial of
degree $k$ giving $p$. Therefore, $\mathcal{Y}$ (defined in
\eqref{incl}) is a principal $S^1$-bundle over $\mathcal{Z}$. Note
that we have a commutative diagram of maps:
\begin{equation*}
\begin{matrix}
\mathcal{Y} & \hookrightarrow & S^5\times \mathbb{P}V_k\\
\downarrow && \downarrow\\
\mathcal{Z}& \hookrightarrow & \CP^{2}\times \mathbb{P}V_k
\end{matrix}
\end{equation*}
where the projection $S^5\times \mathbb{P}V_k\rightarrow
\CP^{2}\times \mathbb{P}V_k$ is $f\times \text{Id}_{\mathbb{P}V_k}$
with $f$ being the quotient by the action of $S^1$.

Let
\begin{equation}\label{projpsi}
\psi : \mathcal{Z}\rightarrow \CP^{2}
\end{equation}
be the projection. From \eqref{volidt} and \eqref{fiveformonp} it
follows that
\begin{equation}\label{integralo}
\omega_\mathrm{WZ} = \frac{2N}{\pi^2}\int_{\mathcal{Y}/\mathbb{P}V_k}
\phi^*\epsilon_5
=  2\pi N\int_{\mathcal{Z}/\mathbb{P}V_k}
                   \psi^*(\omega_2\wedge\omega_2)\, ,
\end{equation}
where $\omega_2$ is the Fubini-Study K\"ahler form on $\CP^{2}$, and
$\int_{\mathcal{Z}/\mathbb{P}V_k}$ is the integral of differential
forms on $\mathcal{Z}$ along the fibres of the projection
$\mathcal{Z}\rightarrow \mathbb{P}V_k$.

We will investigate $\omega_\mathrm{WZ}$ using the identity in
\eqref{integralo}. Since $\omega_2\wedge\omega_2$ is a $(2,2)$-form
on $\CP^{2}$, as well as both $\psi$ and the projection
$\mathcal{Z}\rightarrow \mathbb{P}V_k$ are holomorphic maps, the
fibre integral $\int_{\mathcal{Z}/\mathbb{P}V_k}
\psi^*(\omega_2\wedge\omega_2)$ is of type $(1,1)$. Since
$\omega_2\wedge\omega_2$ is closed we know that
$\int_{\mathcal{Z}/\mathbb{P}V_k} \psi^*(\omega_2\wedge\omega_2)$ is
also closed. Although $\int_{\mathcal{Z}/\mathbb{P}V_k}
\psi^*(\omega_2\wedge\omega_2)$ may have some singularities, it
defines a current on $\mathbb{P}V_k$ of degree two. This means that
$\int_{\mathcal{Z}/\mathbb{P}V_k} \psi^*(\omega_2\wedge\omega_2)$ is
a continuous functional on the space of smooth differential forms on
$\mathbb{P}V_k$ of degree $(k+1)(k+2)-4$; see Ch.1, \S2 of
\cite{Demailly} for currents \footnote{Note however that the form is
not everywhere smooth. This is most easily seen from an explicit
formula for the symplectic form. Let us work in the gauge $z^3=1$ on
$\CP^2$. We find
\begin{equation}\label{expforsympform}
\omega_{M_k} = \frac{N}{4\pi\ir}\int_{C'}
\left(\frac{\dr\zb^1 \wedge \dr\zb^2}{\dr\overline{f}}\right)\wedge
\left(\frac{\dr z^1 \wedge \dr z^2}{\dr f}\right) \times
\frac{1}{(1+ |z^1|^2 + |z^2|^2)^3}\:
\frac{\dr\overline{f} \wedge \dr f}{2\ir} \, ,
\end{equation}
where $\frac{\dr z^1 \wedge \dr z^2}{\dr f}$ is defined to be equal
to $\frac{\dr z^1}{\frac{\p f}{\p z^2}}= -\frac{\dr z^2}{\frac{\p
f}{\p z^1}}$. While this is not manifest, it not difficult to verify
that \eqref{expforsympform} is gauge invariant; i.e.\
\eqref{expforsympform} is invariant under any permutation of the
indices $1,2,3$ of the projective coordinates of $\CP^{2}$. Note
that the symplectic form has singularities at degenerations of the
curve $P(z)=0$, see Appendix \ref{degenerate} for an explicit
example }
 Note that for any smooth differential form $\tau$ on
$\mathbb{P}V_k$ of degree $(k+1)(k+2)-4$ we have
\begin{equation}\label{currentdef}
\int_{\mathbf{P}V_k} \tau\wedge
\left(\int_{\mathcal{Z}/\mathbb{P}V_k}
\psi^*(\omega_2\wedge\omega_2)\right)
= \int_\mathcal{Z}p_2^*\tau \wedge \psi^*(\omega_2\wedge\omega_2)\, ,
\end{equation}
where $p_2$ is the projection of $\mathcal{Z}$ to $\mathbb{P}V_k$.

The cohomology class in $\mathcal{Z}$ defined by the form
$\psi^*(\omega_2\wedge\omega_2)$ lies in $H^2(\mathcal{Z},
\Z)$. Therefore, $\frac{\omega_\mathrm{WZ}}{2\pi}$ is a closed current of
type $(1,1)$, and it defines an element in $H^2(\mathbb{P}V_k, \Z) =
\Z$. In other words, $\omega_\mathrm{WZ}$ gives an integer, which is
the proportionality constant of the cohomology class defined by
$\frac{\omega_\mathrm{WZ}}{2\pi}$.

The above integer may be determined to be $N$ using the arguments
 of \S\ref{cohom}. Indeed, fixing a holomorphic homogeneous
polynomial $P_0$ of degree $k-1$, consider the subspace of
$\mathbb{P}V_k$ that corresponds to the homogeneous polynomials of
the form $P_0P$, where $P$ runs over all nonzero homogeneous
polynomials of degree one. This gives a map from $\mathbb{P}V_1 =
\CP^2$ to $\mathbb{P}V_k$ of degree $1$. From the expression of
$\omega_\mathrm{WZ}$ in \eqref{integralo} we know that the form
$\omega_\mathrm{WZ}$ on $\mathbb{P}V_k$ pulls back to the
$\omega_\mathrm{WZ}$ on $\mathbb{P}V_1 = \CP^2$. On the other hand,
we know that the $\frac{\omega_\mathrm{WZ}}{2\pi}$ on $\mathbb{P}V_1
= \CP^2$ is $N$-times the positive generator of $H^2(\CP^2, \Z) =
\Z$. Therefore, the cohomology class in $H^2(\mathbb{P}V_k, \Z)$
given by $\frac{\omega_\mathrm{WZ}}{2\pi}$ is $N$-times the positive
generator of $H^2(\mathbb{P}V_k, \Z) = \Z$.

Thus, the holomorphic line bundle $\mathcal{O}_{\mathbb{P}V_k}(N)$
is the unique holomorphic line bundle on $\mathbb{P}V_k$ whose first
Chern class coincides with the cohomology class in
$H^2(\mathbb{P}V_k, \Z)$ given by $\frac{\omega_\mathrm{WZ}}{2\pi}$.
Since $\omega_\mathrm{WZ}$ is a current of type $(1,1)$, the
holomorphic line bundle $\mathcal{O}_{\mathbb{P}V_k}(N)$ admits a
Hermitian connection, whose (Hodge type) $(1,0)$-part may have
singularities, such that curvature of the connection coincides with
$\omega_\mathrm{WZ}$.

It thus follows that the holomorphic quantization of $M_k$ is well
defined. As the symplectic form is a closed $(1,1)$ the quantization
may be performed holomorphically; the Hilbert space is simply the
set of all degree $N$ polynomials of the coefficients $c_{n_1, n_1,
n_3}$ (of homogeneous polynomials of degree $k$ in three variables)
with the usual implementation of $U(3)$ generators on this space, in
perfect agreement with \eqref{genfunc}.

\section{Discussion}\label{disc}

In this section we will comment on aspects of the procedure
adopted in our paper, and discuss implications and generalizations
or our results.

\subsection{Gravitons from $D3$-branes}

Our quantization of giant gravitons has reproduced the spectrum of
all $\frac{1}{8}$ BPS states of Yang-Mills theory. Note that at
strong coupling, some of these states are most naturally thought of
as multiparticle superpositions of ordinary (rather than giant)
gravitons. It follows that ordinary gravitons may be obtained from
the quantization of small spherical $D3$-branes, in much the same
way that they may be obtained from the quantization of strings.

Recall that the only properties of the symplectic form $\omega$ (and
hence of the precise nature of the action on the world volume of the
D3-brane) that we needed to derive \eqref{part} were $U(3)$
invariance, smoothness properties and the cohomology class of
$\omega$. Any deformation of $\omega_\mathrm{WZ}+\omega_\mathrm{BI}$
that preserves these properties would give the same result for
\eqref{part}. It is certainly plausible that these three features
are exact features of $\omega$; this would explain why the crude
considerations of this paper (based on a two derivative world volume
action on the D3-brane) manage to reproduce the exact formula even
for gravitons, (a feat that would seem to require considerably
greater precision). Of course supersymmetry underlies this
`miracle', as it appears to permit us to truncate the daunting
problem of the quantization of all solutions to the problem of
quantizing a finite dimensional subspace with rigid topology.

\subsection{Dual giant gravitons}

It is well known that Mikhailov's solutions do not exhaust the list
of $\frac{1}{8}$ D3-brane classical solutions. There exist well half
BPS (and so in particular $\frac{1}{8}$ BPS) D3-brane puffed out
into the $AdS_5$ rather than the $S^5$ directions.

It is familiar from the study of half BPS states that giant
gravitons and dual giant gravitons are not independent solutions,
but are instead dual to each other. We pause here to review this.
The spectrum of half BPS states may be put in one to one
correspondence with Young Tableaux of $SU(N)$. The state
corresponding to a given Young Tableaux may be regarded either as a
collection of as many ordinary giant gravitons as the tableaux has
columns or as a collection of a collection of as many dual giant
gravitons as the Tableaux has rows. In particular a single dual
giant graviton of angular momentum $k$ may equally well be thought
of as  collection of $k$ ordinary giant gravitons, each with unit
angular momentum.

A similar equivalence seems to work for the full spectrum of
$\frac{1}{8}$ BPS states. Recall that a single half BPS dual giant
graviton has a world volume that wraps the $S^3$ (of a given radius)
in $AdS_5$, but is located at a single point on $S^5$. Using the
fact that $\frac{1}{8}$ BPS configurations are necessarily
spherically symmetric on the $S^3$ in $AdS_5$, as they can carry no
$AdS$ angular momentum, the most general configuration of
$\frac{1}{8}$ BPS giant gravitons is a simple superposition of
(arbitrarily $U(3)$ rotated) single dual giant gravitons. Moreover
G. Mandal and N. Suryanarayana \cite{Nemani} have argued that the
bosonic spectrum of single $\frac{1}{8}$ BPS giant graviton turns
out to be that of a 3 dimensional bosonic harmonic oscillator. It
follows that the spectrum of collection of $N$ such dual gravitons
reproduces the spectrum of ordinary giant gravitons derived in this
paper.

Demanding the presence of exactly $N$ dual giant gravitons is not as
unreasonable as it may first seem \cite{Nemani}. Dual gravitons in
the ground state of the Harmonic Oscillator carry no charge, and so
are indistinguishable from nothing. Thus the restriction on number
is really an upper bound, which is also rather intuitive, appearing
to follow from the intuitive requirement that the flux at the center
of $AdS$ be positive.

Modulo cleaning up some loose ends, it thus seems that the partition
function obtained by quantizing dual giant gravitons is identical to
the partition function obtained by quantizing giant gravitons;
further both of these are equal to the partition function over the
classical chiral ring of \cite{index}. Note that, according to this
interpretation, the dual giant gravitons should be identified with
the $N$ bosons in the partition function \eqref{part}.

\subsection{A phase transition for the nucleation of a bubble?}

The authors of \cite{index} have argued that the partition function
\eqref{part} undergoes a phase transition as a function of scaled
chemical potentials, in the large $N$ limit. Only a finite number
(out of the $N$ bosons) are out of their ground state in the `low
temperature' phase while only a finite number of the $N$ bosons
occupy the ground state in the `high temperature' phase. From the
viewpoint of \eqref{part}, the phase transition between these phases
is simply Bose condensation.

We will now investigate the bulk interpretation of these two phases.
The partition function of the low temperature phase is identical to
that of supersymmetric gravitons in $AdS_5 \times S^5$. It follows
that the `low temperature' phase should be identified with a gas of
gravitons in an undeformed ambient $AdS_5 \times S^5$.

On the other hand the  `high temperature' phase should be described
by a bulk solution with vanishing 5 form flux at the origin of
$AdS_5$ \footnote{The origin may be defined as the fixed point of
the $SO(4)$ $AdS_5$ isometry.}; this is the bulk order parameter for
the phase transition.  It thus appears that the bulk dual to the
high temperature phase should be the close analogue of an enhancon
solution \cite{Johnson:1999qt}. It would be fascinating to find the
explicit bulk solution.

\subsection{Extensions for the future}

\subsubsection{Generalization to IIB theory on $AdS_5 \times
L^{abc}$}

Over the last few years a number of authors have discovered an
infinite number of generalizations of the $AdS/CFT$ conjecture
\cite{Gauntlett:2004hh, Gauntlett:2004hs, Martelli:2004wu,
Benvenuti:2004dy, Martelli:2005tp, Martelli:2005wy,
Benvenuti:2005ja, Franco:2005sm, Butti:2005sw, Gauntlett:2005ww,
Martelli:2006yb}. These generalizations establish an equivalence
between a class of $\mathcal{N}$ =1 quiver gauge theory and IIB
theory on $AdS_5 \times L^{abc}$, where $L^{abc}$ is a 5 dimensional
space that may be thought of as the base of a (singular) six
dimensional Calabi-Yau space.

It should be possible to evaluate the spectrum of giant gravitons
(and dual giant gravitons) on $AdS_5 \times L^{abc}$, and to compare
this to the spectrum over the chiral ring of the corresponding
${\cal N}=1$ field theories. We hope to report on this in the near
future.

\subsubsection{Generalization to $\frac{1}{16}$ BPS states}

The spectrum of $\frac{1}{16}$ BPS states in $\mathcal{N}=4$ Yang
Mills theory is much more intricate, and much more poorly
understood, than its $\frac{1}{8}$ BPS counterpart. This problem may
be attacked from three different directions

\begin{enumerate}
\item By an enumeration of the classical cohomology of the
relevant supersymmetry operator (see \eqref{SusyonFields}), a
counting problem whose solution should yield the full spectrum of
$\frac{1}{16}$ BPS states at least weak coupling.
\item By the construction and quantization of all $\frac{1}{16}$
BPS giant gravitons, a procedure that should yield the strong
coupling supersymmetric spectrum of $\mathcal{N}=4$ Yang Mills
theory at least at intermediate energies (energies larger than order
unity but smaller than order $N^2$) and strong coupling.
\item By a study of the supersymmetric black holes of \cite{Gutowski:2004ez,
Gutowski:2004yv, Chong:2005da, Chong:2005hr, Kunduri:2006ek} and
references  therein (valid at strong coupling and energies larger
than or of order $N^2$).
\end{enumerate}

While the three different approaches listed above have logically
distinct domains of validity, we suspect that all three approaches
will give the same answer. In particular it is possible that the
generalization of  the quantization of one sixteenth BPS giant
gravitons could reproduce the full (finite $N$ finite $\lambda$)
$\frac{1}{16}$ BPS spectrum. To this end we have already partially
generalized Mikhailov's classical construction to one sixteenth
giant gravitons. We plan to  study this enlarged solution space and
its quantization, and hope to have reportable results in the not too
distant future.

\section*{Acknowledgements}

We would like to thank A. Dabholkar, A. Dhar, R. Gopakumar, L.
Grant, J. Maldacena, S. Mukhi, A. Nair, S. Nampuri, N. Nitsure, S.
Trivedi and S. Wadia, for interesting discussions. We would
especially like to thank C. Beasley for many helpful comments and G.
Mandal and N. Suryanarayana for explaining their results on dual
giant gravitons prior to publication. The research of DG was
supported in part by DOE grant DE-FG02-91ER40654. The research of SL
was supported in part by NSF Career Grant PHY-0239626 and the Vineer
Bhansali Graduate Travel Fellowship. The research of SM was
supported in part by NSF Career Grant PHY-0239626 and DOE grant
DE-FG01-91ER40654.

\appendix

\section{Geometric quantization}\label{geoquant}

In this appendix, we give a brief review of the main ideas of
geometric quantization, glossing over all subtleties. See, for
instance, \cite{woodhouse, Echeverria-Enriquez:1999jr} for more
details.

\subsection{The set up}

Consider a manifold with a $U(1)$ line bundle, whose curvature,
$\omega$ is an invertible two form called the symplectic form. In
local coordinates, $\omega=\dr\theta =\omega_{ij}\frac {dx^i \wedge
dx^j }{2}$ where $\omega_{ij}$ is an invertible matrix whose inverse
we denote by $\omega^{ij}$.

Given any function $f$ on the manifold, we may be associate to it
the vector field $(X_f)^i=\omega^{ij}\partial_j f$. The Poisson
Bracket of two classical functions, $f_1$ and $f_2$ in phase space,
is given by $\omega^{ij} \partial_i f_1 \partial_j f_2$.

We wish to quantize the classical description, replacing classical
points on phase space with functions on phase space (states), and
replacing real classical functions on phase space with Hermitian
operators acting on states. We demand that this replacement is
linear, maps the classical Poisson Bracket to the quantum commutator
and maps the constant classical function to the constant quantum
(multiplication) operator.

The correspondence $f \rightarrow -\ir\hbar X_f$ is linear and maps
the Poisson bracket to the commutator, but maps the constant
function to the zero operator, and so is unacceptable. It is easy to
check that the alternate map
\begin{equation}\label{map}
f \rightarrow \omega^{ij}\partial_i f \left( \ir\hbar \partial_j +
\theta_j \right) + f \equiv (\ir\hbar) \partial_i f \omega^{ij}
D_j+f\, ,
\end{equation}
satisfies all our conditions, where the covariant derivative is
\begin{equation}\label{partialder}
  D_i= \partial_i -\frac{\ir}{\hbar} \theta_i\, .
\end{equation}
The covariant derivative is well defined only on functions that
transform like charged fields under symplectic gauge transformation
\begin{equation}\label{functiontransf}
\theta_i \rightarrow \theta_i+ \partial_i u\, , \;\;\; \phi
\rightarrow \e^\frac{\ir u}{\hbar} \phi\, .
\end{equation}
Consequently, wavefunctions are sections of charge one under the
symplectic line bundle, \eqref{partialder} is the connection and
$\omega$ is its curvature.

The space of all such sections would form too large a Hilbert Space
(yielding functions of both $x$ and $p$). We should restrict
attention to line bundles that are functions `of $p$ only'. One way
to achieve this is to restrict attention to sections that are
covariantly constant along an (arbitrarily chosen) `polarization'. A
polarization is defined as the collection of  $n$ dimensional vector
spaces generated by locally defined $n$ independent vector fields on
the manifold, with two additional restrictions. The first
restriction that the Lie bracket of any two of these vector fields,
restricted to any point, belongs to the $n$ dimensional subspace
described above; in other words we have a foliation. The second
restriction is that the point wise contraction of any two of these
vectors with the symplectic form vanishes.

More pictorially,  a choice of (real) polarization foliates phase
space into an $n$ parameter set of $n$ dimensional Lagrangian
submanifolds (a submanifold of phase space is Lagrangian if the
restriction of the symplectic form is zero). Given a polarization,
we are instructed to restrict attention to sections that obey
\begin{equation}\label{realpolarization}
 v^i D_i \phi=0\, ,
\end{equation}
where $v^i$ is a tangent vector along the submanifolds. As the
restriction of $\omega$ to these submanifolds is zero, this
condition is integrable. The set of unit charge line bundles that
obey \eqref{realpolarization} constitutes the Hilbert Space of our
system.

One can always choose a symplectic potential that is `adapted' to
the polarization, i.e.\ its components along the Lagrangian
submanifolds are zero. We will always make this choice in what
follows.

Once we have adopted a choice of polarization, not all classical
functions are associated with quantum operators. We require that the
operator defined by \eqref{map} preserves the polarization condition
\eqref{realpolarization}. For any given physical application it is
important that the polarization \eqref{realpolarization} be chosen
such that all important operators pass this test.

We illustrate these ideas with a familiar example. Consider the
quantum mechanics of a single particle in one dimension. Here
$\omega= \dr p \wedge \dr x$. Choosing the polarization $D_x \phi
=0$, and choosing the symplectic potential $\theta= -x\, \dr p$, we
find that our Hilbert space consists of functions of $p$; the
momentum operator maps to multiplication by $p$, while the position
operator maps to $\ir \hbar \partial_p$, in agreement with the usual
formulae.

In order to complete the specification of our Hilbert space we must
define an inner product on states. One may be tempted to use:
\begin{equation*}
\langle\phi|\psi\rangle =
\int{\bar\phi}\psi\frac{\omega^n}{n!(2\pi)^n}\, ,
\end{equation*}
However, this would not map real observables to hermitian operators
if $\theta$ were complex. The way around this is to replace
${\bar\phi}\psi$ with the hermitian structure $(\phi,\psi)$ of the
bundle (i.e.\ a pointwise  inner product for the charged fields that
replaces their simple pointwise product) that is compatible with the
connection. The last requirement translates into the equation
\begin{equation*}
\p_i(\phi,\psi)=(D_{\bar \imath}\phi,\psi)+(\phi,D_i\psi)\, .
\end{equation*}
For the connection \eqref{partialder}, this means:
\begin{equation}\label{hermstr}
\begin{split}
(\phi,\psi)&={\bar\phi}\psi\, W(x)\, ,\\
\p_i W(x) &= \frac{\ir}{\hbar}({\bar\theta_i}-\theta_i) W(x)\, ,
\;\;\;{\overline W} = W\, .
\end{split}
\end{equation}
It must also transform appropriately under gauge transformations.
Note that \eqref{hermstr} only needs to hold for those derivatives
that do not appear in \eqref{realpolarization} and so $W$ is not
completely specified by \eqref{hermstr}.  In the example considered
above, we could choose $W$ to be any unit normalized function of
$x$, so the inner product reduces to the familiar $\int\!\dr p\;
\bar{\phi}(p)\psi(p)$.

While our description of the Hilbert space depends on a choice of
polarization, under some circumstances Hilbert spaces formed with
different polarizations may be shown to be unitarily related. See
Ch.10 of \cite{woodhouse} for details.

We have skipped over several important subtleties in this lightening
review. Perhaps the most important of these is our omission of a
discussion of the so called  `half form' metaplectic correction. We
will not even describe this correction here, referring the
interested reader, once again, to Ch.10 of \cite{woodhouse}.

\subsection{Holomorphic quantization}\label{holquant}

We now turn to the `coherent state' or holomorphic quantization of
phase space. A complex structure on $J$ phase space is said to be
compatible with the symplectic form if $\omega$ is of type (1,1) with
respect to $J$. The key simplification in this case is the ability
to impose holomorphic polarization ${D}_{\bar z^i} \phi =0$; the
Hilbert Space may be identified with the space of square integrable
holomorphic sections.

\subsection{K\"ahler quantization}

When $\omega$ is a nondegenerate
$(1,1)$ form with respect to a complex structure
$J$, and when $\omega$ obeys $\p \omega =\bp \omega=0$, it may be
thought of as a K\"ahler class, and may be derived locally from a
K\"ahler Potential, $\omega=\ir\p\bp K$.

If we use  the K\"ahler form as a symplectic form, we can use the
holomorphic polarization mentioned in the previous subsection. The
symplectic potential $\theta=-\ir\p K$ is adapted to this
polarization, equation \eqref{realpolarization} is solved by
holomorphic sections and \eqref{hermstr} is solved by
$W=\exp\left(-\frac{K}{\hbar}\right)$.

\subsection{Quantization of $\CP^{r-1}$}\label{cpholquant}

Consider the space $\CP^{r-1}$ built out of the $r$ projective
variables $g^\alpha$, with a $(1,1)$ symplectic form whose
cohomology class is that of $(2\pi N)\omega_\mathrm{FS}$. We will
determine the Hilbert space that follows from the holomorphic
quantization of this space. Note that the isomorphism class of a
holomorphic line bundle on $\CP^{r-1}$ is determined by the first
Chern class (as $\CP^{r-1}$ is simply connected and $H^2(\CP^{r-1},
\CO_{\CP^{r-1}}) =0$). The first Chern class of the tautological
line bundle $\mathcal{O}_{\CP^{r-1}}(1)$ on $\CP^{r-1}$ coincides
with the cohomology class given by the Fubini-Study K\"ahler form.
Suppose that we already know that the cohomology class of the
symplectic form on $\CP^{r-1}$ under consideration is $2\pi N$-times
the cohomology class given by the Fubini-Study K\"ahler form. The
space of holomorphic sections of the $N$-th tensor power of the line
bundle $\mathcal{O}_{\CP^{r-1}}(1)$ is the space of homogeneous
polynomials of degree $N$ with variables $g^\alpha$ for
$\alpha=1\ldots r$. Therefore, the holomorphic quantization yields
the Hilbert Space of homogeneous polynomials of degree $N$ with
variables $g^\alpha$.

\section{The $SU(2|3)$ content of $\frac{1}{8}$ cohomology}\label{susy}

$\mathcal{N}=4$ Yang-Mills has 6 scalar fields $\Phi_{ij}$, 4 chiral
fermions $\Psi_{i \alpha}$ and a gauge field $A_{\alpha{\dot
\beta}}$, where lower $SU(4)$ indices $i=1 \ldots 4$ are
antifundamental and upper indices are fundamental. The scalars obey
the complex conjugation rules $\Phi_{ij}^* = \Phi^{ij}$ where
$\Phi^{ij}=\frac{\epsilon^{ijkl} \Phi_{kl}}{2}$. The fermions are
complex and their complex conjugates are ${\overline \Psi}^{i {\dot
\alpha}}$.

The supersymmetry generators $Q^i_\alpha$ act on these fields as
(factors and signs are schematic in the next two equations)
\begin{equation} \label{SusyonFields}
\begin{split}
[Q^i_\alpha, \Phi_{jk}]
     &= \delta^i_j \Psi_{k\alpha} - \delta^i_k \Psi_{j\alpha}\,, \\
\{Q^i_\alpha, \Psi_{j \beta}\} &= 2\ir\, \delta ^i_j f_{\alpha\beta}
     + \ir \epsilon_{\alpha \beta} [\Phi_{jk}, \Phi^{ki}]\,, \\
\{Q^i_\alpha,  {\overline \Psi}^{j}_{\dot \beta}\}
     &= - 2\ir D_{\alpha {\dot \beta}} \Phi^{ij}\,, \\
[Q^i_\alpha, A_{\beta {\dot \gamma}}]
     &= \epsilon_{\alpha \beta} \overline{\Psi}^{i}_{\dot \gamma}\,,
\end{split}
\end{equation}
and the action of ${\overline Q}_i^{\dot\alpha}$ is obtained by taking complex
conjugate.

$Q^1_\alpha$ and ${\overline Q}_{1 {\dot \alpha}}$ generate a
$\mathcal{N}=1$ subalgebra of the $\mathcal{N}=4$ algebra. We will
now display the action of these supercharges on fields in
$\mathcal{N}=1$ language. Let $\Phi^{1\,m+\!1} = \bar{\phi}^m$ so
that $\half \epsilon_{p\,m\,n} \Phi^{m+\!1\,n+\!1} = \phi_p$. (here
the indices $m, n, p$ run from $1 \ldots 3$). Further let
$\Psi_{1\alpha}= \lambda_\alpha$, and $\Psi_{m+\!1\,\alpha}=
\psi_{m\alpha}$. Dropping the superscript on our special
supercharges we have
\begin{equation} \label{nosusy}
\begin{split}
[Q_\alpha, {\bar \phi}^m] &= 0\,,\\
[Q_\alpha, \phi_m] &= \psi_{m\alpha}\,, \\
\{Q_\alpha, \psi_{m \beta}\} &=  \ir \epsilon_{\alpha \beta}
          \epsilon_{mnp} [{\bar \phi}^n, {\bar \phi}^p ] \,,\\
\{Q_\alpha, \lambda_{\beta}\} &= 2\ir f_{\alpha \beta} - \ir
          \epsilon_{\alpha \beta} [\phi_m, \bar{\phi}^m]\,, \\
\{Q_\alpha,  \bar{\psi}^{m}_{\dot \beta}\}
          &= - 2\ir D_{\alpha {\dot \beta}} \bar{\phi}^{m}\,, \\
\{Q_\alpha,  \bar{\lambda}_{\dot \beta}\} &= 0\,,\\
[Q_\alpha, A_{\beta {\dot \gamma}}]
          &= \epsilon_{\alpha \beta} \bar{\lambda}_{\dot \gamma}\,.
\end{split}
\end{equation}
From these equations we identify $\psi$ and $\lambda$ as the
$\mathcal{N}=1$ chiralino and gaugino respectively.

The one eighth BPS states we study in this paper are in one to one
correspondence with states in the cohomology of the operators
$Q_\alpha$. It follows from \eqref{nosusy} that all such operators
are built out of simultaneously commuting `letters' $\bar \phi^m$
and $\bar \lambda_{\dot \alpha}$ (see \cite{index}). The resulting
Hilbert space is in one one correspondence with the Fock space of
$N$ identical, noninteracting particles propagating in the potential
of a 3 bosonic and 2 fermionic dimensional harmonic oscillator.

Notice that $\frac{1}{8}$ BPS states transform in representations of
that part of the superconformal algebra, $PSU(2,\!2|4)$, that
commutes with the superalgebra generated by $Q_\alpha$ and their
Hermitian conjugates. This commuting subalgebra is generated by
${\overline Q}_{m+\!1\,\dot\alpha} \equiv \widetilde{Q}_{m\,\dot\alpha}$
($m=1 \ldots 3$) along with Hermitian conjugates, and is the compact
superalgebra $SU(2|3)$.

In Free Yang-Mills theory, and within the cohomology of $Q_\alpha$,
the letters ${\bar \phi}^m$ and ${\bar \lambda}_{\dot \alpha}$
transform in an irreducible representation of $SU(2|3)$. In
particular
\begin{equation} \label{actqb}
\begin{split}
& [\widetilde{Q}_{m \dot \alpha}, {\bar \phi}^n]
   = \delta_m^n {\bar \lambda}_{\dot \alpha}\,, \\
& \{\widetilde{Q}_{m \dot \alpha}, {\bar \lambda}_{\dot \beta}\} =
0\,.
\end{split}
\end{equation}

In the rest of this appendix we will study the unitary
representations of this compact superalgebra. The only result
derived below, that we use in the bulk of the paper,  is easily
stated. Let $M_{\frac{1}{8}}$ represent the $\frac{1}{8}$ BPS
cohomology, and let $M_{\frac{1}{8}}^S$ represent that part of the
cohomology that is formed only from the fields ${\bar \phi}^m$. We
will demonstrate below that all states in $M_{\frac{1}{8}}$ may be
obtained by the action of (an arbitrary number of applications of)
$\widetilde{Q}^{\dot \alpha}$ on $M_{\frac{1}{8}}^S$. The reader
who feels that \eqref{actqb} already makes this result quite
plausible, and who is otherwise uninterested in the representation
theory of superconformal algebrae, may skip to the next appendix.

\subsection{$PSU(2,\!2|4)$: Algebra and unitary constructions }

In this brief subsection we recall the construction and
representation theory of $PSU(2,\!2|4)$; in order to appreciate how
the $SU(2|3)$ generators we use below are related to the more
familiar symmetry generators of $\mathcal{N}=4$ Yang-Mills.

The superalgebra $SU(4,2|2)$ has a simple unitary implementation on
the Hilbert space of 4 bosonic and 4 fermionic oscillators. Let the
bosonic oscillators be denoted by $a_\alpha$, $a^\alpha$, $b_{\dot
\beta}$, $b^{\dot \beta}$ ($\alpha=1,2$, ${\dot \beta}=1,2$)
(upper/lower indices are creation/annihilation operators). Let the
fermionic oscillators be $\gamma^i, \gamma_i$ $(i=1 \ldots 4)$. The
operators $Q^{i\alpha}$ are implemented by $a^\alpha \gamma^i$.
Similarly $\overline{Q}_i^{\dot \alpha}= b^{\dot \alpha} \gamma_i$.
Superconformal generators are given by the Hermitian conjugates of
these formulae $S_{i \alpha}=a_\alpha \gamma_i$ and
$\overline{S}^i_{\dot \alpha}= b_{\dot \alpha} \gamma^i$. Each of
these operators commutes with the `Supertrace'
\begin{equation*}
ST= a^\alpha a_\alpha - b^{\dot \beta} b_{\dot \beta}
    -\gamma^i \gamma_i +2\, .
\end{equation*}
As a consequence it is consistent to restrict attention to the sub
Hilbert space $ST=0$; we will do this in what follows.

Notice that
\begin{equation}\label{nfcr}
\begin{split}
\{S_{i \alpha}, Q^{j \beta} \} = &
 \{a_\alpha \gamma_i, a^\beta \gamma^j \} \\
= & a_\alpha a^\beta\, \delta_i^j-\delta_\alpha^\beta\, \gamma^j \gamma_i \\
=& J_\alpha^\beta\, \delta_i^j + \delta_\alpha^\beta\,
 R^j_i + \delta^j_i\, \delta^\beta_\alpha\, \frac{\Delta}{2}\,,
\end{split}
\end{equation}
with
\begin{equation}\label{defs}
\begin{split}
&\Delta=\frac{a^\alpha a_\alpha +b^{\dot \alpha} b_{\dot \alpha} +2}{2}\,, \\
&J^\beta_\alpha = \left(a_\alpha a^\beta
                    - \delta_\alpha^\beta\frac{a_\gamma a^\gamma}{2} \right)
                = J^a (T^a)_\alpha^\beta\,, \\
& R^j_i = \left( \gamma_i \gamma^j
              - \delta^j_i\frac{\gamma_k \gamma^k}{4}\right)
        = R^p (\widetilde{T}^p)_i^j\,,
\end{split}
\end{equation}
where $T^a$ and $\widetilde{T}^p$, respectively, are the generators of
$SU(2)$ and $SU(4)$ in the fundamental representation and we have
used $ST=0$ in the last line of \eqref{nfcr}. The operator $\Delta$
in \eqref{nfcr} is the Hamiltonian or generator of scale
transformations.

Irreducible lowest weight representations of $PSU(2,\!2|4)$ contain
a distinguished set of primary states that are all annihilated by
$S_{\alpha i}$ and ${\overline S}_{\dot \beta}^i$. Such states have
definite energy (eigenvalue of $\Delta$) and appear in irreducible
representations of $SU(2) \times SU(2)\times SU(4)$.

Let $R_k$ ($k=1 \ldots 3$) denote the $SU(4)$ Cartan generators
whose $k^{th}$ diagonal element is unity and $(k+1)^{th}$ diagonal
element is minus one. We label representations of $SU(4)$ by the
eigenvalues of $R_1, R_2, R_3$ on the highest weight state in the
representation. Equation
\eqref{nfcr} together with unitarity can be used to
derive inequalities on the quantum numbers of primary states
\begin{equation} \label{unitarity}
\begin{split}
& E \geq E_1\,, \\
& E \geq E_2\,, \\
& E_1=2j_1 +2 -2 \delta_{j_1 0}+ \frac{3 R_1 +2 R_2 + R_3 }{2}\,, \\
& E_2=2j_2+2-2\delta_{j_2 0} +\frac{3R_3 +2 R_2 + R_1}{2}\,.
\end{split}
\end{equation}

\subsection{Oscillator construction and unitarity of $SU(2|3)$ }

The superalgebra $SU(2|3)$ is a subalgebra of $PSU(2,\!2|4)$; this
sub-superalgebra is represented on the subspace of the oscillator
Hilbert space of the previous subsection defined by $a_{\alpha}
|\psi \rangle = \gamma^1 |\psi \rangle = 0$, i.e.\ states
annihilated by $Q^{1\alpha} = a^\alpha\gamma^1$ and $S_{1\alpha} =
a_\alpha\gamma_1$. On this subspace the constraint $ST=0$ reduces to
$ST'= -b^{\dot \beta} b_{\dot \beta} -\gamma^m \gamma_m + 1 =0$. It
is easy to check that
\begin{equation}\label{nfcrn}
\begin{split}
\{ {\widetilde S}^m_{\dot \alpha} , {\widetilde Q}_n^{\dot \beta} \}
=& b_{\dot \alpha} b^{\dot \beta} \delta^m_n-\delta_i^j \gamma_n \gamma^m \\
=& {\overline J}_{\dot \alpha}^{\dot \beta}\, \delta^m_n
- \delta_{\dot\alpha}^{\dot \beta}\, U_n^m
+ \delta_n^m\, \delta^\beta_\alpha\, \frac{\Delta}{3}\,, \\
\end{split}
\end{equation}
with
\begin{equation} \label{defsnn}
\begin{split}
&\Delta=\frac{b^{\dot \alpha} b_{\dot \alpha} +2}{2}\,, \\
&{\overline J}^{\dot \beta}_{\dot \alpha}
  = \left(b_{\dot \alpha} b^{\dot \beta}
  - \delta_{{\dot \alpha}}^{{\dot \beta}}\,
    \frac{b_{\dot \gamma} b^{\dot \gamma}}{2} \right)
  = {\overline J}^a (T^a)^{\dot \beta}_{\dot \alpha}\,, \\
&U_n^m = \left( \gamma_n \gamma^m
              - \frac{\gamma_p \gamma^p}{3}\,\delta^n_m\right)
       = U^b (\widetilde{T}^b)^m_n\,,
\end{split}
\end{equation}
where $T^a$ and $\widetilde{T}^b$ are generators of $SU(2)$ and $SU(3)$
in the fundamental representation.

Let $U_m$  denote the diagonal $SU(3)$ matrices with unity in the
$m^{th}$ row and negative unity in the $(m+1)^{th}$ row
respectively. Note that when $SU(2|3)$ is embedded within
$PSU(2,\!2|4)$ we have
\begin{equation}\label{rel}
U_1=R_2\,, \; U_2=R_3\,.
\end{equation}

All unitary $SU(2|3)$ representations may be obtained by acting on a
set of distinguished primary or lowest weight states with an
arbitrary number of ${\widetilde Q}_{m \dot \alpha}$ generators.
Primary states are annihilated by all ${\widetilde S}^{m}_{\dot
\alpha}$, and appear in representations of $SU(3)\times SU(2) \times
\Delta$ (here $\Delta$ is the energy operator). The requirement of
unitarity, together with \eqref{nfcrn}, yields inequalities on the
energy of primary states of irreducible representations of $SU(2|3)$
as a function of the $SU(3)\times SU(2)$ quantum numbers of these
states. In particular we find
\begin{equation} \label{newunit}
E \geq U_1+2U_2 +3(j_2+1-\delta_{j_2 0})=E_2+ \half(E_2-E_1)\,,
\end{equation}
where we have used \eqref{rel} and \eqref{unitarity}.

\subsection{Representation content of the $\frac{1}{8}$ BPS
cohomology}

Recall that the $\frac{1}{8}$ BPS cohomology may be constructed via
a two step procedure. We first construct a single particle Hilbert
space $\mathcal{H}$, the Hilbert space of a three bosonic and two
fermionic dimensional harmonic oscillator. We then second quantize
this description, promoting states in the single particle Hilbert
space to creation operators, and study the $N$ particle spectrum of
this second quantized formulation.

The first quantized Hilbert space is easily decomposed into
representations of $SU(2|3)$. Let $M_p$ denote the (short) $SU(2|3)$
representation with highest weights $(U_1, U_2)=(p,0)$, ${\overline J}=0$
and $\Delta=p$. Notice these quantum numbers saturate the bound
\eqref{newunit}; a circumstance that is partly responsible for their
simplicity. It is not difficult to explicitly construct the
representation $M_p$ (for instance by using a $p$ flavour version of
the oscillator construction of the previous subsection). It turns
out that $M_p$ decomposes into representations of $SU(3) \times
SU(2) \times \Delta$ as
\begin{equation}\label{mpde}
M_p= (p, 0; 0; p)+ \left(p-1, 0; \half; p+\half\right)
    + ( p-2, 0; 0; p+1)\,,
\end{equation}
where the numbers stand for $(U_1, U_2; {\overline J}; \Delta)$. It then
follows that the single particle Hilbert space $\mathcal{H}$ of the
previous paragraph decomposes into $SU(2|3)$ representations as
\begin{equation}\label{sps}
\mathcal{H}=\sum_{p=0}^\infty M_p\,.
\end{equation}
In more detail, let $a_p$ denote the purely bosonic excitations in
$\mathcal{H}$ with $p$ quanta excited, let $b_p$ denote states in
$\mathcal{H}$ with $p$ bosonic and one fermionic quanta occupied,
and let $c_p$ denote the states in $\mathcal{H}$ with $p$ bosonic
and two fermionic quanta occupied. The decomposition \eqref{mpde},
in terms of states, is simply
\begin{equation*}
M_p=a_p + b_{p-1} + c_{p-2}\,.
\end{equation*}
Note in particular that the purely bosonic $a_p$ are the primary
states (states in the representation $(p, 0; 0; p)$ that are
annihilated by ${\widetilde S}_{\dot \alpha}^m$) in this decompositions;
none of the other states ($b_p$ or $c_p$) are annihilated by all
${\widetilde S}_{\dot \alpha}^m$.

Upon second quantizing the states $a_p, b_p$ and $c_p$ each become
creation operators that inherit their $SU(2|3)$ transformation
properties from the corresponding states. The $\frac{1}{8}$ BPS
cohomology is obtained by acting on a vacuum with $N$ different
creation operators (any of $a_p, b_p, c_p$) on the Fock vacuum. It
follows that ${\widetilde S}_{\dot \alpha}^m$ annihilates states obtained
by acting on the vacuum with $a_p$'s only; further these are the
only states annihilated by  all ${\widetilde S}_{\dot \alpha}^m$. As a
consequence the state obtained by second quantizing the subspace of
$\mathcal{H}$ consisting of the the bosonic harmonic oscillator
(with the fermionic oscillators in their vacuum)  constitute all the
lowest weight (primary) states under $SU(2|3)$; all other states in
the $\frac{1}{8}$ BPS cohomology may be obtained by acting on these
primary states with (sufficient applications of) ${\widetilde Q}_{m \dot
\alpha}$.

\section{Details of the symplectic form and charges}\label{sympdet}

In this Appendix we present the algebraic manipulations that allow
us to replace the symplectic form \eqref{sympforminto}  with
\eqref{sympformintt}. We also present a couple of explicit formulae
for the energy of giant gravitons as a function of the intersecting
polynomial.

\subsection{From the Wess-Zumino coupling}\label{sympdetWZ}

Let us write $\omega_\mathrm{WZ}=\half\,\omega_{ij}\,\dr
c^i\wedge\dr c^j$, where $c^i$ parameterize solutions to the
equations of motion. From \eqref{sympforminto}, we have:
\begin{equation}\label{sympabs}
\begin{split}
\omega_{ij} =&\int\!\!\dr^3\sigma\: \frac{\delta}{\delta c^i}\left[
 A_{\mu_0\mu_1\mu_2\mu_3}
  \frac{\p x^{\mu_1}}{\p\sigma^1}
  \frac{\p x^{\mu_2}}{\p\sigma^2}
  \frac{\p x^{\mu_3}}{\p\sigma^3}\right]
\frac{\delta x^{\mu_0}}{\delta c^j}
   - i\lr j
\\ =& \int\!\!\dr^3\sigma \left\{
\p_\nu A_{\mu_0\mu_1\mu_2\mu_3}
 \frac{\delta x^{\nu}}{\delta c^i}
 \frac{\delta x^{\mu_0}}{\delta c^j}
 \frac{\p x^{\mu_1}}{\p\sigma^1}
 \frac{\p x^{\mu_2}}{\p\sigma^2}
 \frac{\p x^{\mu_3}}{\p\sigma^3}\right.
\\ &+ \left.A_{\mu_0\mu_1\mu_2\mu_3}
 \frac{\delta x^{\mu_0}}{\delta c^j}
 \frac{\p}{\p\sigma^1}\left[\frac{\delta x^{\mu_1}}{\delta c^i}\right]
 \frac{\p x^{\mu_2}}{\p\sigma^2}
 \frac{\p x^{\mu_3}}{\p\sigma^3}
+\mathrm{cyclic}(1,2,3)\right\} -i\lr j\, .
\end{split}
\end{equation}
We deal with the second term by integrating by parts in $\sigma^1$
in the original version, but not the $i\lr j$ version.
\begin{equation*}\begin{split}
\omega_{ij} =&\int\!\!\dr^3\sigma\left\{
 \p_\nu A_{\mu_0\mu_1\mu_2\mu_3}
  \frac{\delta x^{\nu}}{\delta c^i}
  \frac{\delta x^{\mu_0}}{\delta c^j}
  \frac{\p x^{\mu_1}}{\p\sigma^1}
  \frac{\p x^{\mu_2}}{\p\sigma^2}
  \frac{\p x^{\mu_3}}{\p\sigma^3}\right.
\\ &-\p_\nu A_{\mu_0\mu_1\mu_2\mu_3}
  \frac{\delta x^{\nu}}{\delta c^j}
  \frac{\delta x^{\mu_0}}{\delta c^i}
  \frac{\p x^{\mu_1}}{\p\sigma^1}
  \frac{\p x^{\mu_2}}{\p\sigma^2}
  \frac{\p x^{\mu_3}}{\p\sigma^3}
\\ &-\p_\nu A_{\mu_0\mu_1\mu_2\mu_3}
  \frac{\p x^{\nu}}{\p\sigma^1}
  \frac{\delta x^{\mu_0}}{\delta c^j}
  \frac{\delta x^{\mu_1}}{\delta c^i}
  \frac{\p x^{\mu_2}}{\p\sigma^2}
  \frac{\p x^{\mu_3}}{\p\sigma^3}
\\ &- A_{\mu_0\mu_1\mu_2\mu_3}
 \frac{\delta x^{\mu_0}}{\delta c^j}
 \frac{\delta x^{\mu_1}}{\delta c^i}
 \left[\frac{\p^2 x^{\mu_2}}{\p\sigma^1\p\sigma^2}
  \frac{\p x^{\mu_3}}{\p\sigma^3}
 +\frac{\p x^{\mu_2}}{\p\sigma^2}
  \frac{\p x^{\mu_3}}{\p\sigma^1\p\sigma^3}\right]
\\ &-A_{\mu_0\mu_1\mu_2\mu_3}
 \frac{\p}{\p\sigma^1}\left[\frac{\delta x^{\mu_0}}{\delta c^j}\right]
 \frac{\delta x^{\mu_1}}{\delta c^i}
 \frac{\p x^{\mu_2}}{\p\sigma^2}
 \frac{\p x^{\mu_3}}{\p\sigma^3}
\\ &\left.-A_{\mu_0\mu_1\mu_2\mu_3}
 \frac{\delta x^{\mu_0}}{\delta c^i}
 \frac{\p}{\p\sigma^1}\left[\frac{\delta x^{\mu_1}}{\delta c^j}\right]
 \frac{\p x^{\mu_2}}{\p\sigma^2}
 \frac{\p x^{\mu_3}}{\p\sigma^3}
+\mathrm{cyclic}(1,2,3)\right\} ,
\end{split}\end{equation*}
where the second and last terms come from the $i\lr j$ version.

The last two terms combined consist of something symmetric in
$\mu_0,\mu_1$ contracted with $A$, and therefore cancel. Regrouping
terms and relabeling dummy indices:
\begin{alignat}{2}
\omega_{ij} = \int\!\!\dr^3\sigma&\left(
  \frac{\delta x^{\alpha}}{\delta c^i}
  \frac{\delta x^{\beta}}{\delta c^j}\right)&&\left(
  \frac{\p x^{\gamma}}{\p\sigma^1}
  \frac{\p x^{\delta}}{\p\sigma^2}
  \frac{\p x^{\epsilon}}{\p\sigma^3}\right)
  (\p_\alpha A_{\beta\gamma\delta\epsilon}
    +\mathrm{cyclic})
\nonumber\\
 -&\left(\frac{\delta x^{\alpha}}{\delta c^i}
           \frac{\delta x^{\beta}}{\delta c^j}\right)
         &&\left[\frac{\p^2 x^{\mu}}{\p\sigma^1\p\sigma^2}
            \frac{\p x^{\nu}}{\p\sigma^3}
            (A_{\beta\alpha\mu\nu}+A_{\beta\mu\alpha\nu})\right.
\nonumber\\
        &&&\left.\frac{\p^2 x^{\mu}}{\p\sigma^2\p\sigma^3}
            \frac{\p x^{\nu}}{\p\sigma^1}
            (A_{\beta\nu\alpha\mu}+A_{\beta\nu\mu\alpha})\right.
\nonumber\\
        &&&\left.\frac{\p^2 x^{\mu}}{\p\sigma^3\p\sigma^1}
            \frac{\p x^{\nu}}{\p\sigma^2}
            (A_{\beta\alpha\nu\mu}+A_{\beta\mu\nu\alpha})\right]
\nonumber\\
\label{appsympcomp}
 \phantom{\omega_{ij}} = \int\!\!\dr^3\sigma &\left(
  \frac{\delta x^{\alpha}}{\delta c^i}
  \frac{\delta x^{\beta}}{\delta c^j}\right)&&\left(
  \frac{\p x^{\gamma}}{\p\sigma^1}
  \frac{\p x^{\delta}}{\p\sigma^2}
  \frac{\p x^{\epsilon}}{\p\sigma^3}\right)
  F_{\alpha\beta\gamma\delta\epsilon}\, .
\end{alignat}

As there are $N$ units of flux through the sphere (setting the brane
charge and the radius of $S^5$ to one), $F$ is $2\pi N$, times the
volume form of the $S^5$, divided by the volume of $S^5$. Therefore,
the result is $\frac{2N}{\pi^2}$ times the volume swept out by the
two deformations of the surface.

\subsection{From the Born-Infeld term}\label{sympdetBI}

We will denote the complex surface in $\C^3$ that appears in the
$\frac{1}{8}$ BPS solutions of \cite{mikhailov} by $S$ and its
intersection with $S^5$ (that is wrapped by the $D3$-brane) by
$\Sigma$.

The Born-Infeld contribution to the symplectic form can be
computed from its symplectic potential
\begin{equation}\label{thetaBI}
 \theta_\mathrm{BI}= \frac{N}{2\pi^2}
  \int_\Sigma\!\!\dr^3\sigma\,
   \sqrt{-g}\, g^{0\alpha}
   \frac{\partial x^\mu}{\partial \sigma^\alpha}\,
   {G}_{\mu \nu}\,\delta x^\nu\,.
\end{equation}

In a neighbourhood of a particular point of the $D3$-brane surface,
we choose space-time coordinates $(t,x^1,x^2,u^1,u^2,u^3)$ such that
the surface is given by $x^1,x^2$=constant and the contours
$u^i$=constant are perpendicular to it. We fix gauge on the
worldvolume by $\sigma^0=t, \sigma^i=u^i$ so that $x^i$ are the
dynamical fields.

In these coordinates, $\delta x^\nu$ is a deformation perpendicular
to the surface and the perpendicular velocity is
\begin{equation*}
v_\perp^\mu=\frac{\p x^\mu}{\p t}\,.
\end{equation*}
We also have
\begin{equation*}
    g_{00}=-(1-v_\perp^2)\,,
    \qquad g_{0i}=0\,,
    \qquad g_{ij}=G_{ij}\equiv(g_s)_{ij}\,.
\end{equation*}
Therefore, $\theta_\mathrm{BI}$ can be written as
\begin{equation}\label{thetaperp}
  \theta_\mathrm{BI} = \frac{N}{2\pi^2}
     \int_\Sigma\!\!\dr^3\sigma
      \sqrt{g_s}\frac{v_\perp\cdot\delta x}
                     {\sqrt{1-v_\perp^2}}\,.
\end{equation}

In \S6 of \cite{mikhailov}, Mikhailov showed that
\begin{equation*}
\frac{\sqrt{g_s}\dr^3\sigma}{\sqrt{1-v_\perp^2}}
 = 2\,\delta\!\left(|z^i|^2-1\right)
     \dr(\mathrm{vol} S)\: \,,
\end{equation*}
where $\dr(\mathrm{vol} S)$ is the induced volume form of the
complex surface $S$. In addition, the velocity is the perpendicular
component of $e_\parallel$, the unit vector in the direction of the
generator of a simultaneous rotation in the three two-planes of
$\C^3$. Therefore,
\begin{equation}\label{thetaS}
      \theta_\mathrm{BI} = \frac{N}{\pi^2}
        \int_S\!\!\dr(\mathrm{vol} C)\:
        \delta\!\left(|z^i|^2-1\right)
        (e_\parallel\cdot\delta x)\,.
\end{equation}

Let $J$ be the complex structure of $\C^3$. When we are not at the
boundaries of phase space and $\Sigma$ is fully three dimensional,
the tangent space of $\Sigma$ has a two dimensional subspace that is
closed under $J$. Let $e_\psi$ denote the unit vector in $T\Sigma$
that is perpendicular to this subspace.

The only tangent vector to $S$ that is not perpendicular to $\delta
x$ is $J\cdot e_\psi\,$. However, it is perpendicular to
$e_\parallel\,$:
\begin{equation*}
    (J\cdot e_\psi)\cdot e_\parallel = -e_\psi\cdot(J\cdot e_\parallel)
    =e_\psi\cdot e_\perp\,,
\end{equation*}
where $e_\perp$ is the unit vector perpendicular to $S^5$ in $\C^3$,
which is manifestly perpendicular to $e_\psi\,$.

Let $n_1$ and $n_2$ be orthogonal unit vectors perpendicular to $S$,
such that $J\cdot n_1=n_2$ and $J\cdot n_2=-n_1\,$. We have
\begin{equation*}
  e_\parallel\cdot\delta x
   =(e_\parallel\cdot n_1)(\delta x\cdot n_1)
    +(e_\parallel\cdot n_2)(\delta x\cdot n_2)
   =(e_\perp\cdot n_1)(\delta x\cdot n_2)
    -(e_\perp\cdot n_2)(\delta x\cdot n_1)\,.
\end{equation*}

Parameterizing the surface $S$ with coordinates
$\sigma^1\ldots\sigma^4$, we can rewrite \eqref{thetaS} as:
\begin{equation}\label{thetavol}
\begin{split}
  \theta_\mathrm{BI} &= \frac{N}{\pi^2}
     \int_S\!\!\dr^4\sigma\:
        \epsilon_{\mu_1\cdots\mu_6}
        \left[
        \frac{\p x^{\mu_1}}{\p\sigma^1}
        \cdots
        \frac{\p x^{\mu_4}}{\p\sigma^4}
        \right]
        n_1^{\mu_5}n_2^{\mu_6}\,2(n_1)_{[\alpha}(n_2)_{\beta]}
        \left(e_\perp^\alpha\delta x^\beta\right)
         \,\delta\!\left(|z^i|^2-1\right)
\\
    &= \frac{N}{\pi^2}
     \int_S\!\!\dr^4\sigma\:
        \epsilon_{\mu_1\cdots\mu_6}
        \left[
        \frac{\p x^{\mu_1}}{\p\sigma^1}
        \cdots
        \frac{\p x^{\mu_4}}{\p\sigma^4}
        \right]
        e_\perp^{\mu_5}\delta x^{\mu_6}
        \,\delta\!\left(|z^i|^2-1\right)\,.
\end{split}
\end{equation}
This is can be interpreted geometrically as follows:
\begin{itemize}
  \item Compute $\frac{N}{2\pi^2}$ times the volume swept out by the surface
$S$ under deformations by $e_\perp$ and $\delta x$ inside a sphere
of radius $r$;
  \item differentiate with respect to $r$;
  \item set $r=1$.
\end{itemize}

\subsection{Explicit formulae for the energy}

The energy $E=\sum_m L^m$ is of particular interest, and may be
determined to be \cite{mikhailov}
\begin{equation}\label{enofgrav}
 E= \frac{2N}{\omega_3} \int_S \dr(\mathrm{vol} S)\,
\delta\!\!\left( \sum_i |z^i|^2  -1\right) ,
\end{equation}
where $S$ is the 4 real dimensional surface of the curve $P(z^i)=0$
and $\dr(\mathrm{vol} S)$ is the pull back of the volume form on
$\R^6$ onto $S$ and $\omega_3$ is the volume of the unit 3 sphere.
This formula may be rewritten as
\begin{equation}\label{enofgravproc}
E=\frac{2N}{\omega_3} \int\!\! \dr^3z\, \dr^3\zb \:|\partial_i P|^2
\delta(P) \delta ({\bar P}) \delta\!\!\left(\sum_i |z^i|^2 -1
\right) .
\end{equation}

\section{Curves of degree two}\label{degtwoapp}

The family of holomorphic curves
\begin{equation}\label{degtwohom}
c_{ij}z^iz^j=1\,,
\end{equation}
displays more or less all the features that complicate the
quantization of holomorphic curves. As a consequence, a complete and
explicit analysis of the quantization of these curves could be very
valuable. We have not yet performed this (algebraically complicated)
analysis. In this appendix we lay the ground for this analysis by
laying out the issues and setting up the problem.

\subsection{$U(3)$ action}
The matrix
\begin{equation}\label{coefsq}
M_i\mbox{}^k=c_{ij}\cb^{jk}
\end{equation}
is hermitian and can be diagonalized by a $U(3)$ transformation:
$M=$diag$(\lambda_1^2,\lambda_2^2,\lambda_3^2)$. One can show that
the most general symmetric matrix satisfying \eqref{coefsq}
is\footnote{This is an oversimplification. When two of the
eigenvalues are equal there is a more general $c$. This corresponds
to the $U(2)$ subgroup of $U(3)$ that is not fixed by diagonalizing
$M$. A similar remark applies to the situation where all three
eigenvalues are equal.}
\begin{equation*}
c=\mathrm{diag}(\lambda_1\e^{\ir\alpha_1},\lambda_2\e^{\ir\alpha_2}
\lambda_3\e^{\ir\alpha_3})\,.
\end{equation*}
We can then use the remaining $U(1)^3$ symmetry to set the phases to
zero. We conclude that any matrix $c_{ij}$ may be written in the
form
\begin{equation*}
C=U DU^T\,,
\end{equation*}
where $U$ is a $U(3)$ matrix and $D$ is a diagonal matrix with real
positive eigenvalues. Notice that, in this way of writing it, the
12 real degrees of freedom of a complex symmetric matrix have been
distributed into the 9 degrees of freedom of $U$ and the 3 degrees
of freedom of $D$.

\subsection{Holes}

The nearest the curve $\sum_i\lambda_i (x^i)^2=1$ approaches the
origin is the minimum of $\frac{1}{\sqrt{ \lambda_i}}$ for $i=1
\ldots 3$. When $\lambda_i <1$ for $i = 1 \ldots 3$ the curve fails
to intersect the unit sphere; this region of non-intersection
represents a unit cube in the positive octant of $\lambda_1,
\lambda_2, \lambda_3$ space. The complement of this cube in the
positive octant is in one to one correspondence with the
intersections $\lambda_i (x^i)^2=1$ with the unit 5 sphere.

We now study the boundary of this region -- the faces, edges and
vertex of this cube -- in more detail. Consider the face at
$\lambda_1=1$ with $\lambda_2, \lambda_3<1$. Such a curve intersects
the unit 5 sphere at exactly two points $x=\pm 1$, $y=z=0$. This
boundary is an 11 dimensional manifold in the 12 dimensional space
of $c_{ij}s$; the 11 dimensions being spanned by $\lambda_2,
\lambda_3$ and an arbitrary $U(3)$ matrix.

Now let us turn to the edges of the cube, for example
$\lambda_1=\lambda_2=1$. The curve intersects the unit 5 sphere on
such an edge along the circle $x^2+y^2=1$ with $x, y$ real. These
edges constitute a 9 dimensional subspace in the space of
coefficients $c_{ij}$; the 9 dimensions are spanned by $\lambda_3$
together with the elements $U(3)/SO(2)$ (note that an $SO(2)$
subgroup of $U(3)$ acts in a manner so as to leave a curve with
$\lambda_1=\lambda_2$ invariant).

Finally the corner of the cube has
$\lambda_1=\lambda_2=\lambda_3=1$. This curve intersects the unit 5
sphere on the 2 sphere $x^2+y^2+z^2=1$ with all of $x, y, z$ real.
Such curves form a 6 dimensional subspace in the space of
coefficients, the six dimensions parameterize $U(3)/SO(3)$.

It would be fascinating to study $\omega_\mathrm{WZ}$ and
$\omega_\mathrm{full}$ on the $c_{ij}$ space, and in particular to
investigate whether the structure of these forms permits one to
shrink away the cubic hole described in this section in a smooth
manner, as our conjecture suggests should be the case. We will not
address this problem in this paper.

\subsection{Degenerations}\label{degenerate}

In order to study a second kind of potential singularity of the
symplectic form, let us study the limit $c_{ij} \rightarrow \infty$
for all $i, j$. In this limit our curve reduces effectively to a
generic homogeneous degree two polynomial, which may, up to $U(3)$
rotations, be chosen to have the form
\begin{equation}\label{spechom}
x^2+\lambda y^2+\epsilon z^2 = 0\,, \;\;\;\;
0\,\leq\,\lambda,\epsilon\,\leq\,1\,.
\end{equation}

\eqref{spechom} degenerates into two curves when the LHS of
\eqref{degtwohom} factorizes, i.e.\ when $\det c =0$. Note that
$\epsilon=0$ is the degeneration we will concentrate on. It is
interesting to study the symplectic form in the neighbourhood of
this degeneration point.

We use \eqref{expforsympform}:
\begin{equation*}
\omega_{M_k} = \frac{N}{4\pi\ir}\int_{C'}
\left(\frac{\dr\zb^1\wedge \dr\zb^2}{\dr\overline{f}}\right)
\wedge \left(\frac{\dr z^1 \wedge \dr z^2}{\dr f}\right)
\times \frac{1}{(1+ |z^1|^2 + |z^2|^2)^3}\:
\frac{\dr\overline{f} \wedge \dr f}{2\ir} \, ,
\end{equation*}

It is helpful to make the variable change
\begin{equation*}
\alpha=x+\ir\sqrt{\lambda}y\,, \;\;\;\;\;
\beta=x-\ir\sqrt{\lambda}y\,.
\end{equation*}
Working in the gauge $z=1$, we can solve \eqref{spechom} by
$\beta=-\epsilon / \alpha$. We have:
\begin{equation*}\begin{split}
\frac{\dr z^1 \wedge \dr z^2}{\dr f} &\ra
    -\frac{\dr\alpha}{2\ir\sqrt\lambda\frac{\p f}{\p\beta}}
    =-\frac{\dr\alpha}{2\ir\sqrt\lambda\alpha}\,,\\
x&=\frac{\alpha+\beta}{2} =
   \frac{1}{2}\left(\alpha-\frac{\epsilon}{\alpha}\right),\\
y&=\frac{\alpha-\beta}{2\ir\sqrt\lambda} =
   \frac{1}{2\ir\sqrt\lambda}\left(\alpha+\frac{\epsilon}{\alpha}\right),\\
|x|^2+|y|^2+|z|^2 &= 1+\frac{1}{4}\left\{
 \left(\frac{1}{\lambda}+1\right)
\left(|\alpha|^2+\frac{\epsilon^2}{|\alpha|^2}\right)
 +\left(\frac{1}{\lambda}-1\right)\epsilon
    \left(\frac{\alpha}{{\bar\alpha}}+\frac{{\bar\alpha}}{\alpha}\right)
\right\}.
\end{split}\end{equation*}

\eqref{expforsympform} becomes
\begin{equation*}
\omega= \frac{N}{16\pi\ir\lambda}\int\!
\frac{\dr{\bar\alpha}\dr\alpha}{|\alpha|^2} \frac{z^iz^i\zb_k\zb_l}
     {\left[1+\frac{1}{4}\left\{
 \left(\frac{1}{\lambda}+1\right)\left(|\alpha|^2+\frac{\epsilon^2}{|\alpha|^2}\right)
 +\left(\frac{1}{\lambda}-1\right)\epsilon
    \left(\frac{\alpha}{{\bar\alpha}}+\frac{{\bar\alpha}}{\alpha}\right)
\right\}\right]^3 } \frac{\dr\cb^{kl}\wedge\dr c_{ij}}{2\ir}\,.
\end{equation*}
So the most general integral we need to evaluate is:
\begin{equation*}\begin{split}
\frac{N}{8\pi\lambda}
\int \frac{\dr\alpha\dr{\bar\alpha}}{2\ir |\alpha|^6}& \frac{ \alpha^m
{\bar\alpha}^n  }{ \left[1+\frac{1}{4}\left\{
 \left(\frac{1}{\lambda}+1\right)\left(|\alpha|^2+\frac{\epsilon^2}{|\alpha|^2}\right)
 +\left(\frac{1}{\lambda}-1\right)\epsilon
    \left(\frac{\alpha}{{\bar\alpha}}+\frac{{\bar\alpha}}{\alpha}\right)
\right\}\right]^3} \\
&=\frac{N}{8\pi\lambda}
 \int\! \dr r\dr\phi\: \frac{r^{m+n+1}\e^{\ir(m-n)\phi}}{
 \left[r^2+\frac{1}{4}\left\{
  \left(\frac{1}{\lambda}+1\right)\left(r^4+\epsilon^2\right)
  +\left(\frac{1}{\lambda}-1\right)2\epsilon r^2\cos 2\phi
 \right\}\right]^3}\:,
\end{split}\end{equation*}
with $m,n=0,1,2,3,4$.

This becomes considerably easier at the special point $\lambda=1$:
\begin{equation*}
N\int \dr x \frac{x^m\, \delta_{mn}}{(x^2+2x+\epsilon^2)^3}\:.
\end{equation*}
This gives the following beautiful formulae:
\begin{equation}\label{singomega}
 \begin{split}
   \omega_{\overline{11},11}=\omega_{\overline{22},22} & =
       \frac{N}{256\ir}\left(
          \frac{2(2-11\epsilon^2)\sqrt{1-\epsilon^2}
                 -\epsilon^2(4+5\epsilon^2)\log\left[
                               \frac{2-\epsilon^2-2\sqrt{1-\epsilon^2}}
                                    {\epsilon^2}\right]}
               {(1-\epsilon^2)^{5/2}}\right), \\
   \omega_{\overline{22},11}  & =
      -\frac{N}{256\ir}\left(
          \frac{2(2+\epsilon^2)\sqrt{1-\epsilon^2}
                 -\epsilon^2(4-\epsilon^2)\log\left[
                               \frac{2-\epsilon^2-2\sqrt{1-\epsilon^2}}
                                    {\epsilon^2}\right]}
               {(1-\epsilon^2)^{5/2}}\right), \\
   \omega_{\overline{12},12}  & =
       \frac{N}{256\ir}\left(
          \frac{2(2-5\epsilon^2)\sqrt{1-\epsilon^2}
                 -3\epsilon^4\log\left[
                               \frac{2-\epsilon^2-2\sqrt{1-\epsilon^2}}
                                    {\epsilon^2}\right]}
               {(1-\epsilon^2)^{5/2}}\right), \\
   \omega_{\overline{13},13}=\omega_{\overline{23},23} & =
       \frac{N}{64\ir}\left(
          \frac{2(1+2\epsilon^2)\sqrt{1-\epsilon^2}
                 +3\epsilon^2\log\left[
                               \frac{2-\epsilon^2-2\sqrt{1-\epsilon^2}}
                                    {\epsilon^2}\right]}
               {(1-\epsilon^2)^{5/2}}\right), \\
   \omega_{\overline{33},11}=\omega_{\overline{33},22} & =
       \frac{N}{64\ir}\left(
          \frac{6\epsilon\sqrt{1-\epsilon^2}
                 +\epsilon(2+\epsilon^2)\log\left[
                               \frac{2-\epsilon^2-2\sqrt{1-\epsilon^2}}
                                    {\epsilon^2}\right]}
               {(1-\epsilon^2)^{5/2}}\right), \\
   \omega_{\overline{33},33} & =
       -\frac{N}{32\ir}\left(
           \frac{6\sqrt{1-\epsilon^2}
                  +(2+\epsilon^2)\log\left[
                               \frac{2-\epsilon^2-2\sqrt{1-\epsilon^2}}
                                    {\epsilon^2}\right]}
               {(1-\epsilon^2)^{5/2}}\right).
 \end{split}
\end{equation}
All other components can either be determined from symmetry and
reality or are zero. The component $\omega_{\overline{33},33}$ has a
$\log\epsilon$ singularity at $\epsilon=0$. The other non-zero
components have non-analytic $\epsilon^n\log\epsilon$ behaviour, but
are finite. These are mild singularities that can be integrated
over.

\section{Holomorphic surfaces that touch the unit sphere}
\label{touching}

In this subsection we will demonstrate that $\omega_\mathrm{WZ}$
vanishes at every point on the boundary of $H$.

Consider any polynomial $P(z)$ on the boundary of $H$. It follows
that the surface $P(z)=0$ intersects the unit sphere on some
submanifold $M'$, but that there exist infinitesimal deformations of
$P(z)$ under which this intersection goes to zero. In this
subsection we will demonstrate that the submanifold $M'$ cannot have
dimension $3$ (i.e.\ is at most 2 dimensional). As
$\omega_\mathrm{WZ}$ is obtained by integrating a well behaved 3
form over the intersection $M'$, it then follows that
$\omega_\mathrm{WZ}$ vanishes at the boundaries of $H$.

Consider $\R^6$ equipped with the standard inner product structure
defined by
\begin{equation*}
\Vert (x_1,\cdots , x_6)\Vert^2 \, =\, \sum_{i=1}^6 x^2_i\, .
\end{equation*}
Let $S^5\, =\, \{(x_1,\cdots , x_6)\, \mid\,\sum_{i=1}^6 x^2_i=1\}$
be the five--dimensional sphere.

Identify $\R^6$ with $\C^3$ by sending $(x_1,\cdots , x_6)$ to
$(x_1+\ir x_2\, , x_3+\ir x_4\, , x_5+\ir x_6)$. Using this
identification, the operation of multiplication by $\ir$ on $\C^3$
gives a linear operator $J$ on $\R^6$. We have $J^2 \, =\,
-\text{Id}$, and $J$ is orthogonal with respect to the standard
inner product.

Let $N$ denote the section of the normal bundle of $S^5 \, \subset\,
\R^6$ that sends any $\underline{x}\, \in\, S^5$ to the element
$\underline{x}\, \in\, T_{\underline{x}} \R^6\, =\, \R^6$.
Therefore, $N$ is the unit normal vector field that points outwards.

Since $J$ is orthogonal, it follows that $J(N)$ is a vector field on
the manifold $S^5$. Let $\omega$ denote the $C^\infty$ one--form on
$S^5$ defined as follows: for any point $\underline{x}\, \in\, S^5$
and any tangent vector $v\, \in\, T_{\underline{x}} S^5$,
\begin{equation}\label{om}
\omega (v)\, :=\, \langle v\, , J(N (\underline{x}))\rangle\, .
\end{equation}
Let
\begin{equation*}
\mathcal{F}\, :=\, \text{kernel}(\omega)\, \subset\, TS^5
\end{equation*}
be the distribution on $S^5$. So, $\mathcal{F}$ is a $C^\infty$
subbundle of $TS^5$ whose fibre over any $\underline{x}\, \in\, S^5$
is $J(N (\underline{x}))^\perp\bigcap T_{\underline{x}}S^5$.

We will show that for any point $\underline{x}\, \in\, S^5$, the
subspace $\mathcal{F}_{\underline{x}}\, \subset\, T_{\underline{x}}
\R^6\, =\, \R^6$ is closed under the operator $J$. Take any $v\,
\in\, \mathcal{F}_{\underline{x}}$. Since $J$ is orthogonal,
\begin{equation*}
\langle J(v)\, , N(\underline{x})\rangle\, =\, \langle J^2(v)\, ,
J(N(\underline{x}))\rangle\, =\, -\langle v\, ,
J(N(\underline{x}))\rangle\, =\, 0\, ,
\end{equation*}
and $\langle J(v)\, , J(N(\underline{x}))\rangle\, =\, \langle v\, ,
N(\underline{x})\rangle\, =\, 0$. Therefore, $J(v)\, \in\,
\mathcal{F}_{\underline{x}}$, hence the subspace
$\mathcal{F}_{\underline{x}}$ is closed under $J$.

Consider the two--form $\dr\omega$ on $S^5$, where $\omega$ is
defined in \eqref{om}.

\noindent \textbf{Claim.}\, For any point $\underline{x}\, \in\,
S^5$, the restriction of $\dr\omega$ to
$\mathcal{F}_{\underline{x}}$ is a symplectic form.

To prove the claim, consider $U(3)$ as a subgroup of $\text{SO}(6)$
using the identification of $\R^6$ with $\C^3$. This subgroup
commutes with $J$. Since $\omega$ is invariant under the action of
$U(3)$, and $U(3)$ acts transitively on $S^5$, it is enough to check
the above claim for one point of $S^5$. Take $\underline{y}\, =\,
(0,0,0,0,0,1)$. Then $J(N(\underline{y}))\, =\, (0,0,0,0,-1,0)$, and
\begin{equation}\label{is.}
\mathcal{F}_{\underline{y}}\, =\, \R^4\,:= \, (\R, \R,\R,\R,0,0) \,
\subset\, \R^6\,=\, T_{\underline{y}}\R^6\, .
\end{equation}

Consider the one--form $\widetilde{\omega}$ on $\R^6$ that sends any
tangent vector $v\, \in\, T_{\underline{x}}$ to $\langle v\,
,\underline{x}\rangle$. Therefore, $\omega$ is the restriction of
$\widetilde{\omega}$ to $S^5$. The restriction of
$\dr\widetilde{\omega}$ to the subspace
$\mathcal{F}_{\underline{y}}\, =\, \R^4$ in \eqref{is.} coincides
with the symplectic form $2(\dr x_1\bigwedge \dr x_2 +\dr
x_3\bigwedge \dr x_4)$ on $\R^4$.

Therefore, for any $\underline{x}\, \in\, S^5$, the restriction of
$\dr\omega$ to $\mathcal{F}_{\underline{x}}$ is a symplectic form.
Hence the restriction of $\dr\omega\bigwedge \dr\omega$ to
$\mathcal{F}_{\underline{x}}$ is nonzero. Since $\mathcal{F}$ is, by
definition, the kernel of $\omega$, this implies that
\begin{equation}\label{th.}
\theta\, :=\, \omega\wedge \dr\omega\wedge \dr\omega
\end{equation}
is a nowhere vanishing top form on $S^5$.

Let $M$ be a submanifold of $S^5$ such that for each point $m\,
\in\, M$ we have
\begin{equation}\label{inc}
T_m M\, \subset\, \mathcal{F}_m \, \subset\, T_mS^5\, .
\end{equation}
We will show that
\begin{equation*}
\dim M \, \leq\, 2\, .
\end{equation*}

To prove this by contradiction, assume that $\dim M \, =\, 3$. Fix a
point $m_0\, \in\, M$. Let $(t_1,t_2,t_3,t_4,t_5)$, $-\epsilon <t_i
<\epsilon$, be smooth coordinates on $S^5$ defined on an open subset
$U$ containing $m_0$ such that
\begin{equation*}
M\cap U \, =\, \{(t_1,t_2,t_3,t_4,t_5)\, \mid\, t_4\,=\,0 \,=\,
t_5\}\, .
\end{equation*}
Let $\omega\vert_U\, =\, \sum_{i=1}^5 f_i{\rm d}t_i$, where $\omega$ is
the form defined in \eqref{om}, and $f_i$ are some smooth functions
on $U$. From \eqref{inc} and the definition of $\omega$ it follows
that $f_1$, $f_2$ and $f_3$ vanish on $M\bigcap U$.

Since
\begin{equation*}
(\dr\omega)\vert_U\, =\, \sum_{i=1}^5 \sum_{j=1}^5
\frac{\partial{f_i}}{\partial{t_j}}\:  \dr t_j\wedge \dr t_i\, ,
\end{equation*}
and $f_k$, $1\leq k\leq 3$, vanish on $M\bigcap U$, at any point
$m\, \in\, M\bigcap U$,
\begin{equation*}
(\dr\omega) (m)\, =\, \sum_{i=1}^3\sum_{j=4}^5
\frac{\partial{f_i}}{\partial{t_j}}(m)\:  \dr t_j\wedge \dr t_i +
\sum_{i=4}^5\sum_{j=1}^5 \frac{\partial{f_i}}{\partial{t_j}}(m)\:
\dr t_j\wedge \dr t_i\, .
\end{equation*}
From this it follows that the form $\theta$ defined in \eqref{th.}
vanishes at $m$. Indeed, both $\omega (m)$ and $\dr\omega (m)$ are
contained in the ideal generated by $\dr t_4$ and $\dr t_5$ (each
term in their expression contains either $\dr t_4$ or $\dr t_5$).
Hence
\begin{equation*}
\theta(m)\, =\, \omega(m)\wedge (\dr\omega)(m)\wedge
(\dr\omega)(m)\, =\, 0\, .
\end{equation*}

This contradicts the fact that the five--form $\theta$ is nowhere
vanishing. Hence $\dim M \, \leq\, 2$.

If $Z$ is the zero set of a complex polynomial in three variables
such that it can be made disjoint from $S^5$ by arbitrarily small
perturbations, then for any $x\, \in\, Z\bigcap S^5$ the
intersection of $Z$ and $S^5$ is not transversal. This means that
the inclusion
\begin{equation*}
T_x Z \, \subset\, \mathcal{F}_x
\end{equation*}
holds. Hence by the above, $\dim_\R Z\bigcap S^5\, <\, 3$.

\subsection{$\theta_{\rm BI}$ on the boundary}

In \S\S\ref{bound} we have argued that $\theta_{\rm BI}$ vanishes
on the boundary of solution space in a distributional sense. Here we
study the restriction of $\theta_{\rm BI}$ to the boundary in more
detail; in particular we will see that it is zero only as a current
and not pointwise.

The generic surface $P(z)=0$ cuts the unit five sphere (at a
`nonzero angle' on a 3 surface; the the volume $\delta V$ of such a
surface contained within a $\delta r$ shell of the unit sphere is,
consequently, of order $\delta r$.  Consequently contraction of
$\theta_{\rm BI}$ with an arbitrary vector is generically finite and
nonzero. Now let us turn to surfaces $P(z)=0$ where $\rho(P(z))=1$.
Such surfaces are different from generic in two important ways.
First they touch the unit 5-sphere on a $p$ dimensional surface with
$p \leq 2$. Second they touch the unit 5 sphere at `zero angle'
(rather than cutting it at a finite angle); as a consequence the 4
volume contained within the shell of thickness $\delta r$ around the
unit sphere is of order $(\delta r)^{\frac{4 -p}{2}}$.

It follows that the restriction of $\theta_\mathrm{BI}$ to the
boundary of the hole vanishes at those boundary points where the
intersection is zero or one dimensional. We also note that the
boundary surfaces whose intersection with the unit sphere is 2
dimensional form low dimensional submanifolds on the space of all
boundary polynomials (see Appendix \ref{degtwoapp} for an example).
As a consequence $\omega_{\rm BI}$ vanishes as a current on the
boundary, which means that its integral against any genuine form
vanishes.

\section{Symplectic form for linear functions}\label{linear}

\subsection{Coordinates and parametrization}

In this appendix we study the symplectic form for linear
polynomials:
\begin{equation*}
c_i z^i-1=0\,.
\end{equation*}
$U(3)$ rotations may be used to rotate this Polynomial into the
curve
\begin{equation}\label{rotlincurve}
c_0z-1=0\,,
\end{equation}
where $|c_0|^2=|a|^2+|b|^2+|c|^2$. The intersection of
\eqref{rotlincurve} with the unit 5 sphere is a three sphere of
squared radius $1-1/|c_0|^2$. Note that \eqref{rotlincurve} fails to
intersect the unit sphere for $|c_0|^2<1$. Using, for instance
\eqref{enofgrav} we find that the energy of this giant graviton is
\begin{equation}\label{energy}
E=N\left(1-\frac{1}{|c_0|^2}\right).
\end{equation}
A formula that, of course, is valid only for $|c_0|>1$.

We will now determine the symplectic form on the space of linear
polynomials. Using $U(3)$ invariance it will be sufficient to
consider the neighbourhood of the special curve \eqref{rotlincurve}.
Let $\sigma^i$ ($i =1 \ldots 3$) represent any three coordinates on
a unit $S^3$ embedded in $\C^2$ (we will never need to specify what
these are) such that $x=x^0 (\sigma^i)$ and $y=y^0(\sigma^i)$. We
will choose to parameterize points on the target space $S^5$ (this
is our choice of target space variables $x^\mu$ in \eqref{indmet})
by $z$ together with the three coordinates $\sigma^i$, in terms of
which the embedding $\C^3$ coordinates are given by
\begin{equation}\label{fluctuationcurve}
z=r\e^{\ir\theta}\,, \;\;\; (x,y)=\sqrt{1-r^2}(x_0,y_0)\, .
\end{equation}
The target space metric $(G_{\mu \nu}$ in \eqref{indmet}) takes the
form
\begin{equation}\label{sphmet}
\begin{split}\
\dr s^2 &= -\dr t^2
          +\frac{\dr r^2}{ 1-r^2}
          +r^2 \dr\phi^2
          +(1-r^2)(\dr S^3)^2  \\
&= -\dr t^2
   +\frac{\zb^2\dr z^2 +2(2-z\zb)\dr z\dr\zb +z^2\dr\zb^2}
         {4(1-z\zb)}
   + (1-z\zb)(|\dr x_0|^2+|\dr y_0|^2) \, .
\end{split}
\end{equation}
We choose $t, \sigma^i$ as world volume coordinates on the brane
$(\sigma^\alpha$ in \eqref{indmet}). Curves in the neighbourhood of
\eqref{rotlincurve} may be characterized by specifying the function
$z=z(\sigma^i, t)$ (intuitively, $z$ parameterizes the two
transverse fluctuations of the brane away from its ambient $S^3$
shape). The fluctuation $z(\sigma^i, t)$ corresponding to linear
polynomials \eqref{compcurve} is given, to first order in $\delta a,
\delta b$ and $\delta c$ by
\begin{equation}\label{lindelta}
\begin{split}
z&=\frac{\e^{\ir t}}{c_0} -\left( \frac{\e^{\ir t}}{c_0^2}\delta c
    + \frac{\sqrt{1-z\zb}}{c_0}(x_0\delta a+y_0\delta b) \right)
    +\CO(2)\,,\\
{\dot z} &=\frac{\ir\e^{\ir t}}{ c_0}
           +\CO(1)\,,
\\
\p_i z &= \CO(1)\,.
\end{split}
\end{equation}
equation
\eqref{compcurve} together with \eqref{indmet} and  \eqref{sphmet}
may then be used to determine the induced metric on the world volume
of the brane; at $t=0$ and to zeroth order in fluctuations we find :
\begin{equation}\label{indmetfo}
\begin{split} g_{ij} &= (1-r^2)g^{S^3}_{ij} + \CO(2)\,,\\
g_{0i} &= \CO(1)\,, \\
g_{00} &= -(1-r^2)
          + \CO(1)\,.
\end{split}
\end{equation}

\subsection{$\omega_\mathrm{BI}$}

Let $\omega_\mathrm{BI}=\dr\theta_\mathrm{BI}$ denote the
contribution of the Born-Infeld term in the action to the symplectic
form. We have
\begin{equation}\label{bicalc}
\begin{split}
\theta_\mathrm{BI} =& \frac{N}{2\pi^2}\int\!\!\dr^3\sigma\,
        (p_z \,\delta z  + p_\zb \,\delta\zb )\,,
\\ \text{where}\quad
p_z &= \sqrt{-g}g^{00}
      ({\dot z}G_{zz} + {\dot \zb}G_{z\zb})
\\
    &= -\left( {\dot z}\frac{\zb^2}{4}
            + {\dot\zb}\frac{2-z\zb}{4} \right).
\end{split}
\end{equation}
We have dropped factors of $\sqrt{g^{S^3}}$, absorbing them into the
integration measure $\dr^3\sigma$.

Using \eqref{lindelta} and the integrals:
\begin{equation}\label{sphints}
\int\!\!\dr^3\sigma\,1 = 2\pi^2\,,\;\;\;\;\;
\int\!\!\dr^3\sigma\,|x_0|^2=\int\!\!\dr^3\sigma\,|y_0|^2 = \pi^2\,,
\end{equation}
with all other linear and quadratic integrals evaluating to zero, we
find
\begin{equation*}
    \theta_\mathrm{BI}=N\left(\frac{1}{|c_0|^4}-\frac{1}{|c_0|^6}\right)
                       \frac{\cb_0\,\delta c-c_0\,\delta \cb}{2\ir}\,.
\end{equation*}

We can find $\theta_\mathrm{BI}$ at a general point with the replacement
\begin{equation}\label{invtrep}
\cb_0\,\delta c \rightarrow
    \cb^i \,\dr c_i\,,\;\;\;
c_0\,\delta \cb \rightarrow
    c_i \,\dr \cb^i\,,\;\;\;
    |c_0|^2 \rightarrow \cb^ic_i
     \equiv |c|^2 \,,
\end{equation}
yielding
\begin{equation}\label{linthetaBI}
  \theta_\mathrm{BI}=
     N\left(\frac{1}{|c|^4}-\frac{1}{|c|^6}\right)
         \frac{\cb^i \,\dr c_i - c_i \,\dr \cb^i}{2\ir}\,.
\end{equation}
Taking the exterior derivative gives
\begin{equation}\label{linomegaBI}
\omega_\mathrm{BI}=2N\left[
\left(\frac{1}{|c|^4}-\frac{1}{|c|^6}\right)
   \frac{\dr\cb^i\wedge\dr c_i}{2\ir}
+\left(\frac{3}{|c|^6}-\frac{2}{|c|^4}\right)
   \frac{\cb^ic_j}{|c|^2}\frac{\dr\cb^j\wedge\dr c_i}{2\ir}
\right].
\end{equation}

Note that
\begin{equation}\label{omegarewrite}
 \omega_\mathrm{BI} =f_\mathrm{BI}(|c|^2) \frac{\dr\cb^i\wedge\dr c_i}{2\ir}
           + f_\mathrm{BI}'(|c|^2){\cb^ic_j}\frac{\dr\cb^j\wedge\dr c_i }{2\ir}\:;
\qquad f_\mathrm{BI}(|c|^2) =
{2N}\left(\frac{1}{|c|^4}-\frac{1}{|c|^6}\right).
\end{equation}

\subsection{$\omega_\mathrm{WZ}$}

Using the metric \eqref{sphmet}, the volume of the giant graviton is
$2\pi^2(1-r^2)^{3/2}$. To find the volume swept out when it is
deformed, we need only consider the deformations perpendicular to
its surface. About our special point in parameter space, this gives:
\begin{equation*}
\omega_\mathrm{WZ}=\frac{2N}{\pi^2}\int\!\!\dr^3\sigma\:
                      (1-r^2)r\delta r\wedge\delta\theta
= \frac{2N}{\pi^2}\int\!\!\dr^3\sigma\:
        (1-z\zb)\frac{\delta\zb\wedge \delta z}{2\ir}\:.
\end{equation*}
If we use \eqref{lindelta} and \eqref{sphints}, this becomes:
\begin{equation*}
\omega_\mathrm{WZ} = \frac{4N}{|c_0|^4}
\left(1-\frac{1}{|c_0|^2}\right)\frac{\delta\cb\wedge\delta c}{2\ir}
+\frac{2N}{|c_0|^2} \left(1-\frac{1}{|c_0|^2}\right)^2
   \left(\frac{\delta\ab\wedge\delta a}{2\ir}
        +\frac{\delta\bb\wedge\delta b}{2\ir}\right).
\end{equation*}
Making the replacement
\begin{equation}\label{invtrep2}
\delta {\bar c}\wedge\delta c \rightarrow
    \frac{\cb^ic_j}{|c|^2}\dr\cb^j\wedge\dr c_i\,,\qquad
\delta \ab\wedge\delta a+\delta \bb\wedge\delta b \rightarrow
      \dr\cb^i\wedge\dr c_i -\frac{\cb^ic_j}{|c|^2}\dr\cb^j\wedge\dr c_i\,,
\end{equation}
we get the Wess-Zumino contribution to the symplectic form at an
arbitrary point:
\begin{equation}\label{linomegaCS}
\omega_\mathrm{WZ} =2N\left[
\left(\frac{1}{|c|^2}-\frac{2}{|c|^4}+\frac{1}{|c|^6}\right)
   \frac{\dr\cb^i\wedge\dr c_i}{2\ir}
-\left(\frac{1}{|c|^2}-\frac{4}{|c|^4}+\frac{3}{|c|^6}\right)
   \frac{\cb^ic_j}{|c|^2}\frac{\dr\cb^j\wedge\dr c_i}{2\ir}
\right].
\end{equation}
The analogue of \eqref{omegarewrite} also applies to this case upon
defining
\begin{equation}\label{fff}
f_\mathrm{WZ}= {2N}
\left(\frac{1}{|c|^2}-\frac{2}{|c|^4}+\frac{1}{|c|^6}\right) .
\end{equation}

\subsection{$\omega_\mathrm{full}$}

Adding together \eqref{linomegaBI} and \eqref{linomegaCS} gives
\begin{equation}\label{linomegafull}
\omega_\mathrm{full} =2N\left[
\left(\frac{1}{|c|^2}-\frac{1}{|c|^4}\right)
   \frac{\dr\cb^i\wedge\dr c_i}{ 2\ir}
-\left(\frac{1}{|c|^2}-\frac{2}{|c|^4}\right)
   \frac{\cb^ic_j}{|c|^2}\frac{\dr\cb^j\wedge\dr c_i}{ 2\ir}
\right].
\end{equation}
Defining
\begin{equation}\label{ffull}
f_\mathrm{full}={2N} \left(\frac{1}{|c|^2}-\frac{1}{|c|^4}\right),
\end{equation}
the analogue of \eqref{omegarewrite} applies.

\bibliographystyle{utcaps}
\bibliography{gg}

\providecommand{\href}[2]{#2}\begingroup\raggedright\begin{thebibliography}{10}

\bibitem{Sohnius:1981sn}
M.~F. Sohnius and P.~C. West, ``Conformal invariance in N=4 supersymmetric
  Yang-Mills theory,'' {\em Phys. Lett.} {\bf B100} (1981)
245.

\bibitem{Howe:1983sr}
P.~S. Howe, K.~S. Stelle, and P.~K. Townsend, ``Miraculous ultraviolet
  cancellations in supersymmetry made manifest,'' {\em Nucl. Phys.} {\bf B236}
  (1984)
125.

\bibitem{Brink:1982pd}
L.~Brink, O.~Lindgren, and B.~E.~W. Nilsson, ``N=4 Yang-Mills theory on the
  light cone,'' {\em Nucl. Phys.} {\bf B212} (1983)
401.

\bibitem{Seiberg:1988ur}
N.~Seiberg, ``Supersymmetry and nonperturbative beta functions,'' {\em Phys.
  Lett.} {\bf B206} (1988)
75.

\bibitem{Montonen:1977sn}
C.~Montonen and D.~I. Olive, ``Magnetic monopoles as gauge particles?'' {\em
  Phys. Lett.} {\bf B72} (1977)
117.

\bibitem{Goddard:1976qe}
P.~Goddard, J.~Nuyts, and D.~I. Olive, ``Gauge theories and magnetic charge,''
  {\em Nucl. Phys.} {\bf B125} (1977)
1.

\bibitem{Osborn:1979tq}
H.~Osborn, ``Topological charges for N=4 supersymmetric gauge theories and
  monopoles of spin 1,'' {\em Phys. Lett.} {\bf B83} (1979)
321.

\bibitem{Sen:1994yi}
A.~Sen, ``Dyon - monopole bound states, selfdual harmonic forms on the multi -
  monopole moduli space, and SL(2,Z) invariance in string theory,'' {\em Phys.
  Lett.} {\bf B329} (1994) 217--221,
\href{http://arXiv.org/abs/hep-th/9402032}{{\tt hep-th/9402032}}.

\bibitem{Maldacena:1997re}
J.~M. Maldacena, ``The large N limit of superconformal field theories and
  supergravity,'' {\em Adv. Theor. Math. Phys.} {\bf 2} (1998) 231--252,
\href{http://arXiv.org/abs/hep-th/9711200}{{\tt hep-th/9711200}}.

\bibitem{Gubser:1998bc}
S.~S. Gubser, I.~R. Klebanov, and A.~M. Polyakov, ``Gauge theory correlators
  from non-critical string theory,'' {\em Phys. Lett.} {\bf B428} (1998)
  105--114,
\href{http://arXiv.org/abs/hep-th/9802109}{{\tt hep-th/9802109}}.

\bibitem{Witten:1998qj}
E.~Witten, ``Anti-de Sitter space and holography,'' {\em Adv. Theor. Math.
  Phys.} {\bf 2} (1998) 253--291,
\href{http://arXiv.org/abs/hep-th/9802150}{{\tt hep-th/9802150}}.

\bibitem{Aharony:1999ti}
O.~Aharony, S.~S. Gubser, J.~M. Maldacena, H.~Ooguri, and Y.~Oz, ``Large N
  field theories, string theory and gravity,'' {\em Phys. Rept.} {\bf 323}
  (2000) 183--386,
\href{http://arXiv.org/abs/hep-th/9905111}{{\tt hep-th/9905111}}.

\bibitem{Minahan:2002ve}
J.~A. Minahan and K.~Zarembo, ``The Bethe-ansatz for N = 4 super Yang-Mills,''
  {\em JHEP} {\bf 03} (2003) 013,
\href{http://arXiv.org/abs/hep-th/0212208}{{\tt hep-th/0212208}}.

\bibitem{Beisert:2003yb}
N.~Beisert and M.~Staudacher, ``The N = 4 SYM integrable super spin chain,''
  {\em Nucl. Phys.} {\bf B670} (2003) 439--463,
\href{http://arXiv.org/abs/hep-th/0307042}{{\tt hep-th/0307042}}.

\bibitem{Beisert:2004yq}
N.~Beisert, ``Higher-loop integrability in N = 4 gauge theory,'' {\em Comptes
  Rendus Physique} {\bf 5} (2004) 1039--1048,
\href{http://arXiv.org/abs/hep-th/0409147}{{\tt hep-th/0409147}}.

\bibitem{Staudacher:2004tk}
M.~Staudacher, ``The factorized S-matrix of CFT/AdS,'' {\em JHEP} {\bf 05}
  (2005) 054,
\href{http://arXiv.org/abs/hep-th/0412188}{{\tt hep-th/0412188}}.

\bibitem{Beisert:2005tm}
N.~Beisert, ``The su$(2|2)$ dynamic S-matrix,''
\href{http://arXiv.org/abs/hep-th/0511082}{{\tt hep-th/0511082}}.

\bibitem{Janik:2006dc}
R.~A. Janik, ``The AdS(5) x S**5 superstring worldsheet S-matrix and crossing
  symmetry,'' {\em Phys. Rev.} {\bf D73} (2006) 086006,
\href{http://arXiv.org/abs/hep-th/0603038}{{\tt hep-th/0603038}}.

\bibitem{Dolan:2002zh}
F.~A. Dolan and H.~Osborn, ``On short and semi-short representations for four
  dimensional superconformal symmetry,'' {\em Ann. Phys.} {\bf 307} (2003)
  41--89,
\href{http://arXiv.org/abs/hep-th/0209056}{{\tt hep-th/0209056}}.

\bibitem{Ryzhov:2001bp}
A.~V. Ryzhov, ``Quarter BPS operators in N = 4 SYM,'' {\em JHEP} {\bf 11}
  (2001) 046,
\href{http://arXiv.org/abs/hep-th/0109064}{{\tt hep-th/0109064}}.

\bibitem{D'Hoker:2003vf}
E.~D'Hoker, P.~Heslop, P.~Howe, and A.~V. Ryzhov, ``Systematics of quarter BPS
  operators in N = 4 SYM,'' {\em JHEP} {\bf 04} (2003) 038,
\href{http://arXiv.org/abs/hep-th/0301104}{{\tt hep-th/0301104}}.

\bibitem{Beisert:2003tq}
N.~Beisert, C.~Kristjansen, and M.~Staudacher, ``The dilatation operator of N =
  4 super Yang-Mills theory,'' {\em Nucl. Phys.} {\bf B664} (2003) 131--184,
\href{http://arXiv.org/abs/hep-th/0303060}{{\tt hep-th/0303060}}.

\bibitem{index}
J.~Kinney, J.~Maldacena, S.~Minwalla, and S.~Raju, ``An index for 4 dimensional
  super conformal theories,''
\href{http://arXiv.org/abs/hep-th/0510251}{{\tt hep-th/0510251}}.

\bibitem{Myers:1999ps}
R.~C. Myers, ``Dielectric-branes,'' {\em JHEP} {\bf 12} (1999) 022,
\href{http://arXiv.org/abs/hep-th/9910053}{{\tt hep-th/9910053}}.

\bibitem{McGreevy:2000cw}
J.~McGreevy, L.~Susskind, and N.~Toumbas, ``Invasion of the giant gravitons
  from anti-de Sitter space,'' {\em JHEP} {\bf 06} (2000) 008,
\href{http://arXiv.org/abs/hep-th/0003075}{{\tt hep-th/0003075}}.

\bibitem{Grisaru:2000zn}
M.~T. Grisaru, R.~C. Myers, and O.~Tafjord, ``SUSY and Goliath,'' {\em JHEP}
  {\bf 08} (2000) 040,
\href{http://arXiv.org/abs/hep-th/0008015}{{\tt hep-th/0008015}}.

\bibitem{Hashimoto:2000zp}
A.~Hashimoto, S.~Hirano, and N.~Itzhaki, ``Large branes in AdS and their field
  theory dual,'' {\em JHEP} {\bf 08} (2000) 051,
\href{http://arXiv.org/abs/hep-th/0008016}{{\tt hep-th/0008016}}.

\bibitem{mikhailov}
A.~Mikhailov, ``Giant gravitons from holomorphic surfaces,'' {\em JHEP} {\bf
  11} (2000) 027,
\href{http://arXiv.org/abs/hep-th/0010206}{{\tt hep-th/0010206}}.

\bibitem{Beasley}
C.~E. Beasley, ``BPS branes from baryons,'' {\em JHEP} {\bf 11} (2002) 015,
\href{http://arXiv.org/abs/hep-th/0207125}{{\tt hep-th/0207125}}.

\bibitem{Mandal:2005wv}
G.~Mandal, ``Fermions from half-BPS supergravity,'' {\em JHEP} {\bf 08} (2005)
  052,
\href{http://arXiv.org/abs/hep-th/0502104}{{\tt hep-th/0502104}}.

\bibitem{Grant:2005qc}
L.~Grant, L.~Maoz, J.~Marsano, K.~Papadodimas, and V.~S. Rychkov,
  ``Minisuperspace quantization of `bubbling AdS' and free fermion droplets,''
  {\em JHEP} {\bf 08} (2005) 025,
\href{http://arXiv.org/abs/hep-th/0505079}{{\tt hep-th/0505079}}.

\bibitem{Maoz:2005nk}
L.~Maoz and V.~S. Rychkov, ``Geometry quantization from supergravity: The case
  of `bubbling AdS','' {\em JHEP} {\bf 08} (2005) 096,
\href{http://arXiv.org/abs/hep-th/0508059}{{\tt hep-th/0508059}}.

\bibitem{Nemani}
G.~Mandal and N.~V. Suryanarayana, ``Counting 1/8-BPS dual-giants,''
\href{http://arXiv.org/abs/hep-th/0606088}{{\tt hep-th/0606088}}.

\bibitem{Kim:2005mw}
S.~Kim and K.~Lee, ``BPS electromagnetic waves on giant gravitons,'' {\em JHEP}
  {\bf 10} (2005) 111,
\href{http://arXiv.org/abs/hep-th/0502007}{{\tt hep-th/0502007}}.

\bibitem{fuzzy}
G.~Mandal, S.~R. Wadia, and K.~Yogendran, ``Topology change in fuzzy branes.''
  TIFR/TH/01-32, \emph{private communication}.

\bibitem{Berenstein:2005aa}
D.~Berenstein, ``Large N BPS states and emergent quantum gravity,'' {\em JHEP}
  {\bf 01} (2006) 125,
\href{http://arXiv.org/abs/hep-th/0507203}{{\tt hep-th/0507203}}.

\bibitem{Berenstein:2005ek}
D.~Berenstein and D.~H. Correa, ``Emergent geometry from q-deformations of N =
  4 super Yang-Mills,''
\href{http://arXiv.org/abs/hep-th/0511104}{{\tt hep-th/0511104}}.

\bibitem{Berenstein:2005jq}
D.~Berenstein, D.~H. Correa, and S.~E. Vazquez, ``All loop BMN state energies
  from matrices,'' {\em JHEP} {\bf 02} (2006) 048,
\href{http://arXiv.org/abs/hep-th/0509015}{{\tt hep-th/0509015}}.

\bibitem{Berenstein:2006yy}
D.~Berenstein and R.~Cotta, ``Aspects of emergent geometry in the AdS/CFT
  context,''
\href{http://arXiv.org/abs/hep-th/0605220}{{\tt hep-th/0605220}}.

\bibitem{Gutowski:2004ez}
J.~B. Gutowski and H.~S. Reall, ``Supersymmetric AdS(5) black holes,'' {\em
  JHEP} {\bf 02} (2004) 006,
\href{http://arXiv.org/abs/hep-th/0401042}{{\tt hep-th/0401042}}.

\bibitem{Gutowski:2004yv}
J.~B. Gutowski and H.~S. Reall, ``General supersymmetric AdS(5) black holes,''
  {\em JHEP} {\bf 04} (2004) 048,
\href{http://arXiv.org/abs/hep-th/0401129}{{\tt hep-th/0401129}}.

\bibitem{Chong:2005da}
Z.~W. Chong, M.~Cvetic, H.~Lu, and C.~N. Pope, ``Five-dimensional gauged
  supergravity black holes with independent rotation parameters,'' {\em Phys.
  Rev.} {\bf D72} (2005) 041901,
\href{http://arXiv.org/abs/hep-th/0505112}{{\tt hep-th/0505112}}.

\bibitem{Chong:2005hr}
Z.~W. Chong, M.~Cvetic, H.~Lu, and C.~N. Pope, ``General non-extremal rotating
  black holes in minimal five- dimensional gauged supergravity,'' {\em Phys.
  Rev. Lett.} {\bf 95} (2005) 161301,
\href{http://arXiv.org/abs/hep-th/0506029}{{\tt hep-th/0506029}}.

\bibitem{Kunduri:2006ek}
H.~K. Kunduri, J.~Lucietti, and H.~S. Reall, ``Supersymmetric multi-charge
  AdS(5) black holes,'' {\em JHEP} {\bf 04} (2006) 036,
\href{http://arXiv.org/abs/hep-th/0601156}{{\tt hep-th/0601156}}.

\bibitem{Berkooz:2006wc}
M.~Berkooz, D.~Reichmann, and J.~Simon, ``A Fermi surface model for large
  supersymmetric AdS(5) black holes,''
\href{http://arXiv.org/abs/hep-th/0604023}{{\tt hep-th/0604023}}.

\bibitem{gsw2}
M.~B. Green, J.~H. Schwarz, and E.~Witten, {\em Superstring Theory}, vol.~2,
  {\em Loop amplitudes, anomalies and phenomenology}.
\newblock Cambridge University Press, 1987.

\bibitem{Demailly}
J.-P. Demailly, ``Complex Analytic and Differential Geometry.''
  \href{http://www-fourier.ujf-grenoble.fr/~demailly/books.html}{http://www-fo%
urier.ujf-grenoble.fr/$\sim$demailly/books.html}.

\bibitem{Crnkovic:1986ex}
C.~Crnkovic and E.~Witten, ``Covariant description of canonical formalism in
  geometrical theories,'' in {\em Three hundred years of gravitation},
  S.~Hawking and W.~Israel, eds., ch.~16, pp.~676--684.
\newblock Cambridge University Press, 1987.
\newblock Print-86-1309 (PRINCETON).

\bibitem{zuckerman}
G.~J. Zuckerman, ``Action principles and global geometry,'' in {\em
  Mathematical aspects of string theory}, S.-T. Yau, ed., p.~259.
\newblock World Scientific, 1987.

\bibitem{Lee:1990nz}
J.~Lee and R.~M. Wald, ``Local symmetries and constraints,'' {\em J. Math.
  Phys.} {\bf 31} (1990)
725--743.

\bibitem{Milnor}
J.~W. Milnor, {\em Lectures on the h-Cobordism Theorem}.
\newblock Princeton University Press, 1965.

\bibitem{Simons:2004nm}
A.~Simons, A.~Strominger, D.~M. Thompson, and X.~Yin, ``Supersymmetric branes
  in AdS(2) x S**2 x CY(3),'' {\em Phys. Rev.} {\bf D71} (2005) 066008,
\href{http://arXiv.org/abs/hep-th/0406121}{{\tt hep-th/0406121}}.

\bibitem{Gaiotto:2004pc}
D.~Gaiotto, A.~Simons, A.~Strominger, and X.~Yin, ``D0-branes in black hole
  attractors,''
\href{http://arXiv.org/abs/hep-th/0412179}{{\tt hep-th/0412179}}.

\bibitem{Gaiotto:2004ij}
D.~Gaiotto, A.~Strominger, and X.~Yin, ``Superconformal black hole quantum
  mechanics,'' {\em JHEP} {\bf 11} (2005) 017,
\href{http://arXiv.org/abs/hep-th/0412322}{{\tt hep-th/0412322}}.

\bibitem{Das:2000fu}
S.~R. Das, A.~Jevicki, and S.~D. Mathur, ``Giant gravitons, BPS bounds and
  noncommutativity,'' {\em Phys. Rev.} {\bf D63} (2001) 044001,
\href{http://arXiv.org/abs/hep-th/0008088}{{\tt hep-th/0008088}}.

\bibitem{Jevicki:2000ty}
A.~Jevicki, M.~Mihailescu, and S.~Ramgoolam, ``Hidden classical symmetry in
  quantum spaces at roots of unity : From q-sphere to fuzzy sphere,''
\href{http://arXiv.org/abs/hep-th/0008186}{{\tt hep-th/0008186}}.

\bibitem{Das:2000st}
S.~R. Das, A.~Jevicki, and S.~D. Mathur, ``Vibration modes of giant
  gravitons,'' {\em Phys. Rev.} {\bf D63} (2001) 024013,
\href{http://arXiv.org/abs/hep-th/0009019}{{\tt hep-th/0009019}}.

\bibitem{Ho:2000qh}
P.-M. Ho, ``Fuzzy sphere from matrix model,'' {\em JHEP} {\bf 12} (2000) 015,
\href{http://arXiv.org/abs/hep-th/0010165}{{\tt hep-th/0010165}}.

\bibitem{Balasubramanian:2001nh}
V.~Balasubramanian, M.~Berkooz, A.~Naqvi, and M.~J. Strassler, ``Giant
  gravitons in conformal field theory,'' {\em JHEP} {\bf 04} (2002) 034,
\href{http://arXiv.org/abs/hep-th/0107119}{{\tt hep-th/0107119}}.

\bibitem{Corley:2001zk}
S.~Corley, A.~Jevicki, and S.~Ramgoolam, ``Exact correlators of giant gravitons
  from dual N = 4 SYM theory,'' {\em Adv. Theor. Math. Phys.} {\bf 5} (2002)
  809--839,
\href{http://arXiv.org/abs/hep-th/0111222}{{\tt hep-th/0111222}}.

\bibitem{Balasubramanian:2002sa}
V.~Balasubramanian, M.-x. Huang, T.~S. Levi, and A.~Naqvi, ``Open strings from
  N = 4 super Yang-Mills,'' {\em JHEP} {\bf 08} (2002) 037,
\href{http://arXiv.org/abs/hep-th/0204196}{{\tt hep-th/0204196}}.

\bibitem{Berenstein:2004kk}
D.~Berenstein, ``A toy model for the AdS/CFT correspondence,'' {\em JHEP} {\bf
  07} (2004) 018,
\href{http://arXiv.org/abs/hep-th/0403110}{{\tt hep-th/0403110}}.

\bibitem{Lin:2004nb}
H.~Lin, O.~Lunin, and J.~M. Maldacena, ``Bubbling AdS space and 1/2 BPS
  geometries,'' {\em JHEP} {\bf 10} (2004) 025,
\href{http://arXiv.org/abs/hep-th/0409174}{{\tt hep-th/0409174}}.

\bibitem{Dhar:2005qh}
A.~Dhar, ``Bosonization of non-relativstic fermions in 2-dimensions and
  collective field theory,'' {\em JHEP} {\bf 07} (2005) 064,
\href{http://arXiv.org/abs/hep-th/0505084}{{\tt hep-th/0505084}}.

\bibitem{Dhar:2005fg}
A.~Dhar, G.~Mandal, and N.~V. Suryanarayana, ``Exact operator bosonization of
  finite number of fermions in one space dimension,'' {\em JHEP} {\bf 01}
  (2006) 118,
\href{http://arXiv.org/abs/hep-th/0509164}{{\tt hep-th/0509164}}.

\bibitem{Dhar:2005su}
A.~Dhar, G.~Mandal, and M.~Smedback, ``From gravitons to giants,'' {\em JHEP}
  {\bf 03} (2006) 031,
\href{http://arXiv.org/abs/hep-th/0512312}{{\tt hep-th/0512312}}.

\bibitem{Johnson:1999qt}
C.~V. Johnson, A.~W. Peet, and J.~Polchinski, ``Gauge theory and the excision
  of repulson singularities,'' {\em Phys. Rev.} {\bf D61} (2000) 086001,
\href{http://arXiv.org/abs/hep-th/9911161}{{\tt hep-th/9911161}}.

\bibitem{Gauntlett:2004hh}
J.~P. Gauntlett, D.~Martelli, J.~F. Sparks, and D.~Waldram, ``A new infinite
  class of Sasaki-Einstein manifolds,'' {\em Adv. Theor. Math. Phys.} {\bf 8}
  (2006) 987--1000,
\href{http://arXiv.org/abs/hep-th/0403038}{{\tt hep-th/0403038}}.

\bibitem{Gauntlett:2004hs}
J.~P. Gauntlett, D.~Martelli, J.~Sparks, and D.~Waldram, ``Supersymmetric AdS
  backgrounds in string and M-theory,''
\href{http://arXiv.org/abs/hep-th/0411194}{{\tt hep-th/0411194}}.

\bibitem{Martelli:2004wu}
D.~Martelli and J.~Sparks, ``Toric geometry, Sasaki-Einstein manifolds and a
  new infinite class of AdS/CFT duals,'' {\em Commun. Math. Phys.} {\bf 262}
  (2006) 51--89,
\href{http://arXiv.org/abs/hep-th/0411238}{{\tt hep-th/0411238}}.

\bibitem{Benvenuti:2004dy}
S.~Benvenuti, S.~Franco, A.~Hanany, D.~Martelli, and J.~Sparks, ``An infinite
  family of superconformal quiver gauge theories with Sasaki-Einstein duals,''
  {\em JHEP} {\bf 06} (2005) 064,
\href{http://arXiv.org/abs/hep-th/0411264}{{\tt hep-th/0411264}}.

\bibitem{Martelli:2005tp}
D.~Martelli, J.~Sparks, and S.-T. Yau, ``The geometric dual of a-maximisation
  for toric Sasaki- Einstein manifolds,''
\href{http://arXiv.org/abs/hep-th/0503183}{{\tt hep-th/0503183}}.

\bibitem{Martelli:2005wy}
D.~Martelli and J.~Sparks, ``Toric Sasaki-Einstein metrics on S**2 x S**3,''
  {\em Phys. Lett.} {\bf B621} (2005) 208--212,
\href{http://arXiv.org/abs/hep-th/0505027}{{\tt hep-th/0505027}}.

\bibitem{Benvenuti:2005ja}
S.~Benvenuti and M.~Kruczenski, ``From Sasaki-Einstein spaces to quivers via
  BPS geodesics: L(p,q$|$r),'' {\em JHEP} {\bf 04} (2006) 033,
\href{http://arXiv.org/abs/hep-th/0505206}{{\tt hep-th/0505206}}.

\bibitem{Franco:2005sm}
S.~Franco {\em et al.}, ``Gauge theories from toric geometry and brane
  tilings,'' {\em JHEP} {\bf 01} (2006) 128,
\href{http://arXiv.org/abs/hep-th/0505211}{{\tt hep-th/0505211}}.

\bibitem{Butti:2005sw}
A.~Butti, D.~Forcella, and A.~Zaffaroni, ``The dual superconformal theory for
  L(p,q,r) manifolds,'' {\em JHEP} {\bf 09} (2005) 018,
\href{http://arXiv.org/abs/hep-th/0505220}{{\tt hep-th/0505220}}.

\bibitem{Gauntlett:2005ww}
J.~P. Gauntlett, D.~Martelli, J.~Sparks, and D.~Waldram, ``Supersymmetric
  AdS(5) solutions of type IIB supergravity,''
\href{http://arXiv.org/abs/hep-th/0510125}{{\tt hep-th/0510125}}.

\bibitem{Martelli:2006yb}
D.~Martelli, J.~Sparks, and S.-T. Yau, ``Sasaki-Einstein manifolds and volume
  minimisation,''
\href{http://arXiv.org/abs/hep-th/0603021}{{\tt hep-th/0603021}}.

\bibitem{woodhouse}
N.~Woodhouse, {\em Geometric Quantization}.
\newblock Oxford, 1980.

\bibitem{Echeverria-Enriquez:1999jr}
A.~Echeverria-Enriquez, M.~C. Munoz-Lecanda, N.~Roman-Roy, and
  C.~Victoria-Monge, ``Mathematical foundations of geometric quantization,''
  {\em Extracta Math.} {\bf 13} (1998) 135--238,
\href{http://arXiv.org/abs/math-ph/9904008}{{\tt math-ph/9904008}}.

\end{thebibliography}\endgroup

\end{document}